\documentclass[sigconf]{acmart} 
\AtBeginDocument{%
  \providecommand\BibTeX{{%
    Bib\TeX}}}

\copyrightyear{2026}
\acmYear{2026}
\setcopyright{cc}
\setcctype{by-nc-nd}
\acmConference[CCS '26]{Proceedings of the 2026 ACM SIGSAC Conference on Computer and Communications Security}{November 15--19, 2026}{The Hague, Netherlands}
\acmBooktitle{Proceedings of the 2026 ACM SIGSAC Conference on Computer and Communications Security (CCS '26), November 15--19, 2026, The Hague, Netherlands}
\acmDOI{10.1145/3830454.3846617}
\acmISBN{979-8-4007-2871-6/2026/11}
\usepackage{amsmath,amsfonts,amssymb}
\usepackage{algorithm}

\usepackage{hyperref}
\usepackage{booktabs}
\usepackage{diagbox}
\usepackage{pifont}
\usepackage{graphicx}
\newcommand{\TokSet}{\mathsf{TokSet}}
\usepackage{xcolor}
\usepackage{multirow}
\usepackage{multicol}
\usepackage[utf8]{inputenc}
\usepackage{tikz}
\usepackage{textcomp}
\usepackage{float}
\usepackage{tcolorbox}
\usepackage{enumitem}
\usepackage{standalone}
\usepackage{fontawesome5}
\usepackage{algorithmic}
\usepackage{subcaption}
\usepackage{soul}
\usepackage{marvosym}
\usepackage{longtable}

\usepackage[operators,sets]{cryptocode}
\usetikzlibrary{arrows.meta,positioning,fit,calc}
\usetikzlibrary{decorations.pathreplacing}
\usepackage{ulem} 
\def\BibTeX{{\rm B\kern-.05em{\sc i\kern-.025em b}\kern-.08em
    T\kern-.1667em\lower.7ex\hbox{E}\kern-.125emX}}

\mathchardef\mhyphen="2D
\newcommand{\RSP}{\textcolor{rspgray}{\scriptsize [RSP]}}
\newcommand{\RSPA}{\textcolor{rspblue}{\scriptsize [RSP*]}}

\definecolor{rspgray}{RGB}{90,90,90}
\definecolor{rspblue}{RGB}{40,90,170}
\definecolor{zkorange}{RGB}{180,100,20}
\newtheorem{theorem}{\textbf{Theorem}}
\newtheorem{definition}{\textbf{Definition}}

\newcommand{\Alg}[1]{\ensuremath{\mathsf{#1}}}

\newcommand{\Enc}{\Alg{Enc}}

\newcommand{\KDF}{\Alg{KDF}}

\newcommand{\pk}{\ensuremath{\mathit{pk}}}
\newcommand{\sk}{\ensuremath{\mathit{sk}}}

\newcommand{\pid}{\ensuremath{\mathit{PID}}}

\newcommand{\NextLine}[1][1.0]{
	\pgfmathparse{\Y+#1}
	\edef\Y{\pgfmathresult}
}
\newcommand{\linkgame}[2]{\hyperref[#1]{G#2}}

\newcounter{Bdversary}

\newcommand\orcidicon[1]{\href{https://orcid.org/#1}{\mbox{\scalerel*{
\begin{tikzpicture}[yscale=-1,transform shape]
\pic{orcidlogo};
\end{tikzpicture}
}{|}}}}

\newcommand*{\addFileDependency}[1]{
  \typeout{(#1)}

  \IfFileExists{#1}{}{\typeout{No file #1.}}
}

\usepackage{adjustbox}
\usetikzlibrary{calc}
\newcommand{\cmark}{\ding{51}} 
\newcommand{\xmark}{\ding{55}} 
\newcommand{\step}[1]{%
  \tikz[baseline=(n.base)]%
    \node[draw,circle,inner sep=0.6pt,minimum size=2.2ex,font=\scriptsize](n){#1};%
}

\newcommand{\sqcircle}{\tikz[baseline=-0.6ex]\draw[black] (0,0) circle (0.05cm);} 

\usepackage{longtable}
\definecolor{orcidlogocol}{HTML}{A6CE39}
\tikzset{
  orcidlogo/.pic={
    \fill[orcidlogocol] svg{M256,128c0,70.7-57.3,128-128,128C57.3,256,0,198.7,0,128C0,57.3,57.3,0,128,0C198.7,0,256,57.3,256,128z};
    \fill[white] svg{M86.3,186.2H70.9V79.1h15.4v48.4V186.2z}
                 svg{M108.9,79.1h41.6c39.6,0,57,28.3,57,53.6c0,27.5-21.5,53.6-56.8,53.6h-41.8V79.1z M124.3,172.4h24.5c34.9,0,42.9-26.5,42.9-39.7c0-21.5-13.7-39.7-43.7-39.7h-23.7V172.4z}
                 svg{M88.7,56.8c0,5.5-4.5,10.1-10.1,10.1c-5.6,0-10.1-4.6-10.1-10.1c0-5.6,4.5-10.1,10.1-10.1C84.2,46.7,88.7,51.3,88.7,56.8z};
  }
}

\newcommand{\Setup}{\mathsf{Setup}}

\newcommand{\RegisterAndIssue}{\mathsf{RegisterAndIssue}}

\newcommand{\CertInit}{\mathsf{CertInit}}
\newcommand{\ZKRequest}{\mathsf{ZKRequest}}

\newcommand{\OrderProfile}{\mathsf{OrderProfile}}
\newcommand{\MutualAuthAndProvision}{\mathsf{MutualAuthAndProvision}}
\newcommand{\Settle}{\mathsf{Settle}}
\newcommand{\Deanonymise}{\mathsf{Deanonymise}}
\newcommand{\PCert}{\mathsf{PCert}}
\newcommand{\Com}{\mathsf{Com}}

\newcommand{\cred}{\mathsf{cred}}

\newcommand{\Hpid}{\mathsf{Hpid}}
\newcommand{\nonce}{\mathsf{nonce}}
\newcommand{\EncEid}{\mathsf{EncEid}}
\newcommand{\EID}{\mathsf{EID}}
\newcommand{\Ti}{T_i}

\newcommand{\pp}{\mathsf{pp}}

\newcommand{\negl}{\mathsf{negl}}
\newcommand{\PRF}{\mathsf{PRF}}
\renewcommand{\pid}{\mathsf{pid}}

\newcommand{\View}{\mathsf{View}}

\newcommand{\UNLINK}{\mathsf{UNLINK}}

\newcommand{\MNO}{\ensuremath{\mathsf{MNO}}}
\newcommand{\SMDP}{\ensuremath{\mathsf{SMDP}}}
\newcommand{\LEA}{\ensuremath{\mathsf{LEA}}}
\newcommand{\PCA}{\ensuremath{\mathsf{PCA}}}

\usetikzlibrary{arrows.meta,
                quotes}
\tikzset{auto,
                > = Straight Barb,
every path/.style = {semithick}
         }
\usepackage{tabularray}
\UseTblrLibrary{amsmath}
\usepackage{bm}

\definecolor{partyblue}{RGB}{208,230,252}
\definecolor{partyyellow}{RGB}{255,244,201}
\definecolor{partygreen}{RGB}{210,244,222}

\newcommand{\STATEx}{\item[]}

\tikzset{
  lifeline/.style={line width=0.9pt},
  role/.style={rounded corners=2.5pt,inner ysep=2.5pt,inner xsep=6pt,
               font=\bfseries,draw=black,line width=0.6pt,align=center},
  msgbox/.style={draw,rounded corners=2pt,line width=0.6pt,
                 fill=white,align=left,inner sep=6pt,font=\small},
  setupmini/.style={draw,rounded corners=2pt,line width=0.6pt,
                    fill=gray!10,align=left,inner sep=5pt,
                    font=\scriptsize,text width=80pt},
  blackcap/.style={fill=black, minimum width=18pt, minimum height=2pt,  inner sep=1pt,
    outer sep=1pt},bandlabel/.style={anchor=west, font=\footnotesize\bfseries, text=black!80},
  >={Latex[length=2.3mm]}
}

\begin{document}

\title{ZK-eSIM: A Privacy-Centric Zero-Knowledge Approach for eSIM Provisioning}

\author{Liza Ahmad}
\authornote{Both authors contributed equally to this research.}
\affiliation{%
  \institution{University of Sheffield}
  \city{Sheffield}
    \country{United Kingdom}
}
 \email{lahmad2@sheffield.ac.uk}
\author{Quan Shi}
\authornotemark[1]
\affiliation{%
  \institution{National University of Singapore}
  \city{Singapore}
   \country{Singapore}
}
\email{shi.quan@u.nus.edu}
\author{Joshua Haworth}
\authornote{Valuable contribution to the implementation of ZK-eSIM}
\affiliation{%
  \institution{University of Sheffield}
  \city{Sheffield}
    \country{United Kingdom}
}
 \email{jhaworth1@sheffield.ac.uk}

\author{Yilu Dong}
\authornotemark[2]
\affiliation{%
  \institution{Pennsylvania State University}
  \city{University Park}
  \state{PA}
    \country{United States}
 }
 \email{yiludong@psu.edu}
\author{Prosanta Gope\textsuperscript{\Letter}}
\affiliation{%
  \institution{University of Sheffield}
  \city{Sheffield}
    \country{United Kingdom}
 }
 \email{p.gope@sheffield.ac.uk}
\makeatletter
\g@addto@macro\@authornotes{%
  \let\thefootnote\relax\footnotetext{\Letter\ Corresponding author.}}
\makeatother

\author{Behzad Abdolmaleki}
\affiliation{%
  \institution{University of Sheffield}
  \city{Sheffield}
    \country{United Kingdom}
 }
\email{behzad.abdolmaleki@sheffield.ac.uk}

\author{Syed Rafiul Hussain}
\affiliation{%
  \institution{Pennsylvania State University}
  \city{University Park}
  \state{PA}
    \country{United States}
 }
 \email{hussain1@psu.edu}

\date{}

\renewcommand{\shortauthors}{Ahmad et al.}

\begin{abstract}
GSMA Remote SIM Provisioning (RSP) enables over-the-air delivery of eSIM profiles, but it exposes long-lived identifiers during profile ordering and download. In particular, stable device identifiers (e.g., EID), profile identifiers, and long-lived certificate material enable mobile operators and profile-delivery infrastructure to link provisioning events to the same eUICC and, when combined with account records, to the same subscriber. This undermines subscriber anonymity and enables cross-session tracking. We present \emph{ZK-eSIM}, a privacy-preserving redesign that achieves subscriber anonymity and provisioning-session unlinkability while retaining accountable traceability by exception. ZK-eSIM (i) replaces direct disclosure of device identifiers with a zero-knowledge proof of device validity and eligibility; (ii) enforces session unlinkability through short-lived, one-time pseudonymous credentials and per-session identifiers to prevent cross-session tracking; and (iii) provides privacy-preserving accountable traceability through a jointly authorised escrow
mechanism, so that no single
entity can unilaterally deanonymise a user. We formalise a multi-entity, honest-but-curious threat model and prove subscriber anonymity and the unlinkability of provisioning sessions under standard cryptographic assumptions. We implement a Java Card applet on a test eUICC to evaluate performance on commodity hardware with a modified LPA and SM-DP+ server. Our experiments quantify end-to-end cryptographic overhead relative to conventional RSP, confirming that \emph{ZK-eSIM} adds only practical overhead, closing a critical privacy gap while preserving deployability within existing GSMA roles and interfaces.
\end{abstract}



\maketitle
\section{Introduction}

\begin{figure}[t]
    \centering
    \includegraphics[width=1\linewidth, trim=35 70 35 40]{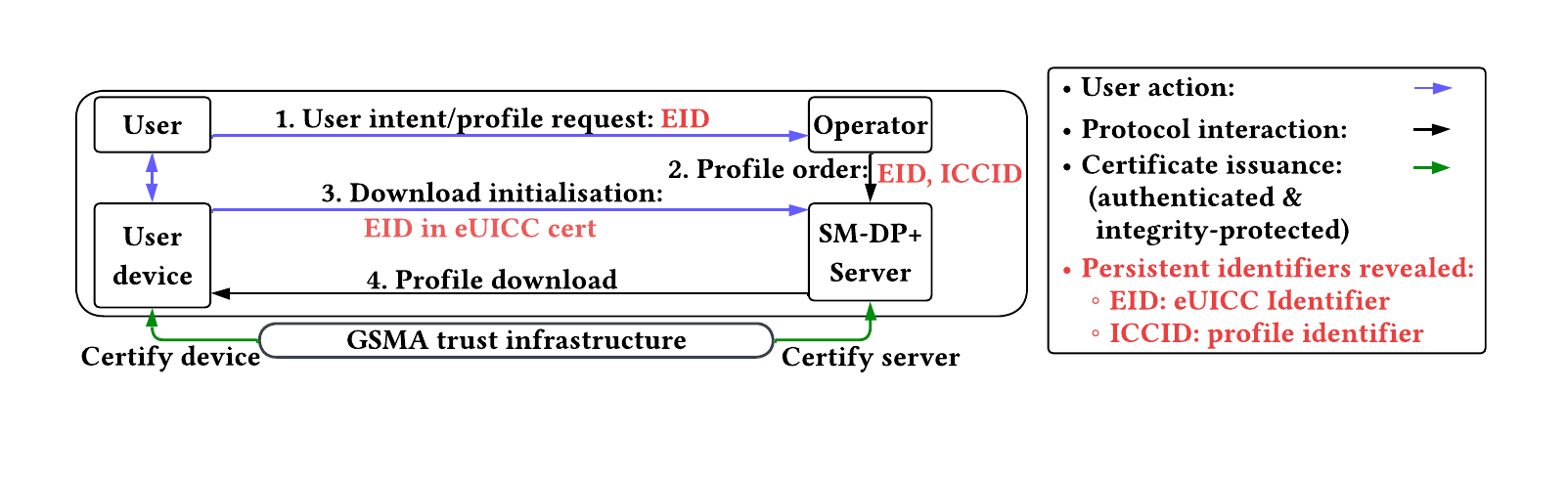}
    \caption{Identifier exposures in the GSMA Consumer Remote SIM Provisioning (RSP) workflow.}
    \label{fig:consumer-RSP}
 \vspace{-1em}
\end{figure}

The embedded SIM (eSIM) is rapidly replacing removable SIM cards as the default method for devices to connect to mobile networks. Unlike physical SIMs, which require manual swapping, an eSIM profile is provisioned digitally and stored on the embedded Universal Integrated Circuit Card (eUICC), a secure chip built directly into smartphones, wearables, and IoT devices. In today's consumer ecosystem, the profile download process follows a standard Remote SIM Provisioning (RSP) workflow defined by the GSM Association (GSMA) \cite{gsma_rsp_2023,yuan_esim_2024}. At a high level, the workflow involves the user device, the mobile operator, a server that prepares and stores the eSIM profile (SM-DP+: the Subscription Manager--Data Preparation+), and the GSMA trust infrastructure (Fig.~\ref{fig:consumer-RSP}).  eSIM adoption is now mainstream: GSMA reports that the number of eSIM-capable smartphones in active use increased from 144 million in 2022 to nearly 600 million in 2024 \cite{gsma_esim_2024}, and major vendors are moving toward eSIM-only devices \cite{apple_apple_2025}. At this scale, the privacy and security properties of the eSIM provisioning protocol become an urgent concern.  
In particular, the current provisioning protocol design may enable tracking of the user's device, location, and data over time, which is at odds with privacy-by-design expectations (e.g., GDPR \cite{european_parliament_and_council_of_the_european_union_regulation_2016} and GSMA's RSP architectural principles \cite{gsma_rsp_2023}). Reported cases of user data leaks and resulting regulatory fines show that these concerns are not purely theoretical \cite{onag_privacy_2025,robuck_fcc_2024}.

\begin{figure}[!t]
    \centering
    \includegraphics[width=1\linewidth, trim=20 40 20 10]{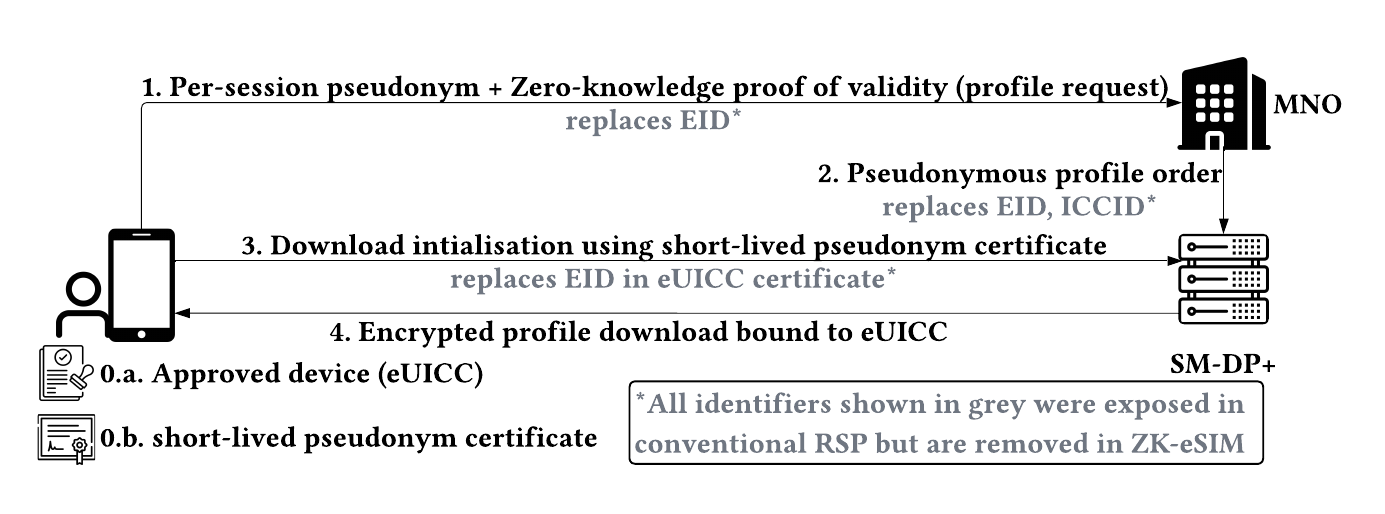}
    \caption{ZK-eSIM provisioning overview. (0.a) An approved eUICC obtains (0.b) a short-lived pseudonym certificate, (1) authenticates to the MNO with a per-session pseudonym and ZK proof, (2) triggers a pseudonymous SM-DP+ order, (3) initialises profile download with the short-lived certificate, and (4) receives an eUICC-bound encrypted profile, without exposing persistent identifiers.
    }

    \label{fig:overview}
     \vspace{-1em}
\end{figure}

To comprehend the source of the risk, it is essential to analyse the identification process of user devices during provisioning. Each eSIM-enabled device is equipped with specific, immutable information, including a permanent device identifier referred to as the EID (eUICC Identifier). This information serves as a basis for the provisioning infrastructure to authenticate the device and facilitate the delivery of new profiles. However, due to the reuse of this identifying information across multiple provisioning sessions, various stakeholders within the infrastructure can consistently recognise the same device during its interactions with the system. As a result, different profile downloads, which may be intended to be independent, can be linked back to the same device and associated with the subscriber's identity.  
The root cause is architectural: conventional RSP operation interlinks \emph{stable identifiers} and exposes them to infrastructure entities (Fig.~\ref{fig:consumer-RSP}). We identify \textbf{three} critical \emph{privacy risks in the conventional RSP protocol}:\\
$\bullet${\textbf{Identity--EID binding at subscription time.}
During profile ordering, the operator and SM-DP+ 
learn the device's
permanent identifier (EID) together with the subscriber's account data,
binding the device to a real person and linking future provisioning events for that EID to the same subscriber~\cite{gsma_sgp22_2023}.\\ 
$\bullet$ \textbf{Linking across multiple downloads.} The current eSIM provisioning protocol reuses the same device identifier for each profile request or download,  allowing the operator or profile server to link otherwise separate profiles to the same device, even when installed for different purposes. Over time, this enables a combined view of the user’s activity and, in shared backends serving multiple operators, can extend across operators \cite{motallebighomi_esimplicity_2025}.\\
$\bullet$ \textbf{Persistent certificate identifiers.} Even if the EID is hidden during provisioning, long-lived certificate data exchanged during authentication can act as a persistent fingerprint, allowing infrastructure parties to recognise the device across sessions \cite{gsma_sgp22_2023,ahmed_security_2024}.

The above risks are amplified by contemporary provisioning practices, as profile switching is no longer a corner case. The most visible example is the rapid growth of travel eSIM usage (up by 85\% in 2025) \cite{juniper_research_travel_2025}. Travellers increasingly create short-lived, data-only travel profiles rather than relying on carrier roaming, which remains significantly more expensive in many regions, such as China and India \cite{travel_and_tour_world_new_2025, treffinger_if_2025}. While the plan is \emph{temporary}, the device’s EID is not: it persists across all installations and profile switches. As a result, reseller platforms and SM-DP+ backends operating shared infrastructures can expose stable mappings between the device EID, the issued profile identifier, and the Integrated Circuit Card Identifier (ICCID), across otherwise independent provisioning events \cite{motallebighomi_esimplicity_2025}.

GSMA standards and prior analyses \cite{gsma_rsp_2023,gsma_sgp22_2023, ahmed_security_2024} focus on delivery security, confidentiality, and the authenticity of profile downloads, rather than on privacy. The key challenge is therefore not secure transport, but privacy by default with accountability by exception. Our privacy goal is \emph{anonymity and unlinkability during provisioning}: provisioning should not reveal a subscriber's real-world identity or permanent device identifiers, and distinct provisioning sessions should not be linkable to the same device or subscriber. This gap directly motivates our three research questions:


\begin{enumerate}[nosep]
\item[\textbf{RQ1:}] \emph{\underline{Subscriber anonymity.} Can a legitimate eUICC 
obtain a profile without disclosing any device identifiers (e.g., EID) to the SM-DP+ or the Mobile Network Operator (MNO) during provisioning?}

\item[\textbf{RQ2:}] \emph{\underline{Provisioning-session unlinkability.} Can provisioning sessions remain unlinkable to each other, even across different operators and multi-tenant SM-DP+ platforms, and resist certificate-chain tracking?}

\item[\textbf{RQ3:}] \emph{\underline{Accountable traceability.} Can we provide accountable traceability
alongside the anonymity and unlinkability guarantees of RQ1--RQ2,
such that on-demand deanonymisation requires joint authorisation
and no single entity can unilaterally compromise privacy?}
\end{enumerate}

To answer \textbf{RQ1--RQ3}, we present \textbf{ZK-eSIM}, a privacy-preserving redesign of the RSP workflow that preserves the existing GSMA roles and interfaces while removing stable identifier exposure from provisioning transcripts (Fig.~\ref{fig:overview}).  Our \textbf{contributions }are:\\
 $\bullet$ \textbf{Anonymous provisioning without persistent device identifiers.}
 We design ZK-eSIM, an end-to-end provisioning workflow that uses fresh, per-session pseudonyms and zero-knowledge proofs of eUICC legitimacy. The MNO and SM-DP+ learn only session-local handles, not persistent identifiers or long-lived identifying certificate-chain attributes (addressing RQ1).\\
  $\bullet$ \textbf{Unlinkability across sessions, operators, and multi-tenant SM-DP+.}
 ZK-eSIM makes protocol-visible credentials and tokens one-time and eliminates certificate-chain fingerprints by using short-lived, session-scoped authentication material.} These mechanisms prevent both application-layer and certificate-based linkage across repeated provisioning events, including those handled by the same SM-DP+ (addressing RQ2).
\\
  $\bullet$ \textbf{Accountable traceability via joint deanonymisation.}
  We design a trace mechanism in which identity recovery requires joint authorisation by the MNO and a designated Law Enforcement Authority (LEA), providing accountability \emph{by exception} while routine provisioning remains anonymous and unlinkable (addressing RQ3).\\
  $\bullet$ \textbf{Evidence for RQ1--RQ3: formalisation and implementation.}
  We formalise the above privacy and traceability properties in a multi-entity threat model and prove ZK-eSIM’s anonymity, provisioning-session unlinkability, and joint-traceability guarantees under standard cryptographic assumptions (Sec.~\ref{sec:security}). We implement and evaluate a prototype on commodity smartphones (including a Java Card 
  applet\footnote{Implementation overview: \url{https://github.com/NaivEPoi/ZK-eSim}}) and \textit{a test eUICC alongside a GSMA-compliant SM-DP+ Server} to demonstrate that ZK-eSIM is practical in real mobile environments (Sec.~\ref{sec:evaluation}, addressing RQ1--RQ3).

\begin{figure}[!t]
  \centering
  \hspace*{-130pt} 
 \resizebox{1.6\linewidth}{!}{\begin{tikzpicture}[x=1pt,y=1pt]

  \def\xA{0}    
  \def\xMNOA{80}
  \def\xMNOB{160}
  \def\xSMDP{240}

  \def\topY{230}
  \def\botY{-430}

  \pgfmathsetmacro{\xTitle}{0.5*(\xA+\xSMDP)}
  \pgfmathsetmacro{\yTitle}{\topY+26}
  \pgfmathsetmacro{\xCompose}{0.5*(\xMNOA+\xSMDP)}

  \node[role, fill=partyblue, minimum width=30pt] (DevA) at (\xA,\topY)
    {\textbf{Device A} \faIcon{mobile-alt}\\\textbf{UE/LPA}\\\textbf{ eUICC}};
  
  \node[role, fill=orange!30, minimum width=30pt] (MNOA) at (\xMNOA,\topY)
    {\textbf{MNO\textbackslash}\\\textbf{MVNO 1}};
  \node[role, fill=partygreen, minimum width=30pt] (MNOB) at (\xMNOB,\topY)
    {\textbf{MNO\textbackslash}\\\textbf{MVNO 2}};

  \node[role, fill=white, draw=black, rounded corners=2pt, minimum width=30pt] (SMDP) at (\xSMDP,\topY)
    {\textbf{SM-DP+}\\{\footnotesize (multi-tenant)}};

  \draw[lifeline] (\xA,\topY-18) -- (\xA,\botY+190);
  \draw[lifeline] (\xMNOA,\topY-+12) -- (\xMNOA,\botY+190);
  \draw[lifeline] (\xMNOB,\topY-+12) -- (\xMNOB,\botY+190);
  \draw[lifeline] (\xSMDP,\topY-+12) -- (\xSMDP,\botY+190);

  bottom caps
  \node[blackcap] at (\xA,\botY+190) {};
  \node[blackcap] at (\xMNOA,\botY+190) {};
  \node[blackcap] at (\xMNOB,\botY+190) {};
  \node[blackcap] at (\xSMDP,\botY+190) {};

  \def\yPhaseZero{210}
  \node[font=\bfseries, anchor=west, fill=white] at (\xA-35
  ,\yPhaseZero-5) {Phase 0 (Pre-RSP): Intent / Contract Subscription};

  \def\yA0{180}
  
  \def\yB0{80}
\draw[dashed,
      rounded corners=6pt,
      line width=0.8pt, draw=red]
  (\xA-30,\yA0+9) rectangle (\xMNOA+85,\yB0+58);

\node[anchor=north west, font=\small\bfseries, text=red!70!black, fill=white]
  at (\xA-35,\yA0+23){R1: Identity--EID binding at subscription time};

\draw[->, line width=0.6pt]
  (\xA,\yA0-3) -- ++(25,0)
  node[pos=0.4, above, font=\small] {\step{1} GetEID}
  -- ++(0,-3) -- ++(-25,0);

  \draw[->, line width=0.6pt] (\xA,\yA0-28) -- (\xMNOA,\yA0-28)
    node[midway, above, font=\small, text width=100pt, align=center]{\step{2} \textcolor{red}{identity},
      \textcolor{red}{billing},
      \textcolor{red}{EID\textsubscript{A}}, profileType
    };
  \draw[->, line width=0.6pt] (\xA,\yA0-40) -- (\xMNOB,\yA0-40)
    node[midway, above, font=\small, text width=210pt, align=center]{%
          \step{3} \textcolor{red}{identity},
      \textcolor{red}{billing},
      \textcolor{red}{EID\textsubscript{A}}, profileType)
    };
\def\yPhaseES9{0}
  \node[font=\bfseries, anchor=west, fill=white, text width=150pt] at (\xA-35,\yPhaseES9+130) {Profile Provisioning Sessions};
\def\ytOne{-50}
  \def\ytTwo{-230}
 \node[anchor=north west, font=\small\bfseries, text=red!70!black, fill=white]
   at (\xA-35,\ytOne+178)
   {R2: Cross-session / cross-operator linkability (SM-DP+ join point)};
 \draw[dashed,
       rounded corners=6pt,
       line width=0.9pt,
       draw=red]
  (\xA-30,\ytOne+167) rectangle (\xSMDP+30,\ytTwo-180+174);
\node[font=\bfseries, anchor=west, fill=white, text width=150pt] at (\xA-30,\yPhaseES9+109) {
Session 1, MNO/MVNO\,1 (4-16):};

\draw[->, line width=0.6pt] (\xMNOA,\ytOne+79+64) -- (\xSMDP,\ytOne+79+64)
  node[midway, above, font=\small, text width=210pt, align=center]{%
    \step{4} DownloadOrder(\textcolor{red}{EID\textsubscript{A}}, profileType=X)
  };
\node[msgbox, anchor=west, text width=75pt, align=center, inner sep=1pt] at (\xSMDP-50,\ytOne+69+66) {%
  \step{5} Reserve \textcolor{red}{ICCID\textsubscript{1}}
};

\draw[->, line width=0.6pt] (\xSMDP,\ytOne+49+69) -- (\xMNOA,\ytOne+49+69)
  node[midway, above, font=\small]{\step{6} \textcolor{red}{ICCID\textsubscript{1}}, MatchingID};

\draw[<-, line width=0.6pt] (\xSMDP,\ytOne+34+70) -- (\xMNOA,\ytOne+34+70)
  node[midway, above, font=\small]{\step{7} ConfirmOrder(\textcolor{red}{ICCID\textsubscript{1}},\textcolor{red}{EID\textsubscript{A}},releaseFlag=true)};

\node[msgbox, anchor=west, text width=70pt, align=center, inner sep=1pt] at (\xSMDP-50,\ytOne+19+72) {%
  \step{8} Map \textcolor{red}{ICCID\textsubscript{1} $\leftrightarrow$ EID\textsubscript{A}}
};
  \draw[->, line width=0.6pt] (\xMNOA,\ytOne+9+74) -- (\xA,\ytOne+9+74)
    node[midway, above, font=\small]{\quad\quad\quad\quad\step{9}\textit{Deliver Activation Code} {\footnotesize (SM-DP+ address, AC Token)}};

 \def\yChal{-128}
\draw[->, line width=0.6pt]
    (\xA, \yChal+73+74) -- ++(25,0)
    node[above, font=\small, text width=460pt, align=left]{%
      \hspace{+180pt}\step{10} GetEUICCInfo, GetEUICCChallenge\\
      
    }  -- ++(0,-3) -- ++(-25,0);

  \draw[->, line width=0.6pt] (\xA,\ytOne-24+75) -- (\xSMDP,\ytOne-24+75)
    node[midway, above, font=\small, text width=300pt, align=center]{%
      \step{11} InitiateAuthentication(euiccChallenge,
      euiccInfo1,
      smdpAddress, ...)
    };

 \def\ySA{-178}
  \draw[->, line width=0.6pt] (\xSMDP, \ySA+92+73.5) -- (\xA, \ySA+92+73.5)
    node[midway, above, font=\small, text width=300pt, align=center]{%
      \step{12} transactionId\textsubscript{1},\;serverSigned1,\;serverSignature1,\;%
      CERT.DPauth.SIG%
    };

     \def\yAS{-203}
  \draw[->, line width=0.6pt]
    (\xA, \yAS+95+75) -- ++(25,0)
    node[above, font=\small, text width=180pt, align=left]{%
      \hspace{+40pt}\step{13} AuthenticateServer(...)\\
      {\hspace{+55pt}\footnotesize eUICC verifies SM-DP+}\\
    }
    -- ++(0,-3) -- ++(-25,0);

\node[font=\small\bfseries, text=red!70!black, fill=white]
  at (\xA+85, \yAS+82+76) {R3: Certificate-chain fingerprinting during authentication};

 \draw[dashed, rounded corners=5pt, line width=0.9pt, draw=red]
   (\xA-20,   \ytOne-78+79) rectangle (\xSMDP+25, \ytOne-140+85);
   
  \def\yACl{-163}
  \draw[->, line width=0.6pt] (\xA, \yACl+12+80) -- (\xSMDP, \yACl+12+80)
    node[midway, above, font=\small, text width=310pt, align=center]{%
      \step{14} AuthenticateClient(euiccSigned1,\;euiccSignature1,\\
      \textcolor{red}{CERT.EUICC.SIG\;(contains EID\textsubscript{A})},\;%
      CERT.EUM.SIG)%
    };

\node[msgbox, anchor=west, text width=70pt, align=center, inner sep=1pt] at (\xSMDP-50,\ytOne-122+83) {%
  \step{15} Verify eUICC cert chain;
  persist \textcolor{red}{EID\textsubscript{A}}
  $\leftrightarrow$ ICCID\textsubscript{1}%
};

\draw[->, line width=0.6pt] (\xSMDP,\ytOne-155+87) -- (\xA,\ytOne-155+87)
  node[midway, above, font=\small]{\step{16} Profile download\;\&\;install (ICCID\textsubscript{1})%
  };

\def\ysTwo{-378}
\node[font=\bfseries, anchor=west, fill=white, text width=150pt] at (\xA-30,\yPhaseES9-128) {
Session 2, MNO/MVNO\,2 (17-22):};

\draw[->, line width=0.6pt] (\xMNOB,\ysTwo+139+96) -- (\xSMDP,\ysTwo+139+96)
  node[midway, above, font=\small, text width=210pt, align=center]{%
    \hspace{-45pt}\step{17} DownloadOrder(\textcolor{red}{EID\textsubscript{A}}, profileType=Y)
  };
\node[msgbox, anchor=west, text width=70pt, align=center, inner sep=1pt] at (\xSMDP-50,\ysTwo+129+97) {%
  \step{18} Reserve \textcolor{red}{ICCID\textsubscript{2}}
};
\draw[->, line width=0.6pt] (\xSMDP,\ysTwo+109+99) -- (\xMNOB,\ysTwo+109+99)
  node[midway, above, font=\small]{\step{19} \textcolor{red}{ICCID\textsubscript{2}}, MatchingID};
\draw[<-, line width=0.6pt] (\xSMDP,\ysTwo+95+99) -- (\xMNOB,\ysTwo+95+99)
  node[midway, above, font=\small]{\hspace{-55pt}\step{20} ConfirmOrder(\textcolor{red}{ICCID\textsubscript{2}},\textcolor{red}{EID\textsubscript{A}},releaseFlag=true)};
\node[msgbox, anchor=west, text width=70pt, align=center, inner sep=1pt] at (\xSMDP-50,\ysTwo+80+101) {%
  \step{21} Map \textcolor{red}{ICCID\textsubscript{2} $\leftrightarrow$ EID\textsubscript{A}}
};
  \node[draw=black!60, fill=black!5, rounded corners=3pt, font=\small,
        text width=280pt, align=left, inner sep=4pt]
    at (0.5*\xA+0.5*\xSMDP, \ysTwo+156) {%
      \step{22} Activation-code delivery, ES9+ mutual authentication
      (\textcolor{red!70!black}{R3}), and profile install proceed identically to steps 9--16;
      SM-DP+ persists \textcolor{red}{EID\textsubscript{A} $\leftrightarrow$ ICCID\textsubscript{2}}.%
    };
\end{tikzpicture}}  
 \caption{Privacy Risks in Conventional RSP (§3 R1--R3)}
	\label{fig:PrivacyRisks}
 \vspace{-1em}
\end{figure}
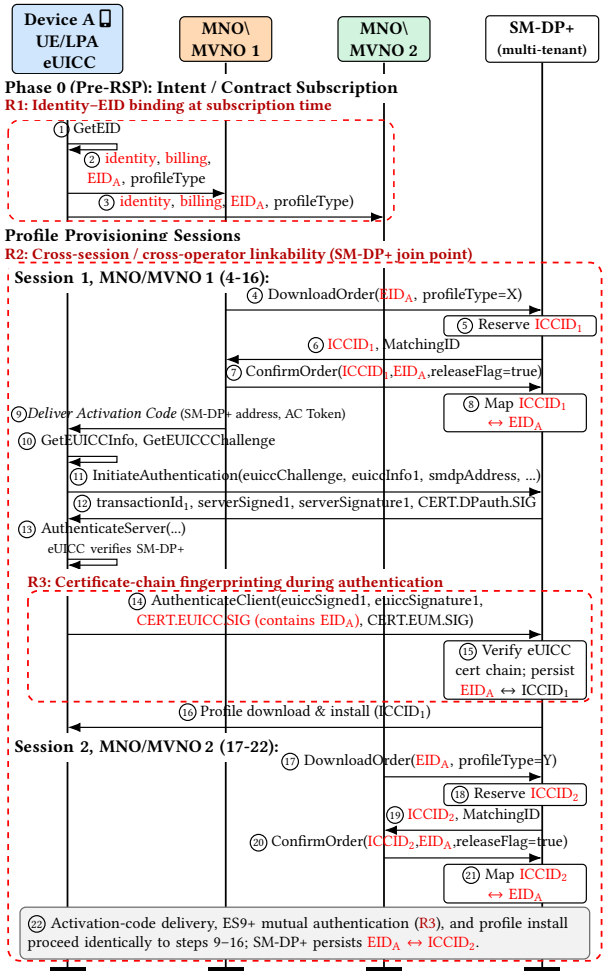
\section{Background and Related Work}

GSMA Consumer RSP enables a subscriber device to download and install operator profiles on an eUICC, a tamper-resistant secure element integrated into the device. The eUICC stores one or more operator profiles and protects the long-term keys and subscription material contained in those profiles. The MNO or mobile virtual network operator (MVNO) authorises profile issuance, while the SM-DP+ prepares, stores, and delivers the protected profile package to the target eUICC, which installs the profile \cite{gsma_rsp_2023,gsma_sgp22_2023}. The RSP trust model is anchored in the GSMA public-key infrastructure: the eUICC presents a certificate chain rooted in the GSMA trust hierarchy, while the SM-DP+ authenticates as an authorised profile-delivery server. During profile download, these certificates enable the SM-DP+ to verify that it is interacting with an eligible eUICC, allow the device/eUICC to authenticate the delivery server, and bind the protected profile package to the intended target \cite{gsma_rsp_2023,gsma_sgp22_2023}.  Several identifiers are central to this workflow. The EID identifies the eUICC. The ICCID identifies an individual profile. The Matching ID provided to the device indicates how to retrieve a pending profile. Finally, the eUICC and SM-DP+ certificates support mutual authentication and profile-package protection under the GSMA trust infrastructure. Fig.~\ref{fig:PrivacyRisks} shows two consumer RSP provisioning sessions for the same device with different operators. The first session corresponds to a profile ordered from MNO/MVNO~1, while the second models a later, independent order from MNO/MVNO~2, for example, when the user purchases a travel eSIM while abroad. We first describe the functional flow; Sec.~\ref{sec:privacy-risks} then analyses the privacy risks from identifier reuse. The device obtains the eUICC identifier \step{1} and provides it during subscription or profile purchase together with the requested profile type \step{2}. The operator then requests profile preparation for the target eUICC \step{4}. The SM-DP+ reserves a profile identifier \step{5}, returns the corresponding profile handle and matching information \step{6}, and finalises the order after operator confirmation \step{7}--\step{8}. The operator then delivers activation material to the device \step{9}. Using the activation material, the device first obtains eUICC information and fresh challenge material from the eUICC \step{10}, then contacts the SM-DP+ to initiate authentication for the download session \step{11}, and verifies the SM-DP+ authentication response \step{12}--\step{13}. The eUICC then authenticates itself to the SM-DP+ using signed session data and its certificate chain \step{14}. Once the SM-DP+ verifies the eUICC and confirms that the pending profile matches the authorised order \step{15}, it delivers the protected profile package for installation on the eUICC \step{16}. The later provisioning session with MNO/MVNO~2 follows the same pattern (\step{3} and \step{17}--\step{22}), resulting in the provisioning of a new profile on the same eUICC.

In practice, the SM-DP+ may be operated by the MNO itself (e.g., Vodafone Idea \cite{gsma_sas_2026}) or outsourced to a multi-tenant provider serving multiple operators (e.g., Workz~\cite{passett_workz_2024}). These deployment choices do not change the provisioning flow in Fig.~\ref{fig:PrivacyRisks}, but they affect which domains observe provisioning state. Existing GSMA specifications \cite{gsma_rsp_2023,gsma_sgp22_2023} focus on correctness, authentication, and secure profile transport, while leaving subscriber anonymity and unlinkability outside the main security goals. A formal analysis by Ahmed \emph{et~al.}~\cite{ahmed_security_2024} confirms this security--privacy split: RSP satisfies delivery guarantees, but does not provide anonymity or unlinkability for subscribers. Accountability-oriented proposals, such as the SIM Profile Transparency Protocol proposed by Ahmed \emph{et~al.}~\cite{ahmed_transparency_2021}, help detect profile misissuance, but still depend on stable identifiers and therefore do not eliminate linkability at the provisioning layer. Moreover, the problem is not confined to provisioning. Rao \emph{et~al.}~\cite{rao_authenticating_2023} show that higher-layer functionality built on top of RSP, including device authentication and network access procedures, inherits the same identifiers and consequently propagates provisioning-time privacy deficits into the operational lifetime of the subscription.

ZK-eSIM composes established cryptographic techniques. Anonymous credentials support unlinkable proofs of eligibility \cite{goos_efficient_2001,hutchison_signature_2004}, blind signatures enable privacy-preserving credential issuance \cite{chaum_blind_1983}, and pseudonym certificates and identity escrow provide session-based authentication and accountable anonymity \cite{brecht_security_2018,goos_identity_2001}. However, these frameworks typically address individual authentication tasks rather than privacy across the complete, stateful RSP workflow, and they do not consider integration with the GSMA trust and operational infrastructure. Existing cellular privacy work instead focuses mainly on protecting subscriber identifiers during 5G network authentication \cite{basin_formal_2018,koutsos_5g-aka_2019}, rather than profile provisioning. ZK-eSIM's contribution is therefore the RSP-specific adaptation and end-to-end composition of established techniques to provide privacy across the GSMA Consumer RSP workflow.


\section{Privacy Risks in Conventional RSP}\label{sec:privacy-risks}

Fig.~\ref{fig:PrivacyRisks} highlights three privacy risks arising from the repeated exposure of stable identifiers throughout the conventional RSP workflow: identity--EID binding at subscription time (\textbf{R1}); cross-session and cross-operator linkability during provisioning (\textbf{R2}); and certificate-chain fingerprinting during authentication (\textbf{R3}). We consider a multi-tenant SM-DP+ deployment, where the same SM-DP+ backend serves multiple MNO/MVNO tenants. The main privacy issue is not just that these identifiers exist, but that they can be reused to connect events that a user would reasonably expect to remain separate. 

\noindent$\bullet$\textbf{ R1:} The earliest risk occurs before profile download begins. During subscription, the operator receives an $\mathsf{EID}$ together with Know Your Customer (KYC), account, and billing data, allowing it to associate a real-world subscriber with a specific eUICC. In Phase~0, the Local Profile Assistant (LPA) retrieves the $\mathsf{EID}$ via $\mathsf{GetEID}$ \step{1} and forwards it during contract subscription with each operator \step{2}, \step{3}. MNO/MVNO~1 or MNO/MVNO~2 therefore learns that $\mathsf{EID\textsubscript{A}}$ belongs to the subscriber purchasing the profile. This binding has standalone privacy implications. Once an $\mathsf{EID}$ is associated with an identity, it can be correlated with other operator-side identifiers and records, including IMEI, IMSI/SUPI, billing data, and account metadata. In multi-device settings, the same process can also reveal a subscriber's device portfolio: for example, if a phone and a smartwatch are provisioned under the same account, their distinct $\mathsf{EID}$s may become linked to the same user. Such records also create an insider-abuse surface: an adversary with access to operator systems could enumerate the $\mathsf{EID}$s associated with a target and use this information to facilitate fraudulent profile replacement or SIM-swap-style account takeover \cite{lee_empirical_2020}.

\noindent$\bullet$\textbf{R2:} The primary privacy risk arises when the same long-lived $\mathsf{EID}$ is reused across otherwise independent profile downloads. Each download creates a fresh $\mathsf{ICCID}$, but the backend observes the same device identifier: the $\mathsf{ICCID}$ identifies the profile being issued, whereas the $\mathsf{EID}$ identifies the eUICC itself. In the first provisioning session, MNO/MVNO~1 sends $\mathsf{DownloadOrder}$ to the SM-DP+ \step{4}. The SM-DP+ reserves $\mathsf{ICCID\textsubscript{1}}$ for that order \step{5} and returns $\mathsf{ICCID\textsubscript{1}}$ together with a $\mathsf{MatchingID}$ \step{6}, which later acts as a handle for retrieving the reserved profile. After receiving $\mathsf{ConfirmOrder}$ \step{7}, the SM-DP+ can persist the mapping $\mathsf{ICCID\textsubscript{1}}$ $\leftrightarrow$ $\mathsf{EID\textsubscript{A}}$ \step{8}. The activation code delivered to the device \step{9} then tells the LPA which SM-DP+ to contact and which reserved profile to claim. The second session through MNO/MVNO~2 repeats this pattern with $\mathsf{EID\textsubscript{A}}$ reappearing in \step{17}--\step{22}, allowing the SM-DP+ to persist a second mapping $\mathsf{ICCID\textsubscript{2}} $$\leftrightarrow$ $\mathsf{EID\textsubscript{A}}$. The SM-DP+ can therefore infer that both profile downloads correspond to the same eUICC, even though they involve different profiles and different operators. This is already a privacy risk even without knowing the subscriber's civil identity: the backend can link purchases, destinations, provisioning times, profile types, and operator choices to the same device. In a multi-tenant SM-DP+ deployment, the risk becomes cross-operator: a single backend may observe provisioning requests from multiple operators and join them using the $\mathsf{EID}$. When combined with \textbf{R1}, the same mechanism becomes subscriber-level tracking by the operator, because the linked device activity can be associated with a real-world user.

\noindent$\bullet$\textbf{ R3:} A natural mitigation would be to minimise or hide the $\mathsf{EID}$ in the profile-ordering flow. However, the authentication phase can reintroduce stable device-linked material. After the device receives the activation code in the first provisioning session, it obtains local eUICC information and a challenge \step{10} and invokes $\mathsf{InitiateAuthentication}$ with the SM-DP+ \step{11}. The SM-DP+ responds with transaction-specific authentication data and its certificate material \step{12}, and $\mathsf{AuthenticateServer}$ lets the eUICC verify that it is communicating with the legitimate SM-DP+ \step{13}. The reciprocal step is more privacy-sensitive. In $\mathsf{AuthenticateClient}$, the eUICC proves itself to the SM-DP+ by presenting signed authentication data and its certificate chain \step{14}. This chain includes $\mathsf{CERT.EUICC.SIG}$, containing stable eUICC identity material associated with $\mathsf{EID\textsubscript{A}}$. The SM-DP+ then verifies the chain and can persist the resulting association between the eUICC identity and the issued profile, here $\mathsf{EID\textsubscript{A}}$ $\leftrightarrow$ $\mathsf{ICCID\textsubscript{1}}$ \step{15}, before the profile is downloaded and installed \step{16}. In the second provisioning session, the same activation-code delivery, mutual authentication, certificate-chain exposure, and profile installation are compressed into \step{22}; the privacy consequence is identical, because the SM-DP+ can again associate the same eUICC identity with $\mathsf{ICCID\textsubscript{2}}$. R3 is not redundant: it explains why R2 cannot be removed by application-layer identifier minimisation alone. If the SM-DP+ can still recognise the same eUICC during mutual authentication, then hiding the $\mathsf{EID}$ in $\mathsf{DownloadOrder}$ does not provide unlinkability. Because \textbf{R1} may already have tied $\mathsf{EID\textsubscript{A}}$ to a subscriber's real identity, the certificate material in \textbf{R3} can carry user-identifying meaning into every authenticated provisioning session \cite{gsma_sgp22_2023,ahmed_security_2024}.

The severity of the privacy risks depends on who operates the SM-DP+. In a \emph{primary} deployment, where the home MNO runs the SM-DP+ end-to-end, identifier reuse remains within the operator's existing administrative boundary; compared with physical SIM issuance, the additional privacy exposure is mainly to insider misuse or data breaches. In a \emph{non-primary} deployment, by contrast, the SM-DP+ is a third-party or multi-tenant service distinct from the home MNO, creating a new observation point that can correlate provisioning events across operators without user visibility. In our work, we focus on the non-primary deployment model where ZK-eSIM is more valuable.

This non-primary setting is already common in practice. Motallebighomi et al.\ \cite{motallebighomi_esimplicity_2025} conduct a large-scale empirical analysis of travel eSIM products and reseller platforms. They find that ``travel eSIM'' connectivity is frequently implemented via home-routed roaming, exporting user traffic and associated metadata to third-country infrastructures, and further show that reseller dashboards can expose sensitive provisioning state, including device identifiers and EID-to-ICCID mappings on commercial SM-DP+ backends \cite{motallebighomi_esimplicity_2025}. Current deployment practices amplify this risk. In practice, SM-DP+ platforms are often multi-tenant services serving hundreds of operators; for example, Workz advertises a cloud platform used by over 100 operators worldwide \cite{passett_workz_2024}. This consolidation turns the SM-DP+ into a high-leverage observation point: a single backend can observe when the same device provisions profiles from multiple operators and at multiple times, enabling cross-operator linkage, coarse travel-history inference, and user association over time. This is a serious threat if provisioning logs are monetised, subpoenaed, or leaked \cite{motallebighomi_esimplicity_2025}. The threat surface is especially misaligned with the user's intent in adopting travel eSIMs. Users increasingly buy travel eSIMs to reduce dependence on home-carrier roaming arrangements and to obtain ``local-like'' connectivity on demand. Yet the provisioning layer reuses identifiers precisely where users expect ephemerality. The EU ``Roam Like at Home'' regime does not cover all countries (e.g., Switzerland), pushing subscribers into expensive day passes, and UK roaming passes (\pounds2--\pounds6/day) often exceed the cost of a travel eSIM (e.g., \pounds7.21 for 5GB in Italy) \cite{verbraucherzentrale_international_2026, deutsch_is_2025}. The market conditions that drive short-lived travel profile adoption also intensify repeated provisioning events, making \textbf{R2} a practical and recurring privacy problem.

\noindent\textbf{Limitations of Strawman Solutions}. We consider two alternatives that might appear to address these risks.
One could attempt to address these risks at the hardware layer by rotating or randomising the EID itself (e.g., via virtual identifiers). However, the EID is a factory-programmed, device-lifetime identifier embedded in the eUICC; modifying it would require a hardware-rooted redesign that is difficult to deploy at ecosystem scale. A user might attempt to achieve unlinkability by provisioning multiple profiles in advance and rotating among them across sessions. This is infeasible: eUICCs support only a limited number of profiles, and each profile remains bound to the same EID during download, reintroducing linkability.
ZK-eSIM therefore keeps the EID and profile-storage model intact and shifts privacy protection entirely to the protocol layer.
\section{System Model, Threat Model, and Design Goals}

\label{sec:designgoals}

We first define the system entities, followed by adversarial capabilities, and security and privacy goals of ZK-eSIM.

\noindent $\bullet$ \textbf{ System Model}\label{sec:system-model}: \emph{ZK-eSIM} consists of five principal entities: UE, MNO, SM-DP+, Pseudonym Certificate Authority (PCA), and LEA. It additionally relies on the existing GSMA PKI as external trust infrastructure. Below, we describe the roles and capabilities of each entity.

\emph{\underline{- User Equipment (UE):}} consists of the LPA and a tamper-resistant eUICC. The LPA mediates profile lifecycle operations, while the eUICC stores profiles and long-term secrets and executes security-sensitive cryptographic operations. We assume the standard GSMA hardware protections and model the LPA--eUICC interface as a local authenticated channel preserving command integrity and response authenticity.

\emph{\underline{- Mobile Network Operator (MNO)}}: provides the subscription service, performs eligibility and subscriber-registration checks, and authorises profile orders with the SM-DP+. Our model also covers MVNO deployments, where the MVNO assumes the operator role while using a host MNO's access infrastructure. End-to-end encryption keeps subscriber data inaccessible to the host MNO. 

\emph{\underline{- Subscription Manager--Data Preparation (SM-DP+) }}: prepares,\\ stores, and delivers operator profiles to authorised eUICCs. It receives profile orders from the MNO and performs authenticated and confidential profile delivery. We consider both dedicated single-tenant and shared multi-tenant SM-DP+ platforms.

\emph{\underline{- Pseudonym Certificate Authority (PCA)}}: is a new entity introduced to address \textbf{R3} (Sec.~\ref{sec:privacy-risks}). It issues short-lived, per-session pseudonym certificates to eligible eUICCs. The PCA operates as an independent Sub-CA under a GSMA Certificate Issuer (CI), allowing each pseudonym certificate to chain to a trust anchor already recognised by UEs and SM-DP+ servers. The certificates contain only the metadata required for validation, such as their validity period.

\emph{\underline{- Law Enforcement Authority (LEA)}}: is an external governance entity involved only in de-pseudonymisation under lawful process. It cannot unilaterally deanonymise a subscriber; de-pseudonymisation requires joint cooperation with an MNO under applicable legal policies, reflecting statutory lawful-interception obligations on service providers~\cite{intoci_p3li5_2023}. For clarity, we model the LEA as a single logical entity. Alternatively, this role may be instantiated by a threshold of independent authorities~\cite{shamir_how_1979,desmedt_threshold_1994}.

\emph{\underline{- Communication and Setup Assumptions}}: All inter-entity communications use TLS: device-facing channels are server-authenticated, while the MNO--SM-DP+ channel is mutually authenticated. All entities additionally have read-only access to a public CRS generated either through an auditable multi-party setup with at least one honest contributor or through a trusted setup whose trapdoor is deleted~\cite{maller_sonic_2019,ben-sasson_secure_2015}.

\noindent\emph{\underline{Deployment Models and Trust Assumptions.}} ZK-eSIM operates among mutually untrusted parties; each entity follows the protocol correctly but may exploit its view. Additionally, ZK-eSIM relies on non-collusion, a practical assumption driven by strict data protection regulations (e.g., GDPR \cite{european_parliament_and_council_of_the_european_union_regulation_2016}) across multi-operator architectures. This aligns with existing privacy-preserving cellular systems (e.g., PGPP~\cite{schmitt_pretty_2021}, LOCA~\cite{luo_loca_2023}, PGUS~\cite{yang_pgus_2025}, LPG~\cite{shi_lpg_2026}). In non-primary deployments (which we adopt in ZK-eSIM), the MNO and SM-DP+ are operated by different organisations, and the PCA may be operated independently, either by the GSMA or by a GSMA-authorised Sub-CA \cite{passett_workz_2024,gsma_security_2019}.  Keeping these records separate reduces data-protection risk and supports GSMA requirements. In primary deployments where an MNO operates its own SM-DP+, we consider them as colluding. \textbf{Appendix~\ref{app:collusion}} provides an analysis of the privacy degradation in this case and in all the remaining pairwise and full-collusion cases.

\noindent$\bullet$\textbf{ Threat Model}\label{sec:threat-model}: Broadly, we  consider two classes of adversaries.

\noindent\underline{\textit{- 
$\mathcal{A}_1$ (Network-level adversary)}}: 
    We adopt the standard Dolev-Yao adversary \cite{dolev_security_1983}, which fully controls the communication channels between entities. The adversary can intercept, delay, replay, modify, or inject messages. The cryptographic keys and the GSMA trust anchors are assumed to remain uncompromised. eUICC‑resident long‑term keys remain hardware‑protected; physical extraction and related hardware attacks are out of scope. This adversary captures risks such as: Man-in-the-middle attacks, passive surveillance, replay attacks, and rogue base stations. Certain threats are out of scope, such as denial-of-service  attacks and global adversaries that can correlate cross-domain traffic.

\noindent\underline{\textit{- $\mathcal{A}_2$ (Honest-but-curious infrastructure adversary):}} We model the MNO, SM-DP+, and PCA as honest-but-curious entities. Each executes its prescribed protocol operations correctly but may analyse its entire legitimate view to deanonymise a subscriber, link multiple provisioning sessions, or derive a stable device identifier. These entities act independently under the non-collusion assumption defined above. We next specify the legitimate view of each entity and the functions for which it is trusted:
\\
\noindent$\circ${ MNO:} observes the subscriber identity, EID, and billing information obtained through the independent onboarding process, together with the provisioning requests, profile orders, and settlement records that it handles. It is trusted to perform eligibility and subscriber-registration checks, authorise profile orders, execute its prescribed provisioning operations, and participate correctly in authorised de-pseudonymisation.  However, because of its honest but curious nature, an MNO may attempt to correlate its onboarding records with provisioning sessions or to link multiple sessions belonging to the same subscriber.

\noindent$\circ${ SM-DP+:} observes profile orders, session-specific authorisation and authentication material, and its profile-delivery and settlement records. It is trusted to prepare and deliver the correct profile, verify the MNO's authorisation material, and enforce one-time token use, but not to refrain from attempting to correlate different provisioning sessions. While conforming to protocol behaviour, the SM-DP+ may attempt to track users or misreport transactions (e.g., inflate the number of provisioned profiles). Settlement integrity is enforced through signed, single-use provisioning tokens and cryptographically bound settlement records, preventing both over-billing and under-reporting.
\\
\noindent$\circ${ PCA:} observes each pseudonym-certificate request, the accompanying eligibility proof, and the short-lived certificate it issues. It is trusted to validate requests, issue certificates only to eligible eUICCs, and protect its signing key, but not to refrain from attempting to link repeated certificate requests.

Under the assumptions above, server‑authenticated TLS neutralises $\mathcal{A}_1$ at the channel level; the privacy violations we study therefore arise under $\mathcal{A}_2$ and depend on \emph{endpoint‑visible} stable identifiers that enable linkability.


\noindent $\bullet$\textbf{ Design Goals of ZK-eSIM}\label{sec:design-goals}: 
ZK-eSIM preserves RSP’s delivery security while ensuring that stable device identifiers are not exposed during provisioning, sessions are unlinkable by default, and identity recovery is possible only via jointly authorised tracing. Our design goals (DG1–DG6) are derived from adversarial threats identified in Sec. \ref{sec:threat-model}. We categorise them into two groups: \emph{baseline security goals} (DG1, DG3), which are retained from RSP, and \emph{novel privacy and accountability goals} (DG2, DG4–DG6).

\noindent \textit{\underline{- DG1: Secure Communication.}}  
For each communicating pair (e.g., UE and SM-DP+), the protocol ensures confidentiality, integrity, entity authentication, and replay protection against a Dolev--Yao adversary controlling the network.

  \noindent\textit{\underline{- DG2: Unlinkable Mutual Authentication.} } 
Mutual authentication must verify endpoints without revealing persistent device identifiers or enabling linkage across sessions. In ZK-eSIM, the UE proves eligibility in zero knowledge and authenticates using short-lived pseudonym certificates that chain to the GSMA CI. Thus, authentication remains verifiable under the GSMA PKI hierarchy while unlinkable to the subscriber’s long-term identity.

 \noindent
 \textit{\underline{- DG3: Profile Confidentiality and Integrity.}}  
  The eSIM profile must be delivered unmodified and confidentially to the eUICC. ZK-eSIM maintains the guarantee by protecting against honest-but-curious infrastructure through authenticated key exchange, encryption, and end-to-end integrity checks. Each profile package is cryptographically bound to the target via an eUICC-held key, so installation on any other device is invalid.

  \noindent
\textit{\underline{- DG4: Subscriber Anonymity (at provisioning).} } 
  No party, internal (MNO, SM-DP+) or external (network adversary), should learn a subscriber’s real-world identity or permanent device identifiers during provisioning. ZK-eSIM achieves this by replacing stable identifiers with ephemeral pseudonyms and unlinkable proofs.
  
  \noindent
 \textit{\underline{- DG5: Provisioning-Session Unlinkability.}}  
  Distinct provisioning sessions must not be linkable to the same subscriber, even when initiated on the same device or across different operators. This prevents tracking and behavioural profiling by MNOs or hosted SM-DP+ platforms. ZK-eSIM enforces unlinkability via Pseudorandom Function (PRF)-generated pseudonyms and ZKPs, across all phases from profile request to installation.

  \noindent
 \textit{\underline{- DG6: Accountable Traceability.}}  
  Strong anonymity and unlinkability must coexist with lawful accountability. ZK-eSIM ensures that deanonymisation requires collaboration between the LEA and an MNO (holding the relevant provisioning record) under legally authorised conditions. No single entity has the unilateral power to deanonymise, ensuring both regulatory compliance and protection against unjustified surveillance.\\
Table~\ref{tab:comparison} illustrates that conventional GSMA Consumer RSP meets DG1 and DG3 but falls short on DG2 and DG4–DG6. This gap motivates our work, ZK‑eSIM, which is designed to close these specific shortcomings.

\begin{table}[t]
  \centering
  \caption{Design Goals: Conventional Consumer RSP vs. ZK‑eSIM}
  \label{tab:comparison}
  \resizebox{\columnwidth}{!}{%
  \begin{tabular}{lcc}
  \toprule
  \textbf{Design Goal} & \textbf{Conventional RSP} & \textbf{ZK-eSIM} \\ \midrule
  DG1: Secure Communication & \cmark & \cmark \\
  DG2: Unlinkable Mutual Authentication & \xmark & \cmark \\
  DG3: Profile Confidentiality \& Integrity & \cmark & \cmark \\
  DG4: Subscriber Anonymity (at Provisioning) & \xmark & \cmark \\
  DG5: Provisioning-Session Unlinkability & \xmark & \cmark \\
  DG6: Privacy-Centric Accountable Traceability (LEA-mediated) & \xmark & \cmark \\
  \bottomrule
  \end{tabular}
  }
 \vspace{-2em}
\end{table}

\section{High-level Overview of ZK-eSIM}\label{sec:tech-overview}

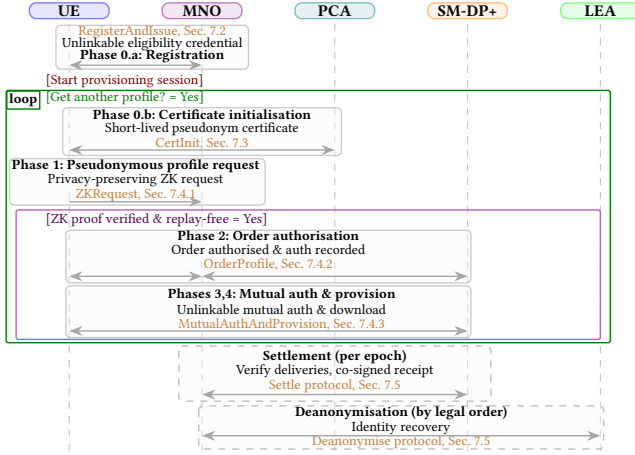
\begin{figure}[t]
  \centering
 \resizebox{1\linewidth}{!}{

\begin{tikzpicture}[
    >=Stealth,
    font=\footnotesize,
    every node/.style={inner sep=1pt},
    header/.style={draw, rounded corners=3pt, minimum height=0.2cm, minimum width=1.2cm, font=\footnotesize\bfseries},
    fragframe/.style={draw, rounded corners=1pt},
    fraglabel/.style={fill=gray!5, font=\scriptsize\bfseries, inner sep=1pt, anchor=north west},
    fragcond/.style={font=\scriptsize},
    arr/.style={semithick, gray!70},
]

\newcommand{\xUE}{0}
\newcommand{\xMNO}{2}
\newcommand{\xPCA}{4}
\newcommand{\xSMDP}{6}
\newcommand{\xLEA}{8}

\newcommand{\sh}{0.176}

\node[header, fill=blue!10,   draw=blue!50]   at (\xUE,0)   {UE};
\node[header, fill=violet!10, draw=violet!50]  at (\xMNO,0)  {MNO};
\node[header, fill=teal!10,   draw=teal!50]    at (\xPCA,0)  {PCA};
\node[header, fill=orange!10, draw=orange!50]  at (\xSMDP,0) {SM-DP+};
\node[header, fill=green!10, draw=green!50]  at (\xLEA,0) {LEA};

\foreach \x in {\xUE, \xMNO, \xPCA, \xSMDP, \xLEA} {
    \draw[dashed, gray!50, thin] (\x,-0.2) -- (\x,-6.7);
}

\draw[rounded corners=2pt, fill=gray!5,fill opacity=0.4, draw=gray!40]
    (-0.2,-0.22) rectangle (2.7,-0.88);
\node[font=\scriptsize\bfseries, anchor=south] at (1.25,-0.82) {Phase 0.a: Registration};
\node[font=\scriptsize, anchor=south] at (1.25,-0.64) {Unlinkable eligibility credential};
\node[font=\scriptsize, anchor=south, text=brown!100] at (1.25,-0.46) {RegisterAndIssue, Sec.\ 7.2};
\draw[<->, arr] (\xUE,-0.82) -- (\xMNO,-0.82);
\node[fragcond, anchor=west, text=red!50!black] at (-0.4,-1.05) {[Start provisioning session]};

\draw[fragframe, green!50!black] (-0.95,-1.2) rectangle (8.15,-5);
\node[fraglabel, draw=green!50!black] at (-0.95,-1.2) {loop};
\node[fragcond, anchor=west, text=green!50!black] at (-0.4,-1.31) {[Get another profile? = Yes]};

\draw[rounded corners=2pt, fill=gray!5, fill opacity=0.4, draw=gray!40]
    (-0.1,-1.460) rectangle (4.1,-2.180);
\node[font=\scriptsize\bfseries, anchor=center] at (2.0,-1.552) {Phase 0.b: Certificate initialisation};
\node[font=\scriptsize, anchor=center] at (2.0,-1.772) {Short-lived pseudonym certificate};
\node[font=\scriptsize, anchor=center, text=brown] at (2.0,-1.992) {CertInit, Sec.\ 7.3};
\draw[<->, arr] (\xUE,-2.098) -- (\xPCA,-2.098);
\draw[rounded corners=2pt, fill=gray!5, fill opacity=0.4, draw=gray!40]
    (-0.90,-2.227) rectangle (2.95,-2.927);
\node[font=\scriptsize\bfseries, anchor=center] at (1.0,-2.331) {Phase 1: Pseudonymous profile request};
\node[font=\scriptsize, anchor=center] at (1.0,-2.551) {Privacy-preserving ZK request};
\node[font=\scriptsize, anchor=center, text=brown] at (1.0,-2.771) {ZKRequest, Sec.\ 7.4.1};
\draw[->, arr] (\xUE,-2.877) -- (\xMNO,-2.877);
\draw[fragframe, violet!60] (-0.8,-3) rectangle (8.0,-4.95);

\node[fragcond, anchor=west, text=violet!60!black] at (-0.4,-3.15) {[ZK proof verified \& replay-free = Yes]};

\draw[rounded corners=2pt, fill=gray!5, fill opacity=0.4, draw=gray!40]
    (-0.05,-3.306) rectangle (6.05,-4.1);
\node[font=\scriptsize\bfseries, anchor=center] at (3.0,-3.380) {Phase 2: Order authorisation};
\node[font=\scriptsize, anchor=center] at (3.0,-3.600) {Order authorised \& auth recorded};
\node[font=\scriptsize, anchor=center, text=brown] at (3.0,-3.820) {OrderProfile, Sec.\ 7.4.2};
\draw[<->, arr] (\xUE,-4.00) -- (\xMNO,-4.00);
\draw[<->, arr] (\xMNO,-4.00) -- (\xSMDP,-4.00);

\draw[rounded corners=2pt, fill=gray!5, fill opacity=0.4, draw=gray!40]
    (-0.05,-4.143) rectangle (6.05,-4.923);
\node[font=\scriptsize\bfseries, anchor=center] at (3.2,-4.277) {Phases 3,4: Mutual auth \& provision};
\node[font=\scriptsize, anchor=center] at (3.2,-4.497) {Unlinkable mutual auth \& download};
\node[font=\scriptsize, anchor=center, text=brown] at (3.2,-4.717) {MutualAuthAndProvision, Sec.\ 7.4.3};
\draw[<->, arr] (\xUE,-4.823) -- (\xSMDP,-4.823);

\draw[rounded corners=2pt, fill=gray!5, fill opacity=0.4, draw=gray!50, dashed]
    (1.65,-5.05) rectangle (6.35,-5.881);
\node[font=\scriptsize\bfseries, anchor=center, text=black] at (4.0,-5.209) {Settlement (per epoch)};
\node[font=\scriptsize, anchor=center] at (4.0,-5.429) {Verify deliveries, co-signed receipt};
\node[font=\scriptsize, anchor=center, text=brown] at (4.0,-5.649) {Settle protocol, Sec.\ 7.5};
\draw[<->, arr] (\xMNO,-5.781) -- (\xSMDP,-5.781);============================================================
\draw[rounded corners=2pt, fill=gray!5, fill opacity=0.4, draw=gray!50, dashed]
    (1.95,-5.951) rectangle (8.05,-6.601);
\node[font=\scriptsize\bfseries, anchor=center, text=black] at (5.0,-6.059) {Deanonymisation (by legal order)};
\node[font=\scriptsize, anchor=center] at (5.0,-6.279) {Identity recovery};
\node[font=\scriptsize, anchor=center, text=brown] at (5.0,-6.499) {Deanonymise protocol, Sec.\ 7.5};
\draw[<->, arr] (\xMNO,-6.401) -- (\xLEA,-6.401);
\end{tikzpicture}}  \caption{High-level phase view of the ZK-eSIM protocol}
	\label{fig:flow}
     \vspace{-2em}
\end{figure}

This section provides a high-level overview of the \emph{ZK-eSIM} workflow (Fig.~\ref{fig:flow}). ZK-eSIM redesigns the GSMA Consumer RSP flow so that, during provisioning, no non-colluding provisioning party can link a provisioning session to both the subscriber record and the eUICC’s long-term identifier. The core idea is to replace cleartext stable identifiers with an unlinkable eligibility credential, a fresh per-session pseudonym certificate, a challenge-derived pseudonym and escrowed EID with a ZKP of correct formation, and one-time authorisation material bound to the session context. Lawful subscriber resolution remains possible, but requires joint LEA--MNO cooperation. The phases of ZK-eSIM are:

\noindent \textbf{\textit{Phase 0.a - Registration and Eligibility }}:
The user enrols once with the MNO and obtains an unlinkable eligibility credential bound to the same eUICC that will later generate per-session keys.

\noindent{\textbf{\textit{Phase 0.b -- Certificate Initialisation}}}:
Before each provisioning session, the eUICC derives a fresh ephemeral key pair and obtains from the PCA a short-lived pseudonym certificate binding the corresponding public key to a bounded validity window. 

\noindent \textbf{\textit{Phase 1 -- Pseudonymous Profile Request}}:
The UE sends to the MNO a privacy-preserving request containing a fresh
challenge-derived per-session pseudonym, an escrow ciphertext EID, a pseudonym certificate, and a ZKP that the pseudonym,
escrow, and previously issued eligibility credential are bound to the same
eUICC and are well formed. The MNO verifies the proof, checks challenge
freshness, and rejects replays without learning a stable identifier.

\noindent \textbf{\textit{Phases 2--4 - Order authorisation, mutual authentication, and provisioning }}:
Upon successful verification, the MNO computes a hashed pseudonym and places an order at the SM-DP+ using only that hashed pseudonym. It also returns to the UE an authorisation bundle consisting of an MNO-issued authorisation credential and a single-use token, both bound to the authorised hashed pseudonym and to the hash of the presented pseudonym certificate. The UE then initiates a fresh provisioning session with the SM-DP+. The SM-DP+ verifies that the hashed pseudonym was authorised, recomputes the certificate hash, checks the authorisation credential and the one-time token, and enforces one-time token use. Only after these checks do the parties complete mutual authentication and an authenticated key exchange, after which the SM-DP+ delivers the encrypted profile for installation on the eUICC. If the user requests another profile, the loop repeats from Phase~0.b with fresh session credentials.

\noindent \textbf{\textit{Supporting Procedure - Settlement}}:
Operational reconciliation runs per epoch between the MNO and the SM‑DP+. Each side maintains a log: the MNO over issued tokens, the SM‑DP+ over redeemed tokens. The SM‑DP+ proves that each redeemed token corresponds to a prior authorisation; both parties co‑sign a settlement receipt. No persistent identifier appears in settlement.

\noindent{\textbf{\textit{Supporting Procedure - Deanonymisation by Exception}}}:
Under a valid legal order, the LEA requests the in-scope escrow ciphertext from the MNO, decrypts it to recover the EID, and returns the result to the MNO for subscriber resolution. Thus, escrow decryption and subscriber resolution are separated, and subscriber deanonymisation requires joint LEA--MNO cooperation.

Phase~0.a establishes eligibility once, decoupled from provisioning. For each new profile, Phases~0.b--4 use a fresh certificate, pseudonym, and one-time authorisation bundle. Under the paper’s threat model, this yields provisioning-session unlinkability for the MNO and SM-DP+ views while preserving operational settlement and lawful traceability by exception. Collectively, these phases realise design goals DG1--DG6 (Sec.~\ref{sec:design-goals}).

\section{Preliminaries for ZK-eSIM}\label{sec:preliminaries}
To keep the paper self-contained, we recall the cryptographic primitives used in our construction. Let $\lambda\in\mathbb{N}$ denote the security parameter and $1^\lambda$ for its unary encoding. For a finite set $S$, $x \sample S$ denotes a uniform sample from $S$. For a (probabilistic) algorithm ${A}$, we write $y \leftarrow $A$(x;r)$ for the result of running $A$ on input~$x$ using randomness $r$; when $r$ is omitted, probabilities are taken over all internal randomness. We use $\{0,1\}^*$ for bitstrings, and $a\gets b$ to denote assignment. All algorithms are probabilistic polynomial-time (PPT) unless stated otherwise. A function $\negl(\lambda)$ is \emph{negligible} if for every $k\in\mathbb{N}$ there exists $\lambda_0$ such that for all $\lambda\ge \lambda_0$, $\negl(\lambda)\le \lambda^{-k}$. 
All notation used in the paper and additional background are deferred to Appendix~\ref{app:additional-preliminaries}.

\subsection{Non-Interactive Zero-Knowledge}

A non-interactive zero-knowledge argument (NIZK) allows a prover ${P}$ to convince a verifier ${V}$ of a statement $x$ without leaking anything beyond its validity, using a single message and CRS~\cite{blum_non-interactive_1988}. We consider an NP-relation $R \subseteq \{0,1\}^* \times \{0,1\}^*$ where $(x,w)\in R$ means witness $w$ proves statement $x$; the corresponding language is $L_R := \{\, x \mid \exists w : (x,w)\in R \,\}$.
A NIZK for $R$ consists of PPT algorithms $(\mathsf{Setup}, \mathsf{Prove}, \mathsf{Verify})$ and simulator $S=(S_1,S_2)$ with the following interfaces: 
$\mathsf{crs} \gets \mathsf{Setup}(1^\lambda)$ outputs a common reference string; 
$\pi \gets \mathsf{Prove}(\mathsf{crs}, x, w)$ generates a proof for statement $x$ with witness $w$ (output $\bot$ if $(x,w)\notin R$); 
$b\gets\mathsf{Verify}(\mathsf{crs}, x, \pi)$ verifies proof $\pi$ for statement $x$, returning $b\in\{0,1\}$; 
$(\mathsf{crs},\mathsf{td}) \gets S_1(1^\lambda)$ generates a simulated CRS (computationally indistinguishable from real setup) with trapdoor $\mathsf{td}$; 
$\pi \gets S_2(\mathsf{td}, x)$ generates a simulated proof using the trapdoor. The NIZK satisfies completeness, zero-knowledge, and knowledge soundness.

\subsection{ZK-eSIM: Definitions}
\label{sec:zk-esim-defs}

\begin{figure*}[t]
\centering
\includegraphics[trim=100 30 100 5, width=0.8\linewidth]{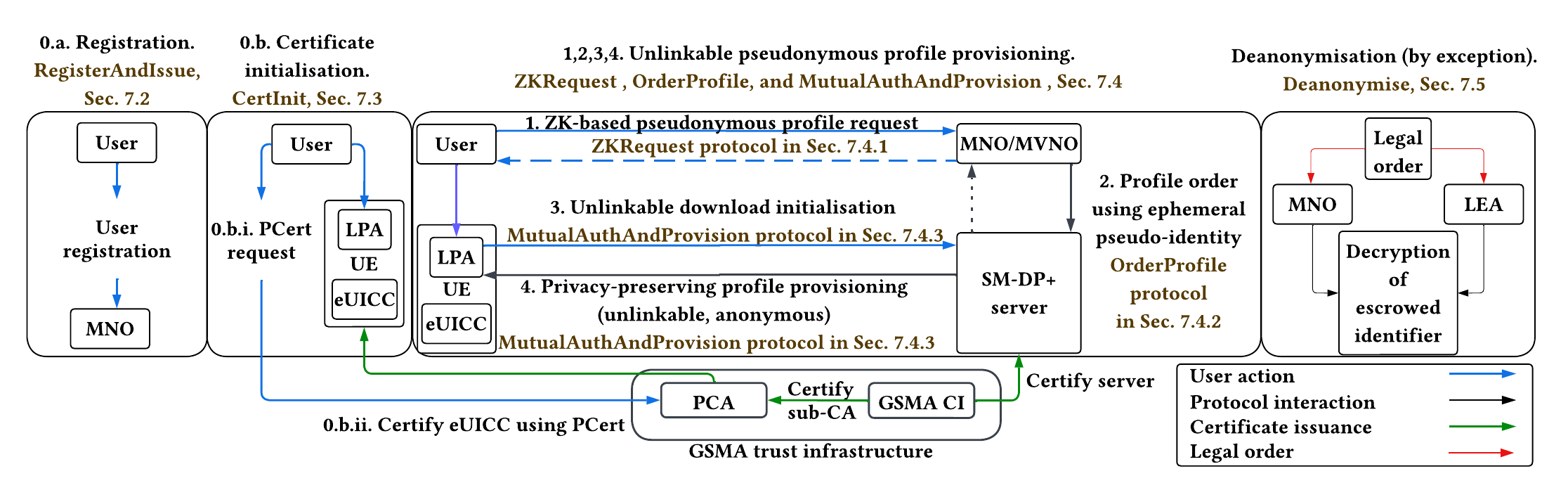}
\caption{System architecture and protocol phases of ZK-eSIM.} 
\label{fig:architecture}
\end{figure*}

ZK‑eSIM is a primitive that enables a user holding an eligible eUICC to be authorised by the MNO and obtain an eSIM profile from SM-DP+ such that the correctness of authorisation and provisioning is verifiable, while the issuance and use phases remain unlinkable and the user’s identifying information is hidden.
We formalise \emph{ZK-eSIM} as a privacy-preserving provisioning scheme that
achieves anonymity at provisioning, multi-session unlinkability, and accountable
traceability. The system is executed by
the entities defined in Sec.~\ref{sec:system-model}. Let $\pp$ denote the \emph{public} parameters output by \textsf{Setup} in Alg. \ref{algo:Setup} and shared by all algorithms. 

\begin{definition}[ZK-eSIM]
\label{def:zk-esim}
A ZK-eSIM scheme is a tuple of PPT algorithms and protocols
$
\Pi=\bigl(\Setup,\RegisterAndIssue,\CertInit,$\\$\ZKRequest,\OrderProfile,\MutualAuthAndProvision,\Settle,$\\$\Deanonymise\bigr),
$ with \(\pp\) as implicit input. Correctness requires that, for any honest execution
with fresh session randomness, \(U\) obtains the intended profile, the corresponding
one-time token is recorded exactly once, and the MNO and SM-DP+ obtain consistent
logs and receipts.
\end{definition}
\noindent$\bullet$ \underline{($\pp$, $L_{\mathsf{auth}}$, $L_{\mathsf{spent}}$, $sk_b$) $\gets \Setup(1^\lambda)$}: 
  the setup algorithm (Alg.~\ref{algo:Setup}) outputs $\pp$, initialises empty \emph{per‑entity} logs $L_{\mathsf{auth}}, L_{\mathsf{spent}}=\emptyset$ (kept private by their respective owners MNO and SM-DP+), and samples a per‑card binding secret $sk_b$ \emph{kept on eUICC and never disclosed}.\\
$\bullet$ \underline{$(\sigma_{\EID}, C_{\EID},r_b) \leftarrow \RegisterAndIssue(\pp;\ \EID,sk_b)$}: 
the one-time registration protocol (Fig.~\ref{fig:UserReg}, Alg.~\ref{algo:RegisterAndIssue}) between $U$ and $\MNO$ takes  $(\EID,sk_b)$, then returns to $U$ an unlinkable eligibility credential $\sigma_{\EID}$, commitment $C_{\EID}$ and randomness $r_b$. $sk_b$, $r_b$ are local to $U$.\\
\noindent$\bullet$ \underline{$(\PCert_U, sk_U, r_\mathsf{seed}) \leftarrow \CertInit(\pp;\ \sigma_{\EID},\EID, sk_b)$}: the certificate initialisation protocol (Fig.~\ref{fig:certInit}, Alg.~\ref{algo:certInit}) between \(U\) and \(\PCA\) samples a fresh session
  key pair \((sk_U,pk_U)\), issues to \(U\) a short-lived pseudonym certificate
  \(\PCert_U\) on \(pk_U\), and returns fresh session randomness
  \(r_\mathsf{seed}\). The values \(sk_b, sk_U, \EID\) are used only locally.\\
$\bullet$ \underline{$(x, \PCert_U, \pi_{\mathsf{req}}, \nonce_{\MNO}) \leftarrow \ZKRequest(\pp; \sigma_{\EID}, sk_b, \EID,$}\\\underline{$ r_\mathsf{seed},  \PCert_U$)}: the request protocol (Fig.~\ref{fig:RSP_Setup_Protocol}, Alg.~\ref{algo:ZKRequest}) run between $U$ and $\MNO$ takes $(\sigma_{\mathsf{EID}},sk_b,\EID,r_\mathsf{seed},\PCert_U)$ as input (where $U$ uses $sk_b$, $r_\mathsf{seed}$ locally, never transmitted), then outputs the statement $x=(pk_{\MNO}, pk_{\mathsf{LEA}}, pk_U, \mathsf{nonce}, \pid, \EncEid )$, $\PCert_U$, $\pi_{\mathsf{req}}$, and $\nonce_{\MNO}$ (stored by MNO),  attesting correctness in zero knowledge and same-eUICC binding without revealing $\EID$.\\ 
 $\bullet$ \underline{$(T_i,\sigma_{\cred}, \Hpid, \pi_{\mathsf{inc}},\mathsf{root}_{\mathsf{auth}},\sigma^{\mathsf{root}}_{\MNO}) \leftarrow \OrderProfile(\pp;$}$\\$ \underline{$ \ x, \pi_{\mathsf{req}}, \PCert_U, \nonce_{\MNO}, L_{\mathsf{auth}})$}: the order-side algorithm (Fig. \ref{fig:RSP_Setup_Protocol}, Alg.~\ref{algo:OrderProfile}) run by $\MNO$ takes $x$ as input, accepts if the certificate chain and the request proof verify and the nonce matches, then appends the hashed pseudonym $\Hpid$ to $L_{\mathsf{auth}}$ (which is stored privately by $\MNO$), places a $\mathsf{DownloadOrder}$ for $\Hpid$ at SM-DP+, and returns to $U$ the authorisation credential $\sigma_{\cred}$, the one-time token $T_i$, and the inclusion proof $\pi_{\mathsf{inc}}$ together with the accumulator
root $\mathsf{root}_{\mathsf{auth}}$ and its MNO signature
$\sigma^{\mathsf{root}}_{\MNO}$.\\
  $\bullet$ \underline{$\mathsf{status} \leftarrow \MutualAuthAndProvision (\pp;\ \PCert_U, sk_U,$} \\
\underline{$ \Hpid, T_i, \sigma_{\cred}, \pi_{\mathsf{inc}}, \mathsf{root}_{\mathsf{auth}}, \sigma^{\mathsf{root}}_{\MNO})$}: the download-side protocol (Fig. \ref{fig:RSP_MutualAuth_Protocol}, Alg.~\ref{algo:MutualAuthAndProvision}) between $U$ and SM-DP+ takes the MNO-issued artifacts together with $(\PCert_U,sk_U)$, performs mutual authentication, verifies $\Hpid\!\in\!L_{\mathsf{auth}}$ and the bindings of $\sigma_{\cred}$ and $T_i$, enforces short-lived certificate use and no-double-spend by inserting $T_i$ into $L_{\mathsf{spent}}$ on accept, delivers and installs the profile on the eUICC, and outputs a single acceptance flag $status\in\{\mathsf{OK},\bot\}$.\\
$\bullet$ \underline{$\mathsf{SR}_\tau \leftarrow \Settle(\pp;\ \tau, L_{\mathsf{spent}}, L_{\mathsf{auth}}, \mathsf{TokSet}_\tau)$}: the settlement protocol (Alg.~\ref{algo:settle}) for epoch $\tau$ between $\MNO$ and SM-DP+ reconciles tokens in $\mathsf{TokSet}_\tau$ against the spent ledger $L_{\mathsf{spent}}[\tau]$ and the authorisation ledger $L_{\mathsf{auth}}[\tau]$, computes a tariff-based settlement amount, and outputs a cross-signed settlement receipt $\mathsf{SR}_\tau$.\\
$\bullet$ \underline{$(\EID', \pi)/\bot \leftarrow \Deanonymise(\EncEid, \textsf{warrant}, sk_{\mathsf{LEA}})$}: the accountable de-anonymisation protocol (Alg.~\ref{algo:deanonymise}) is executed between $\mathsf{LEA}$ (holding $sk_{\mathsf{LEA}}$) and $\mathsf{MNO}$ (holding $\EncEid$). Given a valid warrant defining a lawful scope $S$, $\mathsf{LEA}$ decrypts the escrowed identity and returns $(\EID', \pi)$ to $\mathsf{MNO}$, where $\pi$ proves correct decryption. $\mathsf{MNO}$ verifies $\pi$ and resolves $\EID'$ to the subscriber's KYC record, or outputs $\bot$ if any policy or verification check fails.

  ZK-eSIM achieves provisioning-session unlinkability and unforgeability; \textbf{refer to Sec.~\ref{sec:security} for formal security analysis.}
 

\section{ZK-eSIM Construction}\label{sec:proposed}

 We present the end-to-end construction (Fig.~\ref{fig:architecture}), using the cryptographic interfaces of Sec.~\ref{sec:zk-esim-defs} and the notation in Appendix \ref{app:additional-preliminaries}. The remainder of this section consists of: 1. \textbf{System Setup}, 2. \textbf{Registration}, 3. \textbf{Certificate Initialisation}, 4. \textbf{Unlinkable Pseudonymous Profile Provisioning}, and 5. \textbf{Supporting Procedures}.

\textbf{1. System Setup}\label{sec:system-setup}: $\Setup(1^\lambda)$ initialises the public parameters, cryptographic interfaces and operational anchors in Alg. \ref{algo:Setup}. It also initialises per-entity logs, and seals a per‑card binding secret $sk_b$ inside each eUICC. The MNO creates an append‑only authorisation log: $L_{\mathsf{auth}}$ indexed by hashed subscriber pseudonyms ${\mathsf{Hpid}}$. SM\mbox{-}DP$^+$ initialises $L_{\mathsf{spent}}$, an append‑only record of one‑time tokens used during provisioning to prevent reuse. Both logs are commitment-
backed: each owner signs the current log digest, and short
inclusion proofs can be verified against the signed root. 

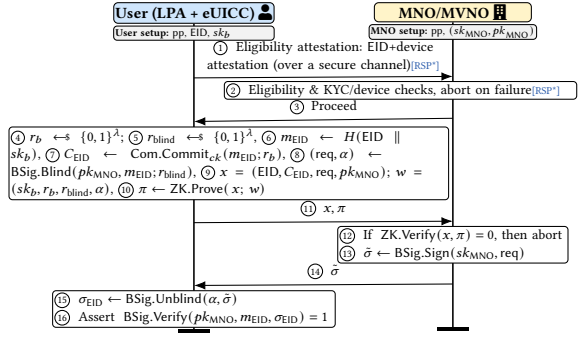
\begin{figure}[]
  \centering
  \resizebox{0.9\columnwidth}{!}{%
    \begin{tikzpicture}[x=1pt,y=1pt]
  \def\xS{0}
  \def\xM{240}
  \def\topY{220}
  \node[role, fill=partyblue, minimum width=90pt, inner sep=1pt, anchor=north west]
    (SNO) at (\xS,\topY)
    {\textbf{User (LPA + eUICC)}~\faUser};
  \node[role, fill=partyyellow, minimum width=90pt, inner sep=1pt, anchor=north east]
    (MNO) at (\xM,\topY)
    {\textbf{MNO/MVNO}~\faBuilding};
  \draw[lifeline] (SNO.south) -- ++(0,-175);
  \draw[lifeline] (MNO.south) -- ++(0,-175);
  \node[setupmini, anchor=north, yshift=-1, text width=90pt, inner sep=1pt]
    (UserSetup) at (SNO.south) {%
    \textbf{User setup: }$\pp$, $\EID$, $sk_b$
  };
  \node[setupmini, anchor=north, yshift=-1, text width=94pt, inner sep=1pt]
    (MNOSetup) at (MNO.south) {%
    \textbf{MNO setup: }$\pp$, $({sk}_{\mathsf{MNO}},{pk}_{\mathsf{MNO}})$
  };
  \node[blackcap] at ($(SNO.south)+(0,-176)$) {};
  \node[blackcap] at ($(MNO.south)+(0,-176)$) {};
  \draw[<-, line width=0.6pt]
    ($(MNO.south)+(0,-31)$) -- ($(SNO.south)+(0,-31)$)
    node[midway, above, font=\small, text width=150pt, align=center]
    {\step{1}\; Eligibility attestation: $\EID$+device attestation (over a secure channel)\RSPA};
  \node[msgbox, text width=200pt, inner sep=1pt]
    (box1a) at ($(MNO.south)+(0-30,-39)$) {%
    \step{2}\; Eligibility \& KYC/device checks,
    abort on failure\RSPA
  };
  \draw[->, line width=0.6pt]
    ($(MNO.south)+(0,-55)$) -- ($(SNO.south)+(0,-56)$)
    node[midway, above, font=\small]
    {\step{3}\; Proceed};
  \node[msgbox, text width=250pt, inner sep=1pt]
    (boxLocal) at ($(SNO.south)+(0+20,-80)$) {%
    \step{4} $r_b \sample \{0,1\}^{\lambda}$; \step{5} $r_{\mathsf{blind}} \sample \{0,1\}^{\lambda}$, \step{6} $m_{\EID} \gets H(\EID\parallel sk_b)$, \step{7} $C_{\mathsf{EID}} \gets \mathsf{Com.Commit}_{ck}(m_{\mathsf{EID}}; r_b)$, \step{8} $(\mathsf{req}, \alpha) \gets \mathsf{BSig.Blind}(pk_{\mathsf{MNO}}, m_{\EID}; r_{\mathsf{blind}})$, \step{9} $x=(\EID, C_{\EID}, \mathsf{req}, pk_{\mathsf{MNO}})$; $w=(sk_b, r_b, r_{\mathsf{blind}}, \alpha)$,
    \step{10} $\pi \leftarrow \mathsf{ZK.Prove}(\, x;\, w)$
  };
  \draw[<-, line width=0.6pt]
    ($(MNO.south)+(0,-114)$) -- ($(SNO.south)+(0,-114)$)
    node[midway, above, font=\small]
    {\step{11}\; $x, \pi$};
  \node[msgbox, text width=130pt, inner sep=1pt]
    (boxVerify) at ($(MNO.south)+(0,-127.5)$) {%
    \step{12}\; If\; $\mathsf{ZK.Verify}( x, \pi)=0$, then abort\\
    \step{13}\; $\tilde{\sigma} \gets \mathsf{BSig.Sign}(sk_{\mathsf{MNO}}, \mathsf{req})$
  };
  \draw[->, line width=0.6pt]
    ($(MNO.south)+(0,-150)$) -- ($(SNO.south)+(0,-150)$)
    node[midway, above, font=\small]
    {\step{14}\; $\tilde{\sigma}$};
  \node[msgbox, text width=160pt, inner sep=1pt]
    (boxUnblind) at ($(SNO.south)+(0,-163.5)$) {%
    \step{15}\; $\sigma_{\EID} \gets \mathsf{BSig.Unblind}(\alpha, \tilde{\sigma})$\\
    \step{16}\; Assert\; $\mathsf{BSig.Verify}(pk_{\mathsf{MNO}}, m_{\EID}, \sigma_{\EID})=1$
  };
\end{tikzpicture}
  }
  \caption{\textit{RegisterAndIssue} Protocol (Alg. \ref{algo:RegisterAndIssue}): 
  User registration with MNO/MVNO over a secure channel to obtain anonymous credential $\sigma_{\EID}$. \RSPA\, denotes steps that follow the RSP flow with modified semantics. Unmarked steps are new.}
    \label{fig:UserReg}
 \vspace{-1em}
\end{figure}

\textbf{2. Registration (Phase 0.a): Eligibility and Accountability Setup (UE $\leftrightarrow$ MNO)}
\label{sec:phase0.a}: The purpose of this phase (Fig.~\ref{fig:UserReg}, Alg.~\ref{algo:RegisterAndIssue}) is not to hide the user’s identity from the issuer at registration; rather, it is to issue a blind, device-bound eligibility credential that can later be used pseudonymously. The flow proceeds as follows: over a secure channel, the user presents \(\mathsf{EID}\) for eligibility/KYC and the MNO authorises continuation.
Next, the eUICC samples fresh randomness \(r_b\) and \(r_{\mathsf{blind}}\), derives a device-bound digest \(m_{\mathsf{EID}}\) from \((\mathsf{EID}, sk_b)\), computes the commitment \(C_{\mathsf{EID}}\) using \(r_b\), and constructs a blinded signing request \(\mathsf{req}\) and (locally) derives the unblinding factor  \(\alpha\) for later use. Here \(sk_b\) is the device‑sealed bootstrap secret (unique to and never leaving the secure element), used only to bind issuance to this eUICC via \(m_{\mathsf{EID}}\).
The zero‑knowledge step attests that the instance satisfies the relation \(\mathcal{R}_{\mathsf{issue}}\): the prover knows \(sk_b, r_b, r_{\mathsf{blind}}\) (and a corresponding \(\alpha\)) such that (i) \(m_{\mathsf{EID}}\) is correctly derived from \(\mathsf{EID}\) and \(sk_b\); (ii) \(C_{\mathsf{EID}}\) is a valid commitment to that \(m_{\mathsf{EID}}\) under randomness \(r_b\); and (iii) \(\mathsf{req}\) is a well‑formed blinded request under \(pk_{\mathsf{MNO}}\) for the same \(m_{\mathsf{EID}}\), while revealing none of these values.
The statement $x$ and ZKP $\pi$ are then transmitted; the MNO verifies \(\pi\) and, if valid, signs the blinded request and returns \(\tilde{\sigma}\).
Unblinding with \(\alpha\) yields \(\sigma_{\mathsf{EID}}\) (a blind‑signed, device‑bound token that can be instantiated as an anonymous credential \cite{looker_bbs_2026} (e.g., BBS+)), which the eUICC verifies under \(pk_{\mathsf{MNO}}\), completing issuance of \((\sigma_{\mathsf{EID}}, C_{\mathsf{EID}})\).

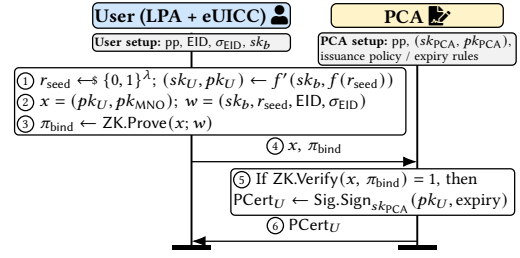
\begin{figure}
  \centering
  \resizebox{0.8\columnwidth}{!}{%
\begin{tikzpicture}[x=1pt,y=1pt]
  \def\xU{40}
  \def\xP{145}
  \def\topY{220}
  \def\botY{-220} 

  \node[role, fill=partyblue,  minimum width=80pt, inner sep=1pt] (USER) at (\xU,\topY) {\textbf{User (LPA + eUICC)}~\faUser};
  \node[role, fill=partyyellow, minimum width=80pt, ,inner sep=1pt] (PCA)  at (\xP,\topY) {\textbf{PCA}~\faFileSignature};

  \draw[lifeline] (\xU,\topY-8) -- (\xU,\botY+332);
  \draw[lifeline] (\xP,\topY-8) -- (\xP,\botY+332);

  \node[setupmini, anchor=north, yshift=-2, text width=90pt, ,inner sep=1pt] at (USER.south) {%
    \textbf{User setup: }$\pp$, $\EID$, $\sigma_{\EID}$, $sk_b$
  };
  \node[setupmini, anchor=north, yshift=-2, text width=90pt, ,inner sep=1pt] at (PCA.south) {%
    \textbf{PCA setup: }$\pp$, $(sk_{\mathsf{PCA}}$, $pk_{\mathsf{PCA}})$, issuance policy / expiry rules
  };

  \node[blackcap] at (\xU,\botY+332) {};
  \node[blackcap] at (\xP,\botY+332) {};
  \node[msgbox, anchor=north east, text width=180pt, inner sep=1pt] at (\xU+100,\topY-24) {%
    \step{1} $r_{\mathsf{seed}} \sample \{0,1\}^\lambda$; $(sk_U, pk_U) \gets f'(sk_b, f(r_{\mathsf{seed}}))$\\
    \step{2} $x=(pk_U, pk_{\MNO})$; $w=(sk_b, r_{\mathsf{seed}}, \mathsf{EID}, \sigma_{\mathsf{EID}})$\\ 
    \step{3} $\pi_{\mathsf{bind}} \leftarrow \textsf{ZK.Prove}(x;w)$
  };
  \draw[->, line width=0.6pt] (\xU,\topY-68) -- (\xP,\topY-68)
        node[midway, above, font=\small]{\step{4} $x,\,\pi_{\mathsf{bind}}$};
  \node[msgbox, anchor=north west, text width=130pt, inner sep=1pt] at (\xP-87,\topY-71) {%
    \step{5} If $\textsf{ZK.Verify}(x,\,\pi_{\mathsf{bind}})=1$, then\\
    $\mathsf{PCert}_U \gets \mathsf{Sig.Sign}_{sk_{\mathsf{PCA}}}(pk_U,\text{expiry})$
  };
  \draw[->, line width=0.6pt] (\xP,\topY-105) -- (\xU,\topY-105)
        node[midway, above, font=\small]{\step{6} $\mathsf{PCert}_U$};
\end{tikzpicture}
  }
    \caption{\textit{CertInit} Protocol (Alg. \ref{algo:certInit}): Pseudonym Certificate Initialisation.}
    \label{fig:certInit}
     \vspace{-1em}
\end{figure}

\textbf{3. Certificate Initialisation (Phase 0.b): Pseudonymous Key Binding (UE $\leftrightarrow$ PCA)}
\label{sec:phase0.b}: This phase (Fig.~\ref{fig:certInit}, Alg.~\ref{algo:certInit}) converts the eligibility credential into a fresh, certified pseudonymous certificate. The eUICC derives a fresh, hardware‑bound public key and proves to the PCA, in zero knowledge, that this key belongs to the same eligibility‑checked device. To preserve multi-session unlinkability, this phase is executed freshly for each provisioning session. The eUICC generates a seed \(r_{\mathsf{seed}}\), derives an ephemeral keypair \((sk_U, pk_U)\) from \(sk_b\) and \(r_{\mathsf{seed}}\), and produces a ZKP \(\pi_{\mathsf{bind}}\) for the relation \(\mathcal{R}_{\mathsf{bind}}\): the prover knows \(sk_b, r_{\mathsf{seed}}, \mathsf{EID}, \sigma_{\mathsf{EID}}\) such that (i) \(pk_U\) is correctly derived within the same eUICC anchored by $sk_b$, (ii) a hash‑based binding \(m_{\mathsf{EID}}\) is consistent with \((\mathsf{EID}, sk_b)\), and (iii) the eligibility credential \(\sigma_{\mathsf{EID}}\) verifies under \(pk_{\MNO}\), all without disclosing these values. The user then transmits \(x\) and $\pi_{\mathsf{bind}}$ to the PCA; upon successful verification, the PCA issues and returns a time‑bounded certificate \(\mathsf{PCert}_U\) over \(pk_U\).

\textbf{4. Unlinkable Pseudonymous Profile Provisioning (Phases 1,2,3,4)}
\label{sec:phase1234}: This section composes Phases~1--4 into an unlinkable provisioning workflow. Our goals are to: (i) prove eligibility in zero knowledge; (ii) let the operator place an order without disclosing the long-lived \(\mathsf{EID}\) at order time; (iii) bind download authorisation to a fresh session and a one-time release token; and (iv) deliver and install the profile without introducing new cross-session linkers. We follow the GSMA Activation Code control flow with modified semantics: the operator-generated Matching ID / activation token is instantiated with the per-session pseudonym \(\Hpid\). This preserves the control-flow shape while replacing the long-lived device identifier with a fresh, session-scoped value.

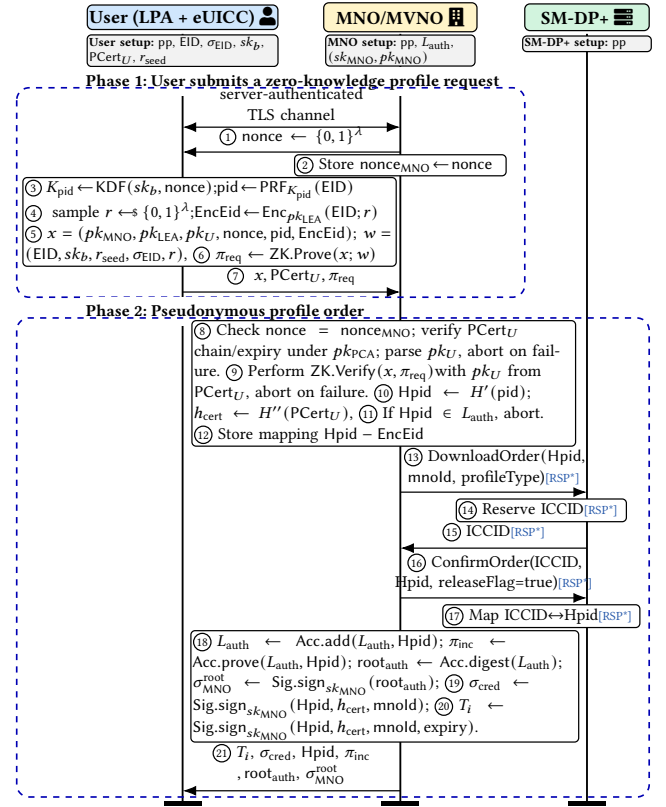
\begin{figure}[]
  \centering
 \resizebox{\columnwidth}{!}{\begin{tikzpicture}[x=1pt,y=1pt]
  \def\xU{40}   
  \def\xM{145}  
  \def\xD{235}  
  \def\topY{280}
  \def\botY{-170}

  \node[role, fill=partyblue, minimum width=60pt , inner sep=1pt] (USER) at (\xU,\topY)
    {\textbf{User (LPA + eUICC)}~\faUser};
  \node[role, fill=partyyellow, minimum width=60pt] (MNO) at (\xM,\topY)
    {\textbf{MNO/MVNO}~\faBuilding};
  \node[role, fill=partygreen, minimum width=60pt] (SDP) at (\xD,\topY)
    {\textbf{SM-DP+}~\faServer};

  \draw[lifeline] (\xU,\topY-8) -- (\xU,\botY+71);
  \draw[lifeline] (\xM,\topY-8) -- (\xM,\botY+71);
  \draw[lifeline] (\xD,\topY-8) -- (\xD,\botY+71);

  \node[setupmini, anchor=north, yshift=-2, inner sep=1pt, text width=90] at (USER.south) {%
    \textbf{User setup: }$\pp$, \(\EID\), \(\sigma_{\EID}\), $sk_b$, \(\PCert_{U}\), \(r_{\mathsf{seed}}\)
  };
  \node[setupmini, anchor=north, yshift=-1, inner sep=1pt, text width=70] at (MNO.south) {%
    \textbf{MNO setup: }$\pp$, \(L_{\text{auth}}\), $(sk_{\MNO}, pk_{\MNO})$
  };
  \node[setupmini, anchor=north, yshift=-2, text width=60pt, inner sep=1pt] at (SDP.south) {%
    \textbf{SM-DP+ setup: }$\pp$
  };

  \node[blackcap] at (\xU,\botY+71) {};
  \node[blackcap] at (\xM,\botY+71) {};
  \node[blackcap] at (\xD,\botY+71) {};

  \def\y{230}

\node[font=\bfseries\small, anchor=west, fill=white]
    at (\xU-50, \y+19)
    {Phase 1: User submits a zero-knowledge profile request};
 
  \draw[dashed,
        rounded corners=6pt,
        line width=0.8pt, draw=blue!70!black]
    (\xU-80, \y+16) rectangle (\xM+60, \y-85);

  \draw[<->, line width=0.6pt] (\xU,\y-3) -- (\xM,\y-3)
    node[midway, above, text width=90pt, font=\small, align=center]{server-authenticated TLS channel};

\draw[<-, line width=0.6pt] (\xU,\y-15) -- (\xM,\y-15)
  node[midway, above, text width=90pt, font=\small, align=center]
  {\step{1} ${\nonce}\leftarrow\{0,1\}^\lambda$};

\node[msgbox, anchor=north, text width=100pt, inner sep=1pt] at (\xM+0,\y-16) {%
  \step{2} Store $\nonce_{\MNO}\!\gets\!\nonce$
};

  \node[msgbox, anchor=north, text width=178pt, inner sep=1pt] at (\xU+13,\y-27) {%
      \step{3}\;\(K_{\pid} \!\gets\! \KDF(sk_b,\mathsf{nonce})\);\(\pid \!\gets\! \PRF_{K_{\pid}}(\mathsf{EID})\)\\
    \step{4}\; sample \(r \sample \{0,1\}^{\lambda}\);\(\EncEid \!\gets\! \Enc_{pk_{\mathsf{LEA}}}(\mathsf{EID}; r)\) \\
    \step{5}
    \(x = (pk_{\MNO}, pk_{\mathsf{LEA}}, pk_U, \mathsf{nonce}, \pid, \EncEid)\); \(w = (\mathsf{EID}, sk_b, r_{\mathsf{seed}}, \sigma_{\mathsf{EID}}, r)\), \step{6} $\pi_{\mathsf{req}} \leftarrow \textsf{ZK.Prove}(x; w)$\\
  };

\node[font=\bfseries\small, anchor=west, fill=white]
    at (\xU-50, \y-92)
    {Phase 2: Pseudonymous profile order};

  \draw[dashed,
        rounded corners=6pt,
        line width=0.8pt, draw=blue!70!black]
    (\xU-80, \y-95) rectangle (\xM+120, \y-326);

  \draw[->, line width=0.6pt] (\xU,\y-82.5) -- (\xM,\y-82.5)
    node[midway, above, text width=90pt, font=\small, align=center]{\step{7}\; \(x, \PCert_U, \pi_{\mathsf{req}}\) };

 \node[msgbox, anchor=north, text width=183pt, inner sep=1pt] at (\xM-8,\y-96) {%
  \step{8} Check $\nonce = \nonce_{\MNO}$; verify $\mathsf{PCert}_U$ chain/expiry under $pk_{\PCA}$; parse $pk_U$, abort on failure.
  \step{9}  Perform $\mathsf{ZK.Verify}(x,\pi_{\mathsf{req}})$with $pk_U$ from $\PCert_U$, abort on failure. \step{10} ${{\mathsf{\Hpid}}} \gets H'(\pid)$; $h_{\mathsf{cert}} \gets H''(\mathsf{PCert}_U)$,
  \step{11} If ${{\mathsf{\Hpid}}} \in L_{\text{auth}}$, abort. \step{12} Store mapping $\Hpid - \EncEid$
};
\draw[->, line width=0.6pt] (\xM,\y-179) -- (\xD,\y-179)
  node[midway, above, text width=100pt, font=\small, align=center]
  {\step{13} DownloadOrder$({\mathsf{Hpid}}$,  $\mathsf{mnoId}$, profileType)\RSPA};
\node[msgbox, anchor=north, text width=80pt, inner sep=1pt] at (\xD-22,\y-182) {%
  \step{14} Reserve {ICCID}\RSPA
};
\draw[<-, line width=0.6pt] (\xM,\y-206) -- (\xD,\y-206)
  node[midway, above, text width=100pt, font=\small, align=center]
  {\step{15} ICCID\RSPA
  };
\draw[->, line width=0.6pt] (\xM,\y-230) -- (\xD,\y-230)
  node[midway, above, text width=140pt, font=\small, align=center]
  {\step{16} ConfirmOrder(ICCID, Hpid, releaseFlag=true)\RSPA};
  \node[msgbox, anchor=north, text width=93pt, inner sep=1pt] at (\xD-22,\y-233) {%
  \step{17} Map ICCID$\leftrightarrow$Hpid\RSPA
};
\node[msgbox, anchor=north, text width=185pt, inner sep=1pt] at (\xM-7.5,\y-245.5) {%
  \step{18} $L_{\text{auth}} \gets \mathsf{Acc.add}(L_{\text{auth}}, {\mathsf{Hpid}})$; $\pi_{\mathsf{inc}} \gets \mathsf{Acc.prove}(L_{\text{auth}}, {\mathsf{Hpid}})$;  $\mathsf{root}_{\text{auth}} \gets \mathsf{Acc.digest}(L_{\text{auth}})$;
  $\sigma^{\text{root}}_{\MNO} \gets \mathsf{Sig.sign}_{sk_{\MNO}}(\mathsf{root}_{\text{auth}})$; \step{19} $\sigma_{\text{cred}} \gets \mathsf{Sig.sign}_{sk_{\MNO}}({\mathsf{Hpid}}, h_{\mathsf{cert}},  \mathsf{mnoId})$; \step{20} $T_i \gets \mathsf{Sig.sign}_{sk_{\MNO}}({\mathsf{Hpid}}, h_{\mathsf{cert}}, \mathsf{mnoId}, \text{expiry})$.
};
  \draw[->, line width=0.6pt] (\xM,\y-323) -- (\xU,\y-323)
    node[midway, above, text width=90pt, font=\small, align=center]{\step{21} $T_i$, $\sigma_{\mathsf{cred}}$, $\Hpid$, $\pi_{\mathsf{inc}}$\\$, \mathsf{root}_{\text{auth}}$, $\sigma^{\text{root}}_{\MNO}$};
\end{tikzpicture}}  \caption{\textit{ZKRequest} (Alg. \ref{algo:ZKRequest}) and \textit{OrderProfile} (Alg. \ref{algo:OrderProfile}) Protocols: Profile Download Initiation. \RSPA\, denotes steps that follow the RSP flow with modified semantics. Unmarked steps are new.}\label{fig:RSP_Setup_Protocol}
\vspace{-1em}
\end{figure}

\textit{- \underline{Phase 1:} ZK-based pseudonymous profile request (U\,$\leftrightarrow$\,MNO)}: This phase (Fig.~\ref{fig:RSP_Setup_Protocol}, Alg.~\ref{algo:ZKRequest}) turns an eligibility‑certified, hardware‑rooted device into a pseudonymous requester over server‑authenticated TLS. The MNO samples a fresh challenge $\nonce$ and sends it, and stores it as $\nonce_{\MNO}$. Inside the eUICC, the device derives a per‑session pseudonym \(\pid\) from \(sk_b\) and \(\mathsf{nonce}\), and in parallel creates an escrow ciphertext \(\EncEid\) of the same \(\mathsf{EID}\) using LEA $pk_{\LEA}$ and fresh randomness \(r\). It then forms the statement \(x\) and witness \(w\) and produces a ZKP \(\pi_{\mathsf{req}}\) for the relation \(\mathcal{R}_{\mathsf{req}}\), asserting knowledge of \((\mathsf{EID}, sk_b, r_{\mathsf{seed}}, \sigma_{\mathsf{EID}}, r)\) such that: (i) the certified key \(pk_U\) corresponds to the same eUICC; (ii) \(\mathsf{EID}\) is bound to \(sk_b\); (iii) the eligibility credential \(\sigma_{\mathsf{EID}}\) verifies under \(pk_{\MNO}\); (iv) \(\pid\) is correctly derived from \((sk_b,\mathsf{nonce})\) and \(\mathsf{EID}\); and (v) \(\EncEid\) encrypts that same \(\mathsf{EID}\) under \(pk_{\mathsf{LEA}}\), without revealing any of these values. The user sends \((x, \mathsf{PCert}_U,\pi_{\mathsf{req}})\) and submits it to the MNO over TLS. Here \(sk_b\) is the device‑sealed bootstrap secret, never exported from the secure element. The MNO learns only a pseudonym and a verifiable proof of hardware continuity and eligibility: unlinkability across sessions follows from the fresh \(\mathsf{nonce}\), while accountability is preserved via the stored mapping \(\mathsf{Hpid}/\EncEid\) and escrow to \(pk_{\mathsf{LEA}}\); thus an eligible device can authenticate pseudonymously yet remain trace‑ready. To keep the core protocol minimal, we treat profile selection and payment as a separate step carried over the same authenticated channel.  Concretely, the client sends a context‑bound, non‑replayable, anonymous proof of payment attesting that the correct tariff was paid to the MNO while revealing nothing about the payer’s identity. A mature body of literature provides suitable instantiations of this primitive~\cite{chaum_blind_1983,ben_sasson_zerocash_2014,nardelli_hitchhikers_2025}. Upon verifying the proof of payment, the MNO treats the user’s request as an authorised order and proceeds with provisioning. This assumption is orthogonal to the ZK request proof and preserves privacy and accountability.

\textit{- \underline{Phase 2:} Pseudonymous profile order (UE $\leftrightarrow$\,MNO\,$\leftrightarrow$\,SM-DP+)} In this phase (Fig.~\ref{fig:RSP_Setup_Protocol}, Alg.~\ref{algo:OrderProfile}), the MNO receives \((x,\PCert_U,\pi_{\mathsf{req}})\) from the user over server-authenticated TLS. It first binds the request to its own challenge and recovers the certified user key by checking \(\nonce=\nonce_{\MNO}\), verifying the chain and expiry of \(\mathsf{PCert}_U\) under \(pk_{\PCA}\), and parsing \(pk_U\); it aborts on failure. The MNO then verifies \(\pi_{\mathsf{req}}\) against \(x\) using the \(pk_U\) recovered from \(\mathsf{PCert}_U\), thereby attesting same-eUICC linkage, eligibility under the MNO's policy, pseudonym correctness, and escrow well-formedness; invalid proofs are rejected. Next, it computes the hashed pseudonym \(\Hpid\) and the certificate hash \(h_{\mathsf{cert}}\), aborts if \(\Hpid \in L_{\mathsf{auth}}\), and stores the mapping \((\Hpid \mapsto \EncEid)\). The MNO then places a $\mathsf{DownloadOrder}$ with the SM-DP+ over an authenticated backend channel. The SM-DP+ reserves a fresh \(ICCID\) and returns it to the MNO, after which the MNO finalises the order with $\mathsf{ConfirmOrder}$. The SM-DP+ then records the internal binding \(ICCID \leftrightarrow \Hpid\). After confirmation, the MNO commits the authorisation decision by updating the accumulator \(L_{\mathsf{auth}}\), producing the inclusion proof \(\pi_{\mathsf{inc}}\), computing the digest \(\mathsf{root}_{\mathsf{auth}}\), and signing it as \(\sigma^{\mathsf{root}}_{\MNO}\). It also issues credentials bound to \((\Hpid,h_{\mathsf{cert}},\mathsf{mnoId})\), namely \(\sigma_{\mathsf{cred}}\), and derives a one-time token \(T_i\) with expiry. These values, together with \((\Hpid,\pi_{\mathsf{inc}},\mathsf{root}_{\mathsf{auth}},\sigma^{\mathsf{root}}_{\MNO})\), are then returned to the user in the final response. ZKRequest and OrderProfile thus convert a ZK-authenticated, pseudonymous request into a minimally identifying yet enforceable authorisation: replay is prevented by checking \(\Hpid \notin L_{\mathsf{auth}}\), the authorisation state is committed via the accumulator root and MNO signature, and subsequent steps can proceed using \(T_i\) and \(\sigma_{\mathsf{cred}}\) without revealing \(\EID\), while accountability and trace-readiness remain intact through the established escrow.
\textit{- \underline{Phases 3 and 4:} Unlinkable Download Initialisation and  Privacy-Preserving Profile Provisioning (LPA/UE/eUICC\,$\leftrightarrow$\,SM-DP+)}
  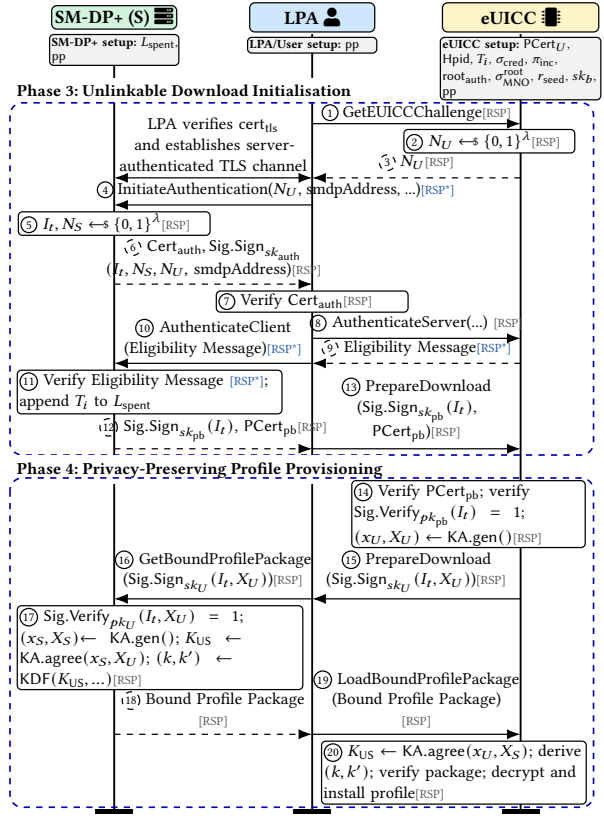
\begin{figure}[!t]
  \centering
 \resizebox{\columnwidth}{!}{\begin{tikzpicture}[x=1pt,y=1pt]

  \def\xS{40}   
  \def\xL{135}  
  \def\xU{235}  
  \def\topY{290}
  \def\botY{-230}

   \node[role, fill=partygreen, minimum width=60pt, inner sep=1pt] (SDP) at (\xS,\topY)
    {\textbf{SM-DP+ (S)}~\faServer};
  \node[role, fill=partyblue, minimum width=60pt] (LPA) at (\xL,\topY)
    {\textbf{LPA}~\faUser};
  \node[role, fill=partyyellow, minimum width=75pt] (EUICC) at (\xU,\topY)
    {\textbf{eUICC}~\faMicrochip};

  \draw[lifeline] (\xS,\topY-8) -- (\xS,\botY+139);
  \draw[lifeline] (\xL,\topY-8) -- (\xL,\botY+139);
  \draw[lifeline] (\xU,\topY-8) -- (\xU,\botY+139);

  \node[setupmini, anchor=north, yshift=-2,  text width=60pt, inner sep=1pt] at (SDP.south) {%
    \textbf{SM-DP+ setup: }$L_{\mathsf{spent}}$, $\pp$
  };
\node[setupmini, anchor=north, yshift=-2,  text width=60pt, inner sep=1pt] at (LPA.south) {%
  \textbf{LPA/User setup: }$\pp$
};
  \node[setupmini, anchor=north, yshift=-2, text width=75pt, inner sep=1pt] at (EUICC.south) {%
    \textbf{eUICC setup: }$\PCert_U$, $\Hpid$, $T_i$, $\sigma_{\mathsf{cred}}$, $\pi_{\mathsf{inc}}$, $\mathsf{root}_{\mathsf{auth}}$, $\sigma^{\mathsf{root}}_{\MNO}$, $r_{\mathsf{seed}}$, $sk_b$, $\pp$
  };

  \node[blackcap] at (\xS,\botY+139) {};
  \node[blackcap] at (\xL,\botY+139) {};
  \node[blackcap] at (\xU,\botY+139) {};

  \def\y{0}
  \node[font=\bfseries\small, anchor=west, fill=white]
  at (\xS-50, \y+255)
  {Phase 3: Unlinkable Download Initialisation};
\draw[dashed,
      rounded corners=6pt,
      line width=0.8pt, draw=blue!70!black]
  (\xS-50, \y+249) rectangle (\xU+35, \y-57+137);

\node[font=\bfseries\small, anchor=west, fill=white]
  at (\xS-50, \y-65+138)
  {Phase 4: Privacy-Preserving Profile Provisioning};

\draw[dashed,
      rounded corners=6pt,
      line width=0.8pt, draw=blue!70!black]
  (\xS-50, \y-70+139) rectangle (\xU+35, \y-89);

   \draw[->, line width=0.6pt, inner sep=1pt] (\xL,\y+226+13) -- (\xU,\y+226+13)
     node[midway, above, font=\small]{\step{1} GetEUICCChallenge\RSP};
\node[msgbox, anchor=north, text width=80pt, inner sep=1pt] at (\xU-15,\y+236.5) {%
    \step{2} $N_U \sample\{0,1\}^\lambda$\RSP
  };

  \draw[->,dashed, line width=0.6pt] (\xU,\y+213) -- (\xL,\y+213)
    node[midway, above, font=\small]{\step{3} $N_U$\RSP};

  \draw[<->, line width=0.6pt] (\xL,\y+213) -- (\xS,\y+213)
    node[midway, above, text width=90pt, font=\small, align=center]{LPA verifies $\mathsf{cert_{tls}}$ and establishes server-authenticated TLS channel};

  \draw[->, line width=0.6pt] (\xL,\y+152+48) -- (\xS,\y+152+48)
    node[midway, above, font=\small]{\hspace{+60pt}\step{4} InitiateAuthentication($N_U$, smdpAddress, ...)\RSPA};

  \node[msgbox, anchor=north, text width=90pt, inner sep=1pt] at (\xS,\y+148+49) {\step{5} 
    $I_t, N_S\sample\{0,1\}^\lambda$\RSP\\
  };

  \draw[->, dashed, line width=0.6pt] (\xS,\y+98+64) -- (\xL,\y+98+64)
    node[midway, above, text width=110pt, font=\small, align=center]{\step{6} $\mathsf{Cert_{\mathsf{auth}}}, \mathsf{Sig.Sign}_{sk_{\mathsf{auth}}}$\\$(I_t, N_S, N_U,$ smdpAddress)\RSP};

  \node[msgbox, anchor=north, text width=90pt, inner sep=1pt] at (\xL+0,\y+95+64) {%
    \step{7} Verify $\mathsf{Cert_{\mathsf{auth}}}$\RSP
  };

  \draw[->, line width=0.6pt] (\xL,\y+49+87) -- (\xU,\y+49+87)
    node[midway, above,  text width=110pt, font=\small, align=center]{
    \step{8} AuthenticateServer(...)
\RSP};

\draw[->, dashed, line width=0.6pt] (\xU,\y+23+101) -- (\xL,\y+23+101)
    node[midway, above, font=\small, text width=100pt, align=center]{%
      \step{9} Eligibility Message\RSPA
    };
 \draw[->, line width=0.6pt] (\xL,\y+23+101) -- (\xS,\y+23+101)
    node[midway, above, font=\small, text width=100pt, align=center]{%
      \step{10} AuthenticateClient\\(Eligibility Message)\RSPA
      };
  \node[msgbox, anchor=north, text width=125pt, inner sep=1pt] at (\xS+17,\y+19+102) {%
    \step{11} Verify Eligibility Message \RSPA; append $T_i$ to $L_{\mathsf{{spent}}}$
  };
\draw[->,dashed, line width=0.6pt] (\xS,\y-53+136) -- (\xL,\y-53+136)
    node[midway, above, text width=110pt, font=\small, align=center]{%
      \step{12} $\mathsf{Sig.Sign}_{sk_{\mathsf{pb}}}(I_t)$, $\PCert_{\mathsf{pb}}$\RSP
    };
 \draw[->, line width=0.6pt] (\xL,\y-53+136) -- (\xU,\y-53+136)
    node[midway, above, text width=100pt, font=\small, align=center]{%
      \step{13} PrepareDownload\\
      ($\mathsf{Sig.Sign}_{sk_{\mathsf{pb}}}(I_t)$, $\PCert_{\mathsf{pb}}$)\RSP
    };
\node[msgbox, anchor=north, text width=110pt, inner sep=1pt] at (\xU-25,\y-73+141) {%
    \step{14} Verify $\PCert_{\mathsf{pb}}$; verify $\mathsf{Sig.Verify}_{pk_{\mathsf{pb}}}(I_t)=1$; $(x_U,X_U)\leftarrow \mathsf{KA.gen}()$\RSP
  };
   \draw[->, line width=0.6pt] (\xU,\y-154+165) -- (\xL,\y-154+165)
    node[midway, above, text width=100pt, font=\small, align=center]{%
      \step{15} PrepareDownload \\
      ($\mathsf{Sig.Sign}_{sk_U}(I_t,X_U)$)\RSP
    };
     \draw[->, line width=0.6pt] (\xL,\y-154+165) -- (\xS,\y-154+165)
    node[midway, above, text width=100pt, font=\small, align=center]{%
      \step{16} GetBoundProfilePackage\\
      ($\mathsf{Sig.Sign}_{sk_U}(I_t,X_U)$)\RSP
    };
\node[msgbox, anchor=north, text width=135pt, inner sep=1pt] at (\xS+22,\y-157+165) {%
    \step{17} $\mathsf{Sig.Verify}_{pk_U}(I_t,X_U)=1$; $(x_S,X_S)$$\leftarrow \mathsf{KA.gen}()$;   $K_{\mathsf{US}}\leftarrow \mathsf{KA.agree}(x_S,X_U)$; $(k,k')\leftarrow \mathsf{KDF}(K_{\mathsf{US}}, ...)$\RSP
  };
  \draw[->, dashed, line width=0.6pt] (\xS,\y-244+190) -- (\xL,\y-244+190)
    node[midway, above, text width=100pt, font=\small, align=center]{%
      \step{18} Bound Profile Package\\\RSP
    };
     \draw[->, line width=0.6pt] (\xL,\y-244+190) -- (\xU,\y-244+190)
    node[midway, above, text width=100pt, font=\small, align=center]{%
      \step{19} LoadBoundProfilePackage\\
      (Bound Profile Package)\\\RSP
    };
  \node[msgbox, anchor=north, text width=125pt, inner sep=1pt] at (\xU-32,\y-248+191) {%
    \step{20} $K_{\mathsf{US}}\leftarrow \mathsf{KA.agree}(x_U,X_S)$;
    derive $(k,k')$;
    verify package;
    decrypt and install profile\RSP
  };

\end{tikzpicture}}  \caption{\emph{ MutualAuthAndProvision} Protocol (Alg. \ref{algo:MutualAuthAndProvision}): Mutual Authentication and Profile Download. \RSP\,denotes unmodified GSMA Consumer RSP steps; 
\RSPA\,denotes steps that follow the same RSP flow with modified semantics. Unmarked steps are new.}
\label{fig:RSP_MutualAuth_Protocol}
\vspace{-1em}
\end{figure}
In these phases (Fig.~\ref{fig:RSP_MutualAuth_Protocol}, Alg.~\ref{algo:MutualAuthAndProvision}), Phase 3 covers server authentication, eligibility validation, and download preparation, while Phase 4 begins with the eUICC's verification of the profile-signer material and continues with the authenticated key establishment and protected profile loading. The LPA first invokes $\mathsf{GetEUICCChallenge}$, causing the eUICC to sample a fresh nonce \(N_U\) and return it. The LPA then verifies $\mathsf{cert_{tls}}$ and establishes a server-authenticated TLS channel to the SM-DP+, before sending $\mathsf{InitiateAuthentication}$. The SM-DP+ samples a fresh transcript identifier \(I_t\) and server nonce \(N_S\), and returns \(\mathsf{Cert_{\mathsf{auth}}}\) together with a signature under \(sk_{\mathsf{auth}}\). After the LPA verifies \(\mathsf{Cert_{\mathsf{auth}}}\), it relays the server-authentication material to the eUICC through $\mathsf{AuthenticateServer}$, thereby binding the session to the fresh challenge pair \((N_U,N_S)\) and transcript \(I_t\). The eUICC then sends an eligibility message = $(
\PCert_U,  \sigma_{\mathsf{cred}},  T_i, \Hpid, \pi_{\mathsf{inc}}, \mathsf{root}_{\mathsf{auth}}\allowbreak, \sigma^{\mathsf{root}}_{\MNO},
 \mathsf{Sig.sign}_{sk_U}(I_t, N_S, N_U, T_i, \Hpid, \PCert_U, \mathsf{smdpAddress}))$ via the LPA, which relays it to the SM-DP+ using $\mathsf{AuthenticateClient}$. The SM-DP+ verifies this eligibility message and atomically appends \(T_i\) to the append-only set \(L_{\mathsf{spent}}\), rejecting reuse. It next returns \((\mathsf{Sig.sign}_{sk_{\mathsf{pb}}}(I_t),Cert_{\mathsf{pb}})\) through the LPA. At the start of Phase 4, the eUICC verifies \(Cert_{\mathsf{pb}}\) and the signature on \(I_t\), thereby authenticating the profile-signing credential for the current transcript, and generates an ephemeral key pair \((x_U,X_U)\). It then signs \((I_t,X_U)\) under \(sk_U\) and returns this authenticated key-share contribution through $\mathsf{PrepareDownload}$ and $\mathsf{GetBoundProfilePackage}$. The SM-DP+ verifies the signature under \(pk_U\), generates its own ephemeral pair \((x_S,X_S)\), computes \(K_{\mathsf{US}}\), and derives \((k,k')\). Finally, the SM-DP+ sends the \emph{BoundProfilePackage}=$(C, t_P, \mathsf{mnoId}, t_{\mathsf{mno}},$\\$
\mathsf{iccid}, t_{\mathsf{iccid}}, X_S, \mathsf{Sig.sign}_{sk_{\mathsf{pb}}}(I_t, X_S, X_U, \mathsf{smdpAddress}))$ via the LPA. Using \(X_S\) carried in that package, the eUICC recomputes \(K_{\mathsf{US}}\), derives the same \((k,k')\), verifies the package, and decrypts and installs the profile. Together, these phases convert the Phase 2 pseudonymous authorisation into a one-time, session-bound profile download: the server is authenticated, client eligibility is validated and spent exactly once, the download keys are established with authenticated eUICC participation, and the resulting profile is delivered under confidentiality and integrity protection.

\textbf{Post-Provisioning:} ZK-eSIM’s privacy claim is intentionally scoped to \emph{provisioning}. Once a profile is installed, ordinary user-initiated lifecycle operations such as $\mathsf{enable}$,$ \mathsf{disable}$, and $\mathsf{delete}$ remain unchanged and are executed locally between the LPA and the eUICC~\cite{gsma_sgp22_2023}. These operations use profile-local state (e.g., ICCID ) and therefore do not re-expose the device-global \(\EID\) to the SM-DP+ or require reconstruction of an \(\EID\)--\(\mathsf{ICCID}\) mapping. Backend-triggered post-installation management can likewise avoid re-exposing \(\EID\). The backend need only address the installed profile via an associated profile-scoped alias recorded at provisioning. Delivery can then proceed by \emph{device pull}, in which the LPA periodically retrieves pending notifications indexed by that profile handle. ZK-eSIM operates at the provisioning layer: it privately installs the eSIM profile and credentials before 5G-AKA begins, after which the UE runs the standard 5G-AKA procedure unchanged, with no additional identifiers, messages, or modifications to 3GPP core functions.

\textbf{5. Supporting Procedures: Settlement and Accountable Deanonymisation}
\label{sec:settle-trace}
We defer full pseudocode to App.~\ref{app:Algos}
(Algos.~\ref{algo:settle} and~\ref{algo:deanonymise}) and describe the
two operational procedures that complete the design: settlement by
one-time tokens and accountable deanonymisation. \emph{For settlement}, the MNO issues each authorised profile order with a
single-use token bound to the corresponding profile identifier and
authenticated by the MNO. When the SM-DP+ provisions the profile, it
records the token as redeemed. At the end of each epoch, both parties
commit to their respective views: the SM-DP+ commits to the set of
redeemed tokens, while the MNO commits to the set of authorised profile
identifiers. The SM-DP+ returns the redeemed-token set together with
evidence that each token was included in its committed view. The MNO
accepts a redeemed token only if it is validly authenticated by the MNO,
appears in the SM-DP+'s committed redemption log, and corresponds to a
profile identifier that the MNO had authorised for that epoch. The final
settlement count therefore includes only tokens that are both redeemed
and authorised. The parties then cross-sign a settlement receipt binding
the epoch, the accepted count, the payable amount, and the two committed
views, yielding an auditable record without revealing any additional
subscriber information. \emph{For accountable deanonymisation}, the eUICC encrypts its EID to the LEA's
public key during the request protocol, and the resulting escrow
ciphertext is stored by the MNO. Under a valid and lawfully scoped
warrant, the LEA obtains the in-scope escrow from the MNO and decrypts it
to recover the EID for the specific provisioning session. The MNO can then
resolve that EID to the corresponding KYC record for the lawful process.
This preserves compatibility with mandatory subscriber registration: the MNO  learns the subscriber's identity during Phase~0.a
(KYC), which lies outside provisioning-layer anonymity. ZK-eSIM removes
only protocol-level cross-session linkability, while requiring LEA
cooperation under a scoped legal order to deanonymise a specific
provisioning session.

\section{Security Analysis}\label{sec:security}

We show that ZK-eSIM satisfies two main security properties:
\emph{provisioning-session unlinkability} and \emph{unforgeability}.
Unlinkability means that infrastructure entities cannot distinguish
whether two accepted provisioning sessions originate from the same
eligible eUICC or from two different eligible eUICCs. Unforgeability
means that an adversary cannot obtain accepted order-side or
download-side provisioning state without legitimate credentials,
fresh challenges, and MNO-issued authorisation material. Formal
game-based definitions are given in \textbf{Appendix}~\ref{app:gamebaseddefinition},
and the proofs are given in \textbf{Appendix}~\ref{app:securityproof}.

\begin{theorem}[Unlinkability]\label{thm:unlink}
If the NIZK system satisfies zero-knowledge, commitments are hiding,
$\KDF$ and $\PRF$ are secure, $H'$ and $H''$ are modelled as random
oracles, encryption provides IND-CPA security, and pseudonym certificates
satisfy ephemeral indistinguishability, then ZK-eSIM achieves
provisioning-session unlinkability for
$\mathcal{E}\in\{\MNO,\SMDP^+,\PCA\}$. For every PPT
adversary $\mathcal{A}$:
$$
\mathsf{Adv}^{\mathsf{UNLINK}\text{-}\mathcal{E}}_{\Pi,\mathcal{A}}(\lambda)
\leq
\mathsf{Adv}^{\mathsf{ZK}}
+
\mathsf{Adv}^{\mathsf{KDF}}
+
\mathsf{Adv}^{\mathsf{PRF}}
+
\mathsf{Adv}^{\mathsf{Cert}}$$\\
$$
+
\mathsf{Adv}^{\mathsf{IND}\text{-}\mathsf{CPA}}
+
\frac{2q_H}{2^{\lambda}},
$$
where $q_H$ is the total number of random-oracle queries to $H'$ and
$H''$.
\end{theorem}
\vspace{-2mm}
\begin{theorem}[MNO-Side Unforgeability]\label{thm:uf-mno}
Under NIZK knowledge soundness, anonymous credential unforgeability,
collision resistance of $H$ and $H'$, and freshness of MNO challenges,
ZK-eSIM satisfies MNO-side unforgeability. For every PPT adversary
$\mathcal{A}$:
$$
\mathsf{Adv}^{\mathsf{UF}\text{-}\MNO}_{\Pi,\mathcal{A}}(\lambda)
\leq
\mathsf{Adv}^{\mathsf{Sound}\text{-}\mathsf{ZK}}
+
\mathsf{Adv}^{\mathsf{UF}\text{-}\mathsf{Cred}}
+
\mathsf{Adv}^{\mathsf{CR}}_{H}
+
\mathsf{Adv}^{\mathsf{CR}}_{H'}
+
\frac{q_N}{2^{\lambda}},
$$
where $q_N$ denotes the number of MNO challenge queries.
\end{theorem}
\vspace{-2mm}
\begin{theorem}[SM-DP+-Side Unforgeability]\label{thm:uf-smdp}
Under EUF-CMA security of the MNO-issued authorisation credential
$\sigma_{\cred}$ and one-time token $\Ti$, collision resistance of
$H'$ and $H''$, and correctness of the stateful spent-token check,
ZK-eSIM satisfies SM-DP+-side unforgeability. For every PPT adversary
$\mathcal{A}$:
$$
\mathsf{Adv}^{\mathsf{UF}\text{-}\SMDP^+}_{\Pi,\mathcal{A}}(\lambda)
\leq
\mathsf{Adv}^{\mathsf{EUF}\text{-}\mathsf{CMA}}_{\sigma_{\cred}}
+
\mathsf{Adv}^{\mathsf{EUF}\text{-}\mathsf{CMA}}_{\Ti}
+$$
$$
\mathsf{Adv}^{\mathsf{CR}}_{H'}
+
\mathsf{Adv}^{\mathsf{CR}}_{H''}
+
\mathsf{Adv}^{\mathsf{DS}}.
$$
Here $\mathsf{Adv}^{\mathsf{DS}}$ denotes the probability that the
stateful spent-token mechanism accepts a token already recorded in
$L_{\mathrm{spent}}$.
\end{theorem}

\section{Implementation and Evaluation}\label{sec:evaluation}


We evaluate ZK-eSIM in two ways: first, by quantifying runtime costs relative to the commercial RSP; second, by conducting a deployment feasibility test on an \textbf{open-source GSMA-compliant testbed.} We provide a deployment feasibility evaluation demonstrating that our scheme can be deployed with minimal adjustments to current standards. 
To facilitate evaluation, we implement a custom Java Card applet packaged inside an eSIM profile that runs the cryptographic operations on a \textbf{test eUICC} (sysmoEUICC1-C2T \cite{welte_sysmocom_2026}). 
Additionally, we modify a GSMA-compliant open-source SM-DP+ server (osmo-smdpp \cite{osmocom_osmocompysim_2026}) and Local Profile Assistant (lpac \cite{estkme_group_github_2025}).


\begin{figure}[t!]
    \centering
    \includegraphics[width=0.6\linewidth]{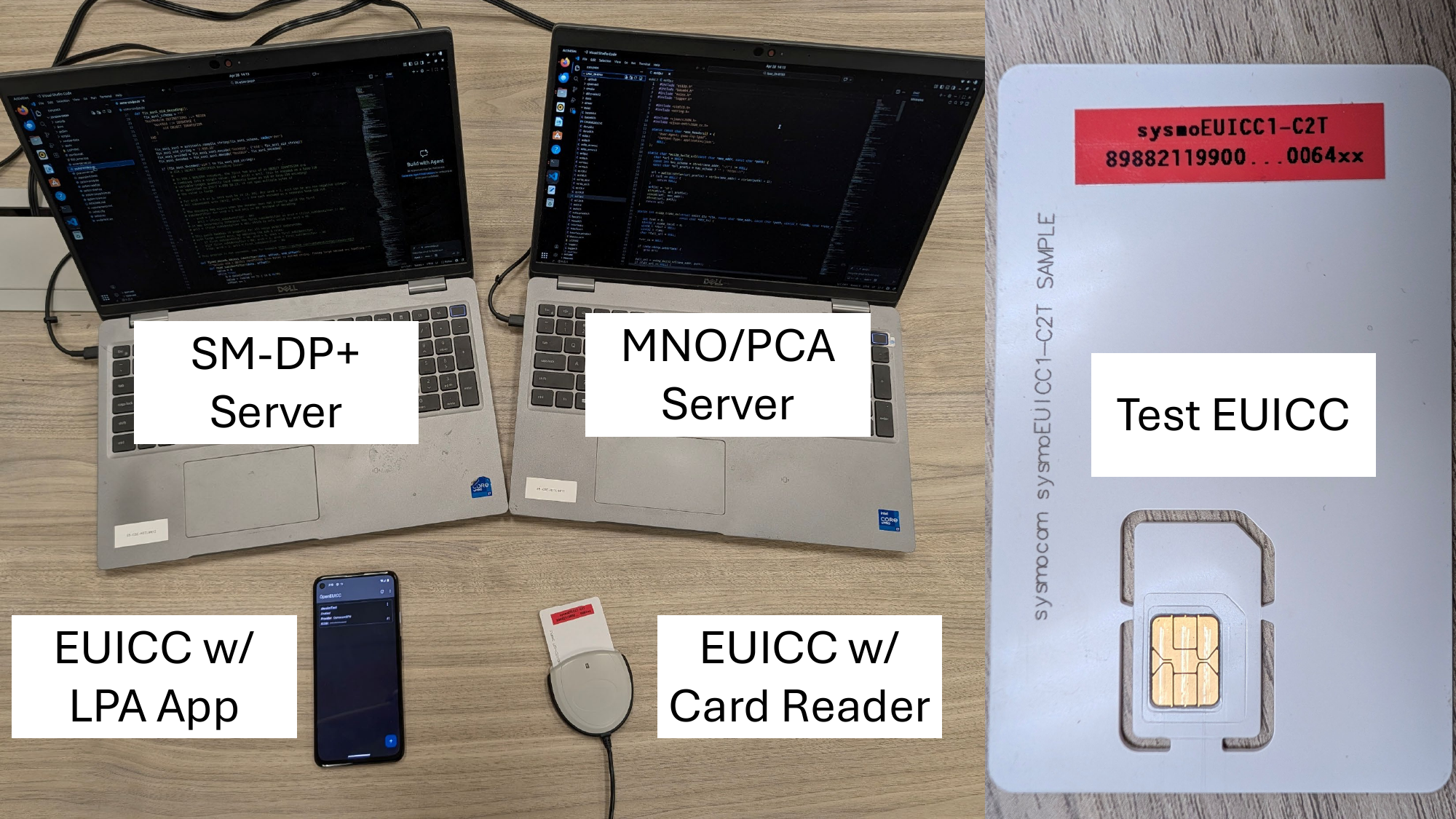}
    \caption{ZK-eSIM Implementation and Evaluation Testbed}
    \label{fig:impl-setup}
    \vspace{-6mm}
\end{figure}
 \vspace{-1mm}

\subsection{Testbed Setup and GSMA Integration} \label{sec:integration}


To evaluate the performance of the proposed ZK-eSIM protocol against the commercial RSP, we establish an experimental testbed illustrated in Fig.~\ref{fig:impl-setup}. To preserve compatibility with current RSP architectures, we extend GSMA-compliant network entities, i.e., the SM-DP+, and the LPA as discussed in §\ref{sec:implementation}. Cryptographic operations are performed with widely used libraries on each network entity, with the JCMathLib \cite{mavroudis_jcmathlib_2020} as a mathematical library in the Java Card applet.  On the eUICC, all operations are performed using the JavaCard and JavaCardX Security libraries \cite{oracle_javacardxsecurity_2015} (with JCMathLib where applicable), and cryptographic operations on the SM-DP+ and MNO servers are performed using Python's cryptographic hazmat library \cite{python_cryptographic_authority_pyca_primitives_2026}. Server-side operations have been performed on a PC running Ubuntu 22.04, with the following specifications: RYZEN 7 8-core CPU, and 64GB RAM.

To complete the ZK-eSIM implementation, we deploy an additional PCA server. In a real-world deployment, the PCA would operate as a sub-CA within the existing GSMA PKI and could be operated by a provider such as Cybertrust, DigiCert, or SEALSQ~\cite{gsma_esim_2026}. In our prototype, we instantiate this role using a dedicated server responsible for certificate issuance and authorisation. The primary role of the PCA in ZK-eSIM is to generate pseudonym certificates for users that can provide a valid proof of eligibility. The provided pseudonym certificate is valid only for a single session of the ZK-eSIM protocol and expires upon completion of profile initialisation.

\subsection{Implementation and Deployability}\label{sec:implementation}
Implementing ZK-eSIM requires extending existing open-source implementations by updating the ASN.1 encodings to add six request messages and their corresponding response messages, introducing four supporting type definitions comprising 17 fields, and extending \textsc{EuiccSigned1} to carry an eligibility bundle.  The changes to the ASN.1 encodings have minimal impact on the GSMA message flow between network entities (SM-DP+, LPA, MNO, and eUICC). We add ZK-eSIM as an alternative message flow within the network entities, following the same messages with the modifications specified in §\ref{sec:proposed}. 

\noindent\underline{ASN.1 Modifications}. We extend the standard GSMA RSP protocol by modifying the ASN.1 level definitions to include a new optional message type (\textsc{EligibilityData}) within the \textsc{EuiccSigned1}. The \textsc{EligibilityData} definition contains: the hashed pseudonym ID ($\mathsf{Hpid}$), the MNO signed credential ($\sigma_\mathsf{cred}$), the one-time authorisation token ($T_i$), the accumulator root ($\mathsf{root}_{\mathsf{auth}}$), the signature over the accumulator root ($\sigma_{^{\mathsf{root}}_{\mathsf{MNO}}}$), and the serialised proof of accumulator inclusion ($\pi_{\mathsf{inc}}$). This enables Phase 2 of ZK-eSIM (pseudonymous profile ordering) with minimal modifications to existing protocol messages. Additionally, due to the lack of a standardised defined communication interface between the eUICC and MNO/PCA servers, we introduce 6 new Request and Response message pairs to facilitate our privacy-preserving zero-knowledge proofs and certificate initialisations.


Modifications to the open-source network entities required adding $\sim$628 LoC (Lines of Code) at the SM-DP+ to enable the alternative workflow using our proposed scheme for encoding and decoding ZK-eSIM messages, allowing ZK-eSIM to be built on top of the existing infrastructure and providing a secondary approach to RSP. Furthermore, $\sim$6.7K LoC are used for the implementation of our custom Java Card Applet (including the JCMathLib and cryptographic functions); however, it should be noted that commercial eSIM profiles may already contain functional applets, which can reduce the number of LoC in a real-world deployment. 

\noindent\underline{Backward Compatibility}. To enable backward compatibility, we develop ZK-eSIM as an alternative privacy-enhanced RSP protocol, which can be implemented alongside current deployments. SM-DP+ providers and MNOs can offer ZK-eSIM as an optional workflow that provides improved privacy over the commercial methods for those who wish to remain anonymous during eSIM profile provisioning, while leaving the commercial offering for users less concerned with privacy. This is enabled by the fact that all encoding-based changes are introduced as optional values within the existing ASN.1 format and require only minimal modifications to existing functions to enable the correct processing of new information elements. The main point at which ZK-eSIM deviates from existing deployments is the pseudonym certificate, which requires changes at the PCA (or may require a separate service offering) to allow the ZK-eSIM deployment.

\subsection{End-to-End Cost Comparison}

\begin{table}[t!]
    \centering
    \scalebox{0.7}{
    \begin{tabular}{|c|c|c|c|}
    \hline
    \diagbox{\textbf{Scheme}}{\textbf{Cost}} 
        & \multicolumn{2}{c|}{\textbf{Runtime Cost (ms)}} 
        & \textbf{End-to-End Cost (ms)} \\ \hline

    \multirow{4}{*}{\textbf{Commercial RSP}\cite{gsma_rsp_2023}} 
        & Registration & N/A & \multirow{4}{*}{$ 53660\; $} \\ \cline{2-3}
        & Certificate Initialisation & 7$\; (\pm 1 )$&  \\ \cline{2-3}
        & Order Profile & $1073 \; $&  \\ \cline{2-3}
        & Profile Download & $52580 \; $&  \\ \hline

    \multirow{4}{*}{\textbf{ZK-eSIM}} 
        & Registration & 9964$\; (\pm 918)$& \multirow{4}{*}{$98199 \; $} \\ \cline{2-3}
        & Certificate Initialisation & $ 10448 \; (\pm \; 620)$&  \\ \cline{2-3}
        & Order Profile & $ 18432\; (\pm \; 764)$ &  \\ \cline{2-3}
        & Profile Download & $ 59355\; (\pm \;955)$&  \\ \hline
    \end{tabular}
    }
     \caption{End-to-End cost of commercial RSP and the proposed ZK-eSIM averaged over 25 Runtimes}
     \vspace{-7.5 mm}
    \label{tab:cost-results}
\end{table}

\underline{End-to-End Cost.} Table \ref{tab:cost-results} presents the mean and standard deviation (denoted by $\pm$) for the execution time of each phase in both the commercial and the proposed ZK-eSIM schemes. The reported values capture the full end-to-end computational and communication costs of cryptographic operations and message flows for each protocol phase. All operations are executed on their respective devices in the experimental setup. ECDSA and SHA256 have been used during the certificate initialisation phase to align with the GSMA RSP specification \cite{gsma_rsp_2023}. Additionally, for this evaluation, we use an EC-based Schnorr proof \cite{brassard_efficient_1990} to allow the user to prove their identity without leaking their $\EID$ (simplified in implementation for a Javacard environment), SHA256-based commitments during the certificate initialisation phase, and ECDH \cite{koblitz_elliptic_1987} for key derivation and agreement. In terms of computational performance, ZK-eSIM exhibits an approximate increase in cryptographic processing time of 83\% (from $53.660 \; \text{seconds}$ vs. $98.199 \; \text{seconds}$). One contributing factor to this increase is the registration phase, which is not included in the cost of the commercial RSP because this step occurs prior to protocol initialisation. The cost is included in the ZK-eSIM calculation due to the introduced cryptographic primitives that demonstrate the overhead costs associated with improvements to user privacy when a user registers with a new MNO. Considering only the phases within the commercial RSP, the cost increases by $\sim $ 64\%.
This increase results from the integration of additional cryptographic primitives, which introduce additional communication overhead costs between the UE and SM-DP+ and MNO servers. The added overhead is accompanied by a significant enhancement in user privacy compared to the commercial RSP, achieved by anonymising user identifiers. Protocol phases containing zero-knowledge proofs contribute the most to the cost, increasing by $10.441$ seconds and $17.359$ seconds for certificate initialisation and profile ordering, respectively. Both of these phases involve ZKP generation and verification, which incur relatively high runtimes compared to the commercial method, which has no proof computations and offers little to no privacy guarantees. Overall, these results indicate that although ZK-eSIM incurs higher computational costs (as evidenced by the end-to-end execution time) than the commercial RSP, it considerably improves user privacy. This is achieved by integrating ZKPs with additional cryptographic primitives, mitigating inherent user-identifier leakage in RSP. Combined with the analysis in Table \ref{tab:comparison}, we note that ZK-eSIM requires longer runtimes but provides greater privacy than the commercial RSP, justifying the additional cost for those who value privacy. 

%
\begin{figure}[t!]
    \centering
    \setlength{\abovecaptionskip}{3pt}
    \includegraphics[width=\linewidth]{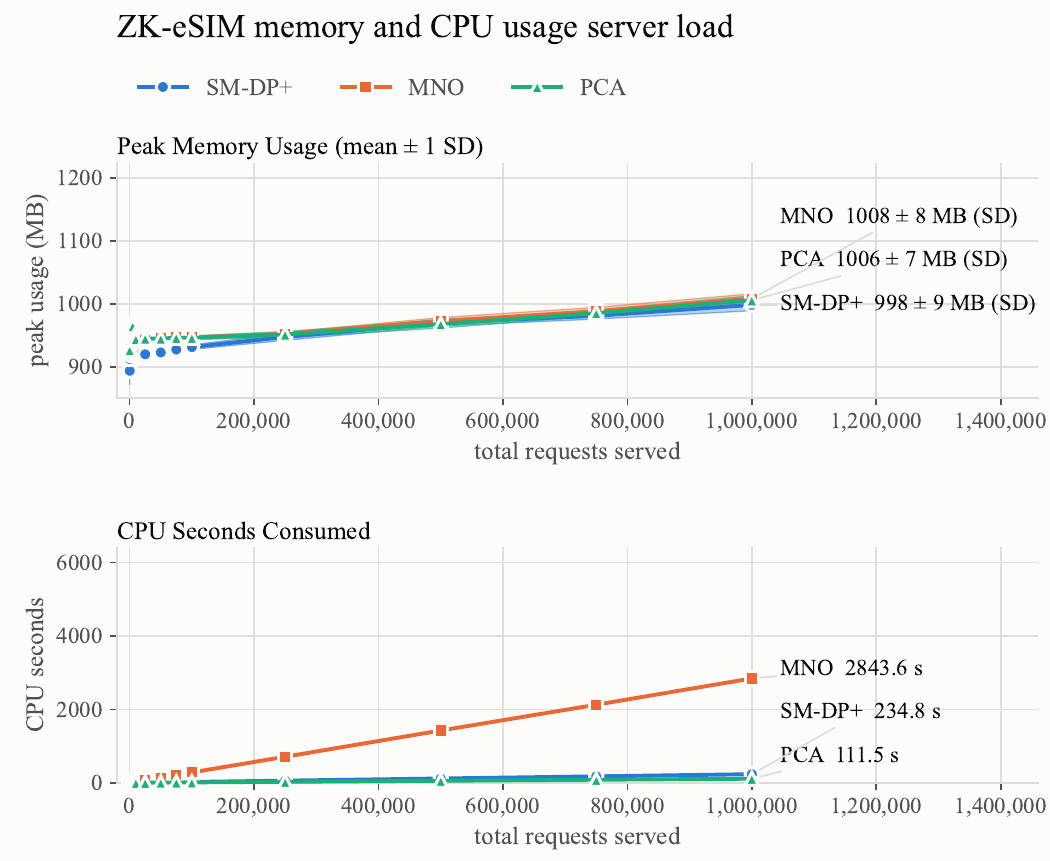}
    \caption{Server performance under load in range from 1 to 1,000,000 concurrent download requests}
    \label{fig:server-results}
\end{figure}

\subsection{Server Scalability of ZK-eSIM}
Figure~\ref{fig:server-results} presents the peak memory usage (and standard deviation of memory usage across the session) and cumulative CPU-seconds consumed by each server as the total number of requests served scales towards 1,000,000 concurrent download requests. Peak memory usage remains reasonable for large scale and closely clustered across all three roles, converging to $998 \; \pm 9\; \mathrm{MB}$ at the SM-DP+, $1006 \pm 7 \; \mathrm{MB}$ at the MNO, and $1008 \pm 8 \; \mathrm{MB}$ at the PCA. The small increase in memory usage indicates that the ZK-eSIM scheme incurs no significant per-request memory accumulation, since the zero-knowledge proof verification operates on fixed-size cryptographic material.

The CPU-seconds consumed reveal the same asymmetry observed in the per-download timings, with the majority of the processing cost concentrated at the MNO. Serving 1,000,000 requests, the MNO consumes $2843.6 \; \mathrm{s}$ in CPU time on average, an order of magnitude greater than the $234.8 \; \mathrm{s}$ at the SM-DP+ and the $111.5 \; \mathrm{s}$ at the PCA. This reflects the MNO's role in performing the EC-based Schnorr identity proof verification, which incurs a higher computational load while ensuring that the network learns no permanent device or subscriber identifiers.

Overall, the server load scales linearly with the number of requests served, with memory usage remaining bounded across all roles and CPU consumption remaining minimal at the SM-DP+ and PCA. As with the per-download analysis, the bulk of the cost is produced at the MNO, where the core identity privacy processing occurs, and this additional cost coincides with significant user privacy enhancements delivered through the anonymisation of user identifiers.
 


\section{Discussion}
ZK-eSIM primarily focuses on securing provisioning-layer privacy but does not address cross-protocol risks that persist once a profile is installed. Users may still face jurisdictional exposure when traffic is routed through unexpected foreign networks, when silent background communication initiated by embedded profile commands goes unnoticed, when fragile lifecycle operations silently fail, and when private 5G deployments shift control to administrators with weaker oversight \cite{motallebighomi_esimplicity_2025}. 
Also, IP address deanonymisation is out of scope for this paper. 
Extending ZK-eSIM to support routing transparency, IP address anonymisation, constrained profile behaviour, and resilient lifecycle management remains future work.
\section{Conclusion}
\label{sec:conclusion}

\emph{ZK-eSIM} replaces the persistent identifiers exposed during GSMA Consumer RSP with ZKPs, single-use pseudonym certificates, and per-session pseudonyms, achieving provable subscriber anonymity and provisioning-session unlinkability while retaining accountable traceability. A Java Card applet prototype on a test eUICC, integrated with a modified open-source SM-DP+ and LPA, shows that these guarantees are practical within the existing GSMA ecosystem.

\bibliographystyle{ACM-Reference-Format}
\balance
\bibliography{references}
\clearpage
\appendix

\section{Additional Preliminaries}\label{app:additional-preliminaries}

\subsubsection{Pedersen Commitment}
The Pedersen commitment scheme is constructed within a cyclic group $G=\langle g\rangle$ of prime order $p$, where the discrete logarithm problem is computationally hard. The scheme consists of the following algorithms~\cite{pedersen_non-interactive_1991}:

\begin{itemize}[left=0pt]
  \item $\pp \gets \mathsf{Ped.Setup}(1^\lambda)$: Sample a prime-order group $G=\langle g\rangle$ with $|G|=p$. Choose $\alpha \gets \mathbb{Z}_p$ uniformly, set $u \gets g^\alpha$, and \emph{erase} $\alpha$. Output public parameters $\pp=(G,p,g,u)$.
  \item $cm \gets \mathsf{Ped.Commit}(m;r)$: For $m \in \mathbb{Z}_p$ and $r \gets \mathbb{Z}_p$, output
  $
     cm \;=\; C(m;r) \;=\; g^{r}\,u^{m} \in G.
  $
  \item $\{0,1\} \gets \mathsf{Ped.Open}(cm,m,r)$: Accept iff $cm = g^{r}u^{m}$ in $G$.

\end{itemize}

The scheme must satisfy:
\begin{itemize}[left=0pt]
  \item \emph{Perfect hiding.} For any $m_0,m_1\in\mathbb{Z}_p$, the distributions $\{C(m_0;r)\}_{r\gets\mathbb{Z}_p}$ and $\{C(m_1;r)\}_{r\gets\mathbb{Z}_p}$ are identical; the commitment leaks no information about $m$.
  \item \emph{Computational binding (under DLP and unknown $\log_g u$).} If an adversary produces two valid openings
  \(
  cm=g^{r}u^{m}=g^{r'}u^{m'}
  \)
  with $(m,r)\neq(m',r')$, then one can recover
  $
    \alpha \;=\; \log_g u \;=\; (r-r')\cdot(m'-m)^{-1} \bmod p,
  $  which breaks DLP. Thus forging two openings is infeasible.
 
\end{itemize}

\subsection{ZK-eSIM Construction: Notation and Interfaces}\label{app:notation}
\subsubsection{Table of Notations}
We write $\mathsf{ZK.Prove}/\mathsf{ZK.Verify}$ for non-interactive zero-knowledge proofs of knowledge, $\mathsf{PRF}_k(\cdot)$ for a pseudorandom function keyed by $k$, $\mathsf{Enc}/\mathsf{Dec}$ for a public-key encryption scheme, $\mathsf{Com}(m;r)$ for a statistically hiding, computationally binding commitment scheme with public parameters.
 
\begin{table}[h!]
\centering
\small
\caption{Notation}\label{tab:notations}
\begin{tabular}{@{}p{0.96\columnwidth}@{}}
\toprule
\textbf{Symbol — meaning}\\
\midrule
$x \in \{\MNO,\PCA,\SMDP,\mathsf{LEA},U,\mathsf{LPA}\}$ — Principals: MNO, PCA, SM-DP+, LEA, eUICC (U), LPA\\
$(sk_x,\,pk_x)$ - Entity $x$’s long‑term keypair\\
$(sk_{\mathsf{tls}},\,Cert_{\mathsf{tls}})$ - SM‑DP$^+$ TLS endpoint key/cert\\
$(sk_{\mathsf{auth}},\,Cert_{\mathsf{auth}})$ - SM‑DP$^+$ server‑auth key/cert\\
$(sk_{\mathsf{pb}},\,Cert_{\mathsf{pb}})$ - SM‑DP$^+$ profile‑signer/binder key/cert\\
$sk_b$ - eUICC‑sealed binding secret\\[2pt]
 
$\PCert_U$ - Pseudonym cert (PCA sig on $(pk_U,\text{expiry})$)\\
$m_{\mathsf{EID}}$ - bound EID digest\\
$C_{\mathsf{EID}}$ - commitment to bound EID digest\\
$\sigma_{\mathsf{EID}}$ - MNO blind signature on $m_{\mathsf{EID}}$\\
$\EncEid$ - EID escrow\\
$K_{\pid}$ - pseudonym key\\
$\pid$ - per‑session pseudonym\\
$\Hpid$ - hashed pseudonym\\
$h_{\mathsf{cert}}$ - pseudonym cert hash\\
 
$L_{\text{auth}},\,L_{\text{spent}}$ - MNO authorisation log/SM‑DP$^+$ spent‑token log\\
$\mathsf{root}_{\text{auth}},\,\mathsf{root}_{\text{spent}}$ - Accumulator digests of the above\\
$\pi_{\mathsf{inc}}$ - Inclusion proof for $\Hpid$ in $L_{\mathsf{auth}}$\\
$\omega^{\text{spent}}_i$ - Witness that token $T_i$ is in $L_{\mathsf{spent}}$\\
$T_i$ - One‑time release/authorisation token\\
$\sigma_{\mathsf{cred}}$ - MNO authorisation over $(\Hpid,h_{\mathsf{cert}},\mathsf{mnoId})$\\[2pt]
 
$I_t$ - Transcript nonce (server‑chosen)\\
$N_U,\,N_S$ - UE/server nonces\\
$\nonce$ - Single‑use server challenge\\
$\mathsf{smdpAddress}$ - Server identifier (from $Cert_{\mathsf{tls}}$)\\
$\mathsf{serverOID}$ - Server OID (from $Cert_{\mathsf{auth}}$,$Cert_{\mathsf{pb}}$)\\[2pt]
 
$(x_U,X_U),(x_S,X_S)$ - AKE secrets/publics for UE and SM-DP+;\\
$(k,k')$ - Session keys\\
$P$ - Profile package; $t_P,t_{\mathsf{mno}},t_{\mathsf{iccid}}$ are MACs under $k'$\\
 
$\mathsf{TokSet}_\tau$ - Tokens redeemed in epoch $\tau$; $c_\tau$ = accepted count\\
$\mathsf{SR}_\tau$ - Dual‑signed settlement receipt for epoch $\tau$\\
\bottomrule
\end{tabular}
\end{table}
 
\makeatletter
\newenvironment{algorithmbreak}[1][]{%
  \par\addvspace{\floatsep}%
  \hrule height.8pt\vskip2pt%
  \refstepcounter{algorithm}%
  \noindent{\small\textbf{Algorithm \thealgorithm:} #1}\par
  \vskip1pt\hrule height.4pt\vskip3pt%
}{%
  \vskip3pt\hrule height.8pt%
  \par\addvspace{\floatsep}%
}
\makeatother
 
\providecommand{\longstate}[1]{\STATE $\displaystyle\begin{aligned}[t]#1\end{aligned}$}

\section{Algorithms}\label{app:Algos}
 
In this section, we list the following algorithms: \emph{Setup, RegisterAndIssue, CertInit, ZKRequest, OrderProfile, MutualAuthAndProvision, Settle, and Deanonymise}.

\begin{algorithmbreak}[\textsf{Setup}]
\label{algo:Setup}
\small
\begin{algorithmic}[1]
\REQUIRE Security parameter $1^\lambda$.
\MNO\ has $(sk_{\MNO}, pk_{\MNO})$; \PCA\ has $(sk_{\PCA}, pk_{\PCA})$; LEA has $(sk_{\mathsf{LEA}}, pk_{\mathsf{LEA}})$, SM-DP+ sets long-term credentials: $\mathbf{(sk_{tls}, Cert_{tls})}$ (the key and certificate for TLS authentication), $\mathbf{(sk_{auth}, Cert_{auth})}$ (the key and certificate for server authentication), and $\mathbf{(sk_{pb}, Cert_{pb})}$ (the key and certificate for profile binding)
\STATE $CRS \leftarrow \mathsf{ZK}.\Setup(1^\lambda)$; $ck \leftarrow \Com.\Setup(1^\lambda)$
\STATE Fix a domain-separated hash family $H, H', H''$
\STATE Fix the functions $f,f'$ where $f:\{0,1\}^* \times \mathbb{N} \to \{0,1\}^*$ is deterministic with $|f(x,\ell)|=\ell$. We fix a randomness‑budget function $\rho:\mathbb{N}\to\mathbb{N}$ in the public parameters and, when needed, set $s \gets f(r_{\mathsf{seed}},\rho(\lambda))$ with $r_{\mathsf{seed}} \sample \{0,1\}^{\lambda}$, and $f':\mathcal{K}_b \times \{0,1\}^{\rho(\lambda)} \to \mathcal{SK}_U \times \mathcal{PK}_U$ is deterministic and outputs a valid keypair, and $\mathcal{K}_b$ denote the space of sealed per‑card binding secrets; $\mathcal{SK}_U$ and $\mathcal{PK}_U$ the secret/public key spaces of the user key type. We write $\{0,1\}^\ell$ for $\ell$‑bit strings and $\{0,1\}^*$ for arbitrary‑length strings.
\STATE Fix algorithm families: $\PRF$, $\KDF$, $\mathsf{Sig}$, $\mathsf{PKE}$, $\mathsf{BSig}$, $\mathsf{SE}$, $\mathsf{MAC}$, $\mathsf{Acc}$, $\mathsf{Com}$ with interfaces: $\mathsf{Sig}$ = $(\mathsf{Sig.sign},\,\mathsf{Sig.verify})$ - Digital signature, $\mathsf{PKE}$ = $(\mathsf{PKE.enc},\,\mathsf{PKE.dec})$ - Public-key encryption, $\mathsf{SE}$ = $(\mathsf{SE.E},\,\mathsf{SE.D})$ - Symmetric encryption, $\mathsf{MAC}$ = $(\mathsf{MAC.tag},\,\mathsf{MAC.verify})$ - Message authentication code, $\mathsf{BSig}$ = $(\mathsf{BSig.blind},\,\mathsf{BSig.sign},\,\mathsf{BSig.unblind},\,\mathsf{BSig.verify})$ - Blind signatures, $\mathsf{KA}$ = $(\mathsf{KA.gen}, \mathsf{KA.agree})$ - Key agreement, $\mathsf{Acc}$ = $(\mathsf{Acc.init},\,\mathsf{Acc.add},\,\mathsf{Acc.digest},\,\mathsf{Acc.prove},\,\mathsf{Acc.verify})$ - Append-only accumulator, $\Com$ = $(\Com.\Setup,\Com.\mathsf{Commit},\Com.\mathsf{Open},\Com.\mathsf{Verify})$ - commitment.
\STATE Load GSMA CI roots and PCA chain; install required certs in the LPA (system anchor).
\STATE Fix an operator‑identifier namespace and let $\mathsf{mnoId}$ denote the issuing MNO identifier.
\STATE Publish registry entries $(pk_{\MNO}, pk_{\PCA}, pk_{\mathsf{LEA}})$ under CI roots; record certificate policies.
\STATE \textbf{Initialise local state:} MNO sets $L_{\text{auth}}\gets\emptyset$; SM-DP+ sets $L_{\text{spent}}\gets\emptyset$; each eUICC samples and maintains sealed: $sk_b \sample \mathcal{K}_b$
\longstate{\pp \leftarrow \big(CRS, ck, H, H', H'', \rho, f, f', \PRF, \KDF,\\ \mathsf{Sig}, \mathsf{PKE}, \mathsf{BSig}, \mathsf{KA}, \mathsf{SE}, \mathsf{MAC}, \mathsf{Acc}, \mathsf{Com},\\ \mathsf{CI~roots}, \mathsf{PCA~chain}, \mathsf{mnoId}, pk_{\MNO}, pk_{\PCA}, pk_{\mathsf{LEA}} \big)}
\RETURN Public parameters $\pp$, $L_{\mathsf{auth}}$, $L_{\mathsf{spent}}$, $sk_b$
\end{algorithmic}
\end{algorithmbreak}

\begin{algorithmbreak}[\textsf{RegisterAndIssue}]
\label{algo:RegisterAndIssue}
\small
\begin{algorithmic}[1]
\REQUIRE UE (eUICC) holds sealed $sk_b$ and identifier $\mathsf{EID}$; public parameters $\pp$.
\STATEx; any $UE\leftrightarrow$ message is relayed by the LPA.
\STATE \textbf{Eligibility attestation over a secure channel:} $UE \to \MNO$: $\mathsf{EID}$ + device attestation.
\STATE \textbf{Eligibility \& KYC/device checks:} MNO verifies eligibility. If checks fail, \textbf{abort}.
\STATE $\MNO \to UE$: \textbf{Proceed}.
\STATE \textbf{Local preparation (inside eUICC):} UE: Sample $r_b \sample \{0,1\}^{\lambda}$ and $r_{\mathsf{blind}} \sample \{0,1\}^{\lambda}$, $m_{\mathsf{EID}} \gets H(\mathsf{EID} \parallel sk_b)$, $C_{\mathsf{EID}} \gets \mathsf{Com}.\mathsf{Commit}_{ck}(m_{\mathsf{EID}}; r_b)$, $(\mathsf{req}, \alpha) \gets \mathsf{BSig.blind}(pk_{\mathsf{MNO}}, m_{\mathsf{EID}}; r_{\mathsf{blind}})$.
\STATE \textbf{Relation $\mathcal{R}_{\mathsf{issue}}$ (predicate over $(x,w)$):}
\STATE \quad \emph{Statement} $x=(\mathsf{EID}, C_{\mathsf{EID}}, \mathsf{req}, pk_{\mathsf{MNO}})$; \quad \emph{Witness} $w=(sk_b, r_b, r_{\mathsf{blind}}, \alpha)$.
\STATE \quad \emph{Require:} $m_{\mathsf{EID}} = H(\mathsf{EID}\parallel sk_b)$ \textit{(binding)} $\land$ $C_{\mathsf{EID}} = \mathsf{Com}.\mathsf{Commit}_{ck}(m_{\mathsf{EID}}; r_b)$ \textit{(commit correctness)} $\land$ $\exists\,\alpha:\ (\mathsf{req}, \alpha)=\mathsf{BSig.blind}(pk_{\mathsf{MNO}}, m_{\mathsf{EID}}; r_{\mathsf{blind}})$ \textit{(well-formed blind request)}
\STATE \textbf{Generate proof:} $\pi \leftarrow \textsf{ZK.Prove}(x; w)$.
\STATE $UE \to \mathsf{MNO}$: $(x, \pi)$.
\STATE \textbf{MNO verifies and issues:}
\STATE \quad \textbf{If } $\textsf{ZK.Verify}(x, \pi)=0$ \textbf{, then abort}.
\STATE \quad $\tilde{\sigma} \gets \mathsf{BSig.sign}(sk_{\mathsf{MNO}}, \mathsf{req})$; \quad $\mathsf{MNO} \to UE$: $\tilde{\sigma}$.
\STATE \textbf{Unblind and check (inside eUICC):}
\STATE \quad $\sigma_{\mathsf{EID}} \gets \mathsf{BSig.unblind}(\alpha, \tilde{\sigma})$.
\STATE \quad \textbf{Assert } $\mathsf{BSig.verify}(pk_{\mathsf{MNO}}, m_{\mathsf{EID}}, \sigma_{\mathsf{EID}})=1$.
\end{algorithmic}
\end{algorithmbreak}

\begin{algorithmbreak}[\textsf{CertInit}]
\label{algo:certInit}
\small
\begin{algorithmic}[1]
\REQUIRE $UE$ holds eligibility credential $\sigma_{\mathsf{EID}}$, $\EID$ and sealed $sk_b$ (local state, never transmitted); public parameters $\pp$.
\STATE \textbf{Ephemeral key (inside eUICC):} Sample $r_{\mathsf{seed}} \sample\{0,1\}^\lambda$; $(sk_U, pk_U) \gets f'(sk_b, f(r_{\mathsf{seed}}))$. For example, in a pairing setting, we let $e:\mathbb{G}_1\times\mathbb{G}_2\!\to\!\mathbb{G}_T$ be a bilinear map of prime order $q$ with generators $g_1,g_2$, and let $a=sk_b\in\mathbb{Z}_q^*$. Set $\alpha\gets H_{\mathsf{fld}}(r_{\mathsf{seed}})\in\mathbb{Z}_q^*$. Then $f(r_{\mathsf{seed}})=\alpha$ and $f'(a,\alpha)=(sk_U,pk_U)$ with $sk_U=a\alpha\bmod q$ and $pk_U=g_1^{\,sk_U}$; equivalently $e(pk_U,g_2)=e(g_1^{\,\alpha},g_2^{\,a})$.
\STATE \textbf{Relation $\mathcal{R}_{\mathsf{bind}}$ (predicate over $(x,w)$):}
\STATE \quad \emph{Statement} $x=(pk_U, pk_{\mathsf{MNO}})$;\quad \emph{Witness} $w=(sk_b, r_{\mathsf{seed}}, \mathsf{EID}, \sigma_{\mathsf{EID}})$.
\STATE \quad \emph{Require:} $\exists\, sk_U\in\mathcal{K}:\ f'(sk_b, f(r_{\mathsf{seed}}))=(sk_U, pk_U)$ \textit{(same-eUICC)} $\land$ $m_{\mathsf{EID}} = H(\mathsf{EID} \parallel sk_b)$ \textit{(binding)} $\land$ $\mathsf{BSig.verify}(pk_{\mathsf{MNO}},\,\sigma_{\mathsf{EID}},\,m_{\mathsf{EID}})=1$ \textit{(eligibility)}
\STATE \textbf{Generate proof:} $\pi_{\mathsf{bind}} \leftarrow \textsf{ZK.Prove}(x;w)$.
\STATE \textbf{Request \& issuance:}
\STATE \quad $UE \rightarrow \mathsf{PCA} : (x, \pi_{\mathsf{bind}})$.
\STATE \quad \textbf{if} {$\textsf{ZK.Verify}(x,\pi_{\mathsf{bind}})=0$}, \textbf{then abort}
\STATE \quad $\mathsf{PCA}$ issues $\mathsf{PCert}_U \gets \mathsf{Sig.sign}_{sk_{\mathsf{PCA}}}(pk_U,\text{expiry})$.
\STATE $\mathsf{PCA}$ $\to$ $UE$: $\PCert_U$
\STATE $UE$ retains $sk_U$ and $r_\mathsf{seed}$
\STATE \textbf{return to UE:} $\mathsf{PCert}_U$.
\end{algorithmic}
\end{algorithmbreak}

\begin{algorithmbreak}[\textsf{ZKRequest}]
\label{algo:ZKRequest}
\small
\begin{algorithmic}[1]
\REQUIRE $U$ holds $\mathsf{PCert}_U$ on $(pk_U,\text{expiry})$, eligibility credential $\sigma_{\mathsf{EID}}$, $\EID$ and sealed eUICC state including $sk_b,r_\mathsf{seed}$; public parameters $\pp$.
\STATE \textbf{Transport:} Establish server-authenticated TLS.
\STATE \textbf{Challenge:} $\mathsf{MNO} \rightarrow U$: globally unique and single‑use: $\mathsf{nonce} \sample \{0,1\}^\lambda$, $\mathsf{MNO}$ stores $\mathsf{nonce_{MNO}} \gets \nonce$
\STATE \textbf{Pseudonym (inside eUICC):} $K_{\mathsf{pid}} \gets \KDF(sk_b,\mathsf{nonce})$;\quad $\pid \gets \PRF_{K_{\pid}}(\mathsf{EID})$.
\STATE \textbf{Escrow encryption (inside eUICC):} Sample $r \sample \{0,1\}^\lambda$;\quad $\EncEid \gets \mathsf{PKE.Enc}_{pk_{\mathsf{LEA}}}(\mathsf{EID}; r)$.
\STATE \textbf{Relation $\mathcal{R}_{\mathsf{req}}$ (predicate over $(x,w)$):}
\STATE \quad \emph{Statement} $x=(pk_{\mathsf{MNO}}, pk_{\mathsf{LEA}}, pk_U, \mathsf{nonce}, \pid, \EncEid )$; \emph{Witness} $w=(\mathsf{EID}, sk_b, r_{\mathsf{seed}}, \sigma_{\mathsf{EID}}, r)$.
\STATE \quad \emph{Require:} $\exists\, sk_U\in\mathcal{K}:\ f'(sk_b, f(r_{\mathsf{seed}}))=(sk_U, pk_U)$ \textit{(same-eUICC)} $\land$ $m_{\mathsf{EID}} = H(\mathsf{EID} \parallel sk_b)$ \textit{(binding)} $\land$ $\mathsf{BSig.verify}(pk_{\mathsf{MNO}},\, \sigma_{\mathsf{EID}},\, m_{\mathsf{EID}})=1$ \textit{(eligibility under MNO)} $\land$ $\pid = \PRF_{\KDF(sk_b,\mathsf{nonce})}(\mathsf{EID})$ \textit{(pseudonym correctness)} $\land$ $\EncEid = \mathsf{PKE.Enc}_{pk_{\mathsf{LEA}}}(\mathsf{EID}; r)$ \textit{(ciphertext binds to same $\mathsf{EID}$)}
\STATE \textbf{Generate proof:} $\pi_{\mathsf{req}} \leftarrow \textsf{ZK.Prove}(x; w)$.
\STATE \textbf{return to MNO:} $x, \PCert_U , \pi_{req}$
\end{algorithmic}
\end{algorithmbreak}

\begin{algorithmbreak}[\textsf{OrderProfile}]
\label{algo:OrderProfile}
\small
\begin{algorithmic}[1]
\REQUIRE $\mathsf{nonce}_{\mathsf{MNO}}$; $\pi_{\mathsf{req}}; \PCert_U$; $L_{\mathsf{auth}}$; $x=(pk_{\mathsf{MNO}}, pk_{\mathsf{LEA}}, pk_U, \mathsf{nonce}, \pid, \EncEid );$ Public parameters $\pp$.
\STATE \textbf{Sanity \& freshness (bind to server challenge):}
\STATE \quad \textbf{If } $\mathsf{nonce}\neq \mathsf{nonce}_{\mathsf{MNO}}$ \textbf{ then abort}.
\STATE \textbf{Verification at $\mathsf{MNO}$:} Check $\mathsf{PCert}_U$ chain and expiry under $pk_{\mathsf{PCA}}$; parse $pk_U$. \textbf{If} check fails \textbf{then abort}; Accept iff $\textsf{ZK.Verify}(x,\pi_{\mathsf{req}})=1$ with this $pk_U$ and received $(\mathsf{nonce},\pid,\EncEid)$; otherwise \textbf{abort}.
\STATE \quad \emph{Attests same-eUICC linkage, eligibility under $\mathsf{MNO}$, pseudonym correctness, and well-formed $\EncEid$.}
\STATE \textbf{Compute identifiers and defend against replay:} ${\Hpid} \gets H'(\pid)$;\quad $h_{\mathsf{cert}} \gets H''(\mathsf{PCert}_U)$; \textbf{If} ${\Hpid} \in L_{\mathsf{auth}}$, then \textbf{abort}.
\STATE MNO stores a mapping $\Hpid \mapsto \EncEid$.
\STATE \textbf{Place order with SM-DP+ (two-round exchange):}
\STATE \quad $\mathsf{MNO} \rightarrow \text{SM-DP+} : \texttt{DownloadOrder}({\Hpid},\; \mathsf{mnoId},\; \text{profileType})$.
\STATE \quad \emph{SM-DP+ reserves a fresh $ICCID$.}
\STATE \quad $\text{SM-DP+} \rightarrow \mathsf{MNO} : ICCID$.
\STATE \quad $\mathsf{MNO} \rightarrow \text{SM-DP+} : \texttt{ConfirmOrder}(ICCID,\; {\Hpid},\; \text{releaseFlag}=\text{true})$.
\STATE \quad \emph{SM-DP+ maps $ICCID \leftrightarrow {\Hpid}$ and acknowledges.}
\STATE \textbf{MNO authorises \& publishes log root:}
\longstate{L_{\mathsf{auth}} \gets \mathsf{Acc.add}(L_{\mathsf{auth}}, {\Hpid});\ \pi_{\mathsf{inc}} \gets \mathsf{Acc.prove}(L_{\mathsf{auth}}, {\Hpid});\\ \mathsf{root}_{\mathsf{auth}} \gets \mathsf{Acc.digest}(L_{\mathsf{auth}}).}
\STATE \quad $\sigma^{\mathsf{root}}_{\MNO} \gets \mathsf{Sig.sign}_{sk_{\MNO}}(\mathsf{root}_{\mathsf{auth}})$.
\STATE \textbf{MNO issues single-use tokens/credentials (bound to $({\Hpid}, h_{\mathsf{cert}})$):}
\STATE \quad $\sigma_{\mathsf{cred}} \gets \mathsf{Sig.sign}_{sk_{\MNO}}({\Hpid}, h_{\mathsf{cert}}, \mathsf{mnoId})$.
\STATE \quad $T_i \gets \mathsf{Sig.sign}_{sk_{\MNO}}({\Hpid}, h_{\mathsf{cert}}, \mathsf{mnoId}, \text{expiry})$.
\STATE \textbf{Deliver to user securely on success:}
\STATE \quad $\MNO \rightarrow UE : (T_i, \sigma_{\mathsf{cred}}, \Hpid, \pi_{\mathsf{inc}}, \mathsf{root}_{\mathsf{auth}}, \sigma^{\mathsf{root}}_{\MNO})$.
\STATE \textbf{return to U:} $T_i, \sigma_{\mathsf{cred}}, \Hpid, \pi_{\mathsf{inc}}, \mathsf{root}_{\mathsf{auth}}, \sigma^{\mathsf{root}}_{\MNO}$.
\end{algorithmic}
\end{algorithmbreak}

\begin{algorithmbreak}[\textsf{MutualAuthAndProvision}]
\label{algo:MutualAuthAndProvision}
\small
\begin{algorithmic}[1]
\REQUIRE $U$ holds $\PCert_U$, $sk_U$, $\Hpid$, $T_i$, $\sigma_{\mathsf{cred}}$, $\pi_{\mathsf{inc}}$, $\mathsf{root}_{\mathsf{auth}}$, $\sigma^{\mathsf{root}}_{\MNO}$; Public parameters $\pp$.
\STATEx failed verification : $\mathsf{status}\leftarrow\perp$, abort.
\STATE \textbf{Channel Setup (LPA$\leftrightarrow$Server)} \COMMENT{Server-auth TLS}:
\STATE eUICC samples $N_U \sample \{0,1\}^\lambda$; eUICC $\rightarrow$ LPA: $N_U$. \label{line:nu}
\STATE LPA verifies $\mathsf{cert\_tls}$; on success obtain $\mathsf{smdpAddress}$. LPA $\rightarrow$ SM-DP+: $(N_U, \mathsf{smdpAddress})$.
\STATE SM-DP+ samples transcript nonce $I_t \sample \{0,1\}^\lambda$ and server nonce $N_S \sample\{0,1\}^\lambda$, then $\text{SM-DP+} \rightarrow$ LPA:
\longstate{\big(\mathsf{Cert_{\mathsf{auth}}},\ \mathsf{Sig.sign}_{sk_{\mathsf{auth}}}(I_t, N_S, N_U, \mathsf{smdpAddress})\big);}
\STATE \quad LPA verifies.
\STATE LPA $\rightarrow U$: $\big(\mathsf{Cert_{\mathsf{auth}}}, \mathsf{Sig.sign}_{sk_{\mathsf{auth}}}(I_t, N_S, N_U, \mathsf{smdpAddress})\big)$. $U$ verifies $\mathsf{Cert_{\mathsf{auth}}}$ and obtains $\mathsf{serverOID}$ from $\mathsf{Cert_{\mathsf{auth}}}$.
\STATE \textbf{Eligibility bundle (channel-bound, replay‑protected)}: \label{phase:elig}
\STATE $U$ computes $\mathsf{h_{cert}} := H''(\PCert_U)$.
\STATE $U \rightarrow$ LPA the \emph{eligibility bundle}:
\longstate{(\PCert_U, \sigma_{\mathsf{cred}}, T_i, \Hpid, \pi_{\mathsf{inc}}, \mathsf{root}_{\mathsf{auth}}, \sigma^{\mathsf{root}}_{\MNO},\\ \mathsf{Sig.sign}_{sk_U}(I_t, N_S, N_U, T_i, \Hpid, \PCert_U, \mathsf{smdpAddress}))}
\STATE \quad LPA forwards to SM-DP+.
\STATE \textbf{Server-side verification and token spend}:
\STATE Parse $pk_U$ from $\PCert_U$ and verify the chain under $pk_{\PCA}$; set $h_{\mathsf{cert}} \gets H''(\PCert_U)$.
\STATE Verify all of the following:
\longstate{\mathsf{Sig.verify}_{pk_U}(I_t, N_S, N_U, T_i, \Hpid, \PCert_U, \mathsf{smdpAddress}) = 1\\ \land\ \mathsf{Sig.verify}_{pk_{\MNO}}(\mathsf{root}_{\mathsf{auth}}, \sigma^{\MNO}_{\mathsf{root}}) = 1\\ \land\ \mathsf{Acc.verify}(\mathsf{root}_{\mathsf{auth}}, \Hpid, \pi_{\mathsf{inc}})=1\\ \land\ \mathsf{Sig.verify}_{pk_{\MNO}}((\Hpid, h_{\mathsf{cert}}, \mathsf{mnoId}), \sigma_{\mathsf{cred}})=1.}
\STATE Verify $T_i$ is a signature by $pk_{\MNO}$ on $(\Hpid, h_{\mathsf{cert}}, \mathsf{mnoId}, \text{expiry})$ and that $\texttt{now}<\texttt{expiry}$.
\STATE One-time use: if $T_i \notin L_{\text{spent}}$ then append $T_i$ to $L_{\mathsf{spent}}$, else \textbf{abort}.
\STATE \textbf{Profile-signer introduction (bind signer to transcript)}:
\STATE $\text{SM-DP+} \rightarrow$ LPA: $\mathsf{Sig.sign}_{sk_{\mathsf{pb}}}(I_t),~ \mathsf{Cert_{\mathsf{pb}}}$.
\STATE LPA $\rightarrow U$: same. $U$ verifies $\mathsf{Cert_{\mathsf{pb}}}$; let $\mathsf{serverOID}_{\mathsf{pb}} \gets \mathsf{ServerOID}(\mathsf{Cert_{\mathsf{pb}}})$; \textbf{require} $\mathsf{serverOID}_{\mathsf{pb}} = \mathsf{serverOID}$; then verify $\mathsf{Sig.verify}_{pk_{\mathsf{pb}}}(I_t)=1$.
\STATE \textbf{AKE with explicit authentication and context separation:}
\STATE $U$: $(x_U, X_U) \gets \mathsf{KA.gen}()$.
\STATE $UE \rightarrow$ LPA: $\mathsf{Sig.sign}_{sk_U}(I_t, X_U)$; LPA $\rightarrow \text{SM-DP+}$: forward.
\STATE SM-DP+ verifies $\mathsf{Sig.verify}_{pk_U}(I_t, X_U)=1$.
\STATE $(x_S, X_S) \gets \mathsf{KA.gen}()$; compute $K_{US} \gets \mathsf{KA.agree}(x_S, X_U)$; derive $(k, k') := \mathsf{KDF}(K_{US}, \mathsf{serverOID}, U)$.
\STATE \textbf{Profile delivery (confidentiality, and integrity)}:
\STATE SM-DP+ prepares profile package $P$ and associated metadata $(\mathsf{mnoId}, \mathsf{iccid})$.
\STATE Compute ciphertext and tags:
\longstate{C \gets \mathsf{SE.E}(P;\, k),\quad t_P \gets \mathsf{MAC}(C;\, k'),\\ t_{\mathsf{mno}} \gets \mathsf{MAC}(\mathsf{mnoId};\, k'),\quad t_{\mathsf{iccid}} \gets \mathsf{MAC}(\mathsf{iccid};\, k').}
\STATE $\text{SM-DP+} \rightarrow \text{LPA}$:
\longstate{(C, t_P, \mathsf{mnoId}, t_{\mathsf{mno}}, \mathsf{iccid}, t_{\mathsf{iccid}}, X_S,\\ \mathsf{Sig.sign}_{sk_{\mathsf{pb}}}(I_t, X_S, X_U, \mathsf{smdpAddress})).}
\STATE \textbf{Client-side key derivation, verification, install:}
\STATE $U$ verifies $\mathsf{Sig.verify}_{pk_{\mathsf{pb}}}(I_t, X_S, X_U, \mathsf{smdpAddress})=1$.
\STATE Decrypt and install: $P \gets \mathsf{SE.D}(C;\, k)$; install $P$ on eUICC.
\STATE $\mathsf{status} \leftarrow \texttt{OK}$ \textbf{iff} $\big(\text{all verifications passed}~\wedge~P~\text{installed}\big)$.
\STATE \textbf{return} $\textsf{status} \in \{\texttt{OK}, \perp\}$.
\end{algorithmic}
\end{algorithmbreak}

\begin{algorithmbreak}[\textsf{Settle}]
\label{algo:settle}
\small
\begin{algorithmic}[1]
\REQUIRE Epoch $\tau$ has ended; SM-DP+ maintains $L_{\mathsf{spent}}[\tau]$; $\MNO$ maintains $L_{\mathsf{auth}}[\tau]$; $\mathsf{TokSet_{\tau}} \gets \{\,T_i\,\}$ redeemed during epoch $\tau$; $\pp$.
\STATE \textbf{Provider commitments (epoch-bound):}
\STATE \quad $\mathsf{root}^{\tau}_{\mathsf{spent}} \gets \mathsf{Acc}.\mathsf{digest}(L_{\mathsf{spent}}[\tau])$; $\sigma^{\tau}_{\SMDP} \gets \mathsf{Sig}.sign_{sk_{\SMDP}}(\tau\|\mathsf{root}^{\tau}_{\mathsf{spent}})$.
\STATE \quad $\mathsf{root}^{\tau}_{\mathsf{auth}} \gets \mathsf{Acc}.\mathsf{digest}(L_{\mathsf{auth}}[\tau])$; $\sigma^{\tau}_{\MNO} \gets \mathsf{Sig.sign}_{sk_{\MNO}}(\tau\|\mathsf{root}^{\tau}_{\mathsf{auth}})$.
\STATE \textbf{Inclusion witnesses (from SM-DP+):}
\STATE \quad For each $T_i \in \mathsf{TokSet}_\tau$: $\omega^{\text{spent}}_i \gets \mathsf{Acc.prove}(L_{\mathsf{spent}}[\tau], T_i)$.
\STATE \textbf{Bundle from SM-DP+ to $\MNO$:}
\STATE \quad SM-DP+ $\to \MNO$: $\Big(\mathsf{TokSet}_\tau, \{\omega^{\text{spent}}_i\}_{T_i\in\TokSet_\tau}, \mathsf{root}^{\tau}_{\mathsf{spent}}, \sigma^{\tau}_{\SMDP}\Big)$.
\STATE \textbf{MNO-side verification \& filtering}:
\STATE \quad \textbf{Verify} $\mathsf{Sig.verify}_{pk_{\SMDP}}(\tau\|\mathsf{root}^{\tau}_{\mathsf{spent}}, \sigma^{\tau}_{\SMDP}) = 1$.
\STATE \quad \textbf{Compute (or verify)} $\mathsf{root}^{\tau}_{\mathsf{auth}}$ and keep $\sigma^{\tau}_{\MNO}$ for the receipt.
\FOR{each $T_i \in \TokSet_\tau$}
  \STATE \textbf{Signature check on token:} \ \textbf{if} $\mathsf{Sig.verify}_{pk_{\MNO}}(T_i)=0$ \textbf{ then continue} \COMMENT{reject $T_i$}
  \STATE \textbf{Spent-membership:} \ \textbf{if} $\mathsf{Acc.verify}(\mathsf{root}^{\tau}_{\mathsf{spent}}, T_i, \omega^{\text{spent}}_i)=0$ \textbf{ then continue}
  \STATE \textbf{Authorisation linkage:}
  \STATE \quad Parse payload to recover $\Hpid_i$; \ \textbf{if} $\Hpid_i \notin L_{\mathsf{auth}}[\tau]$ \textbf{ then continue}
  \STATE Mark $T_i$ as \emph{accepted}.
\ENDFOR
\STATE \textbf{Pricing \& tally}
\STATE \quad $c_\tau \leftarrow \#\{\text{accepted } T_i\}$; \quad $\textsf{amount} \leftarrow \mathsf{TariffCalc}(c_\tau)$.
\STATE \textbf{Mutual settlement receipt (cross-signing)}
\longstate{\mathsf{SR}_\tau^{\MNO} \leftarrow \mathsf{Sig.sign}_{sk_{\MNO}}\!\big(\tau, c_\tau, \mathsf{amount}, \mathsf{root}^{\tau}_{\mathsf{auth}},\\ \sigma^{\tau}_{\MNO}, \mathsf{root}^{\tau}_{\mathsf{spent}}, \sigma^{\tau}_{\SMDP}\big)}
\longstate{\mathsf{SR}_\tau^{\SMDP} \leftarrow \mathsf{Sig.sign}_{sk_{\SMDP}}\!\big(\tau, c_\tau, \mathsf{amount}, \mathsf{root}^{\tau}_{\mathsf{auth}},\\ \sigma^{\tau}_{\MNO}, \mathsf{root}^{\tau}_{\mathsf{spent}}, \sigma^{\tau}_{\SMDP}\big)}
\STATE \quad $\mathsf{SR}_\tau \leftarrow \mathsf{SR}_\tau^{\MNO} \,\|\, \mathsf{SR}_\tau^{\SMDP}$.
\RETURN $\mathsf{SR}_\tau$.
\end{algorithmic}
\end{algorithmbreak}

\begin{algorithmbreak}[\textsf{Deanonymise}]
\label{algo:deanonymise}
\small
\begin{algorithmic}[1]
\REQUIRE Valid legal warrant specifying a lawful scope $S$; $\mathsf{MNO}$ retains $\mathsf{EncEid}$; $\mathsf{LEA}$ stores secret tracing key $sk_{\mathsf{LEA}}$.
\STATE \textbf{Lawful trigger \& scope (Secure channel)}
\STATE \quad $\mathsf{LEA}$ registers the legal order and the lawful scope $S$ (e.g., a time window, case ID, target set).
\STATE \quad $\mathsf{LEA} \rightarrow \mathsf{MNO}$: request the unique escrow ciphertext satisfying $S$.
\STATE \quad $\mathsf{MNO}$ evaluates $S$, and returns $\EncEid$.
\STATE \textbf{Policy-gated unsealing at $\mathsf{LEA}$}
\STATE \quad $\mathsf{EID}' \leftarrow \mathsf{PKE.Dec}_{sk_{\mathsf{LEA}}}(\EncEid)$, $\pi \gets \mathsf{PKE}.dec_{sk_{\mathsf{LEA}}}({\EncEid})=\EID^{\prime}$
\STATE \textbf{Scoped identity resolution at $\mathsf{MNO}$}
\STATE \quad $\mathsf{LEA} \rightarrow \mathsf{MNO}$: $(\EncEid, \mathsf{EID}',\pi)$
\STATE \quad $\mathsf{MNO}$ verifies that $\EncEid$ is the escrow returned under $S$; resolve $\mathsf{EID}' \mapsto \textsf{KYC}$, else $\bot$
\STATE \quad \textbf{Return} $(\mathsf{EID}',\pi)$ to the lawful process.
\end{algorithmic}
\end{algorithmbreak}

\section{Game-based Definition}
\label{app:gamebaseddefinition}

We define the security of ZK-eSIM through two families of games:
provisioning-session unlinkability and unforgeability. The former captures
whether an infrastructure entity can link two accepted sessions to the
same eUICC. The latter captures whether an adversary can make an honest
MNO or SM-DP+ accept an unauthorised request.

\begin{definition}[Provisioning-Session Unlinkability]
For a PPT adversary $\mathcal{A}$ and an entity
$\mathcal{E}\in\{\MNO,\SMDP^+,\PCA\}$, define
$$
\mathsf{Adv}^{\mathsf{UNLINK}\text{-}\mathcal{E}}_{\Pi,\mathcal{A}}(\lambda)
:=
\Big|
\Pr[
\UNLINK_{\Pi,\mathcal{A}}^{\mathcal{E}}(\lambda)=1
]
-\tfrac{1}{2}
\Big|.
$$
We say that $\Pi$ provides \emph{provisioning-session unlinkability}
if this advantage is negligible in $\lambda$ for every PPT
$\mathcal{A}$ and every
$\mathcal{E}\in\{\MNO,$\\$\SMDP^+,\PCA\}$.
\end{definition}

\begin{definition}[Unforgeability]
Let
$\mathsf{UF}^{\MNO}_{\Pi,\mathcal{A}}(\lambda)$ and
$\mathsf{UF}^{\SMDP^+}_{\Pi,\mathcal{A}}$\\$(\lambda)$ denote the
order-side and download-side unforgeability games in Fig.~\ref{fig:uf}.
Define
$$
\mathsf{Adv}^{\mathsf{UF}\text{-}\MNO}_{\Pi,\mathcal{A}}(\lambda)
:=
\Pr[
\mathsf{UF}^{\MNO}_{\Pi,\mathcal{A}}(\lambda)=1
],
$$
and
$$
\mathsf{Adv}^{\mathsf{UF}\text{-}\SMDP^+}_{\Pi,\mathcal{A}}(\lambda)
:=
\Pr[
\mathsf{UF}^{\SMDP^+}_{\Pi,\mathcal{A}}(\lambda)=1
].
$$
We say that $\Pi$ is \emph{unforgeable} if both advantages are negligible
in $\lambda$ for every PPT adversary $\mathcal{A}$.
\end{definition}

\begin{figure}[H]
\centering
\renewcommand{\arraystretch}{1}
\fbox{
\begin{minipage}{0.96\linewidth}
\noindent\uline{Game}
$\UNLINK^{\mathcal{E}}_{\Pi,\mathcal{A}}(\lambda)$
for $\mathcal{E}\in\{\MNO,\SMDP^+,\PCA\}$
\sqcircle\; $\pp\leftarrow\Setup(1^\lambda)$.
Initialise $L_{\mathrm{auth}},L_{\mathrm{spent}}\leftarrow\emptyset$.
Let $(U_0,\EID_0)$ and $(U_1,\EID_1)$ be two eligible registered
devices with $\EID_0\neq\EID_1$.
\sqcircle\; $\mathcal{A}$ chooses $(U_0,\EID_0)$ and $(U_1,\EID_1)$.
The challenger samples $\mathsf{b}\xleftarrow{\$}\{0,1\}$ and defines
\[
\mathrm{idx}_s(\mathsf{b}) :=
\begin{cases}
0 & \text{if } \mathsf{b}=0,\\
s-1 & \text{if } \mathsf{b}=1,
\end{cases}
\quad s\in\{1,2\}.
\]
\sqcircle\; For each $s\in\{1,2\}$, the challenger runs an accepted
provisioning session for
$(U_{\mathrm{idx}_s(\mathsf{b})},\EID_{\mathrm{idx}_s(\mathsf{b})})$
and gives $\mathcal{A}$ the corresponding entity view
$\View^{\mathcal{E}}_s$.
\sqcircle\; The views are defined as follows, where
$\Hpid_s:=H'(\pid_s)$ and
$h_{\mathit{cert},s}:=H''(\PCert_{U,s})$:
$$
\View^{\MNO}_s :=
(\pid_s,\EncEid_s,\pi_{\mathrm{req},s},\PCert_{U,s},
h_{\mathit{cert},s},
$$
$$
\Hpid_s,\sigma_{\cred,s},T_{i,s},
L_{\mathrm{auth}}).
$$
$$
\View^{\SMDP^+}_s :=
(\PCert_{U,s},h_{\mathit{cert},s},\Hpid_s,T_{i,s},\sigma_{\cred,s},
\pi_{\mathrm{inc},s},
$$
$$
\mathsf{root}_{\mathrm{auth},s},
\sigma^{\mathsf{root}}_{\MNO,s},I_{t,s},N_{S,s},N_{U,s},
L_{\mathrm{spent}}).
$$
$$
\View^{\PCA}_s :=
(\pk_{U,s},\pi_{\mathrm{bind},s},\PCert_{U,s},\mathsf{expiry}_s).
$$
\sqcircle\; $\mathcal{A}$ receives
$(\View^{\mathcal{E}}_1,\View^{\mathcal{E}}_2)$ and outputs
$\mathsf{b}'\in\{0,1\}$.
\sqcircle\; The game outputs $1$ iff $\mathsf{b}'=\mathsf{b}$.
\end{minipage}
}
\caption{Provisioning-session unlinkability game for ZK-eSIM.}
\label{fig:unlink-unified}
\end{figure}

\begin{figure}[H]
\centering
\renewcommand{\arraystretch}{1}
\fbox{
\begin{minipage}{0.96\linewidth}
\noindent\uline{Game}
$\mathsf{UF}^{\mathcal{E}}_{\Pi,\mathcal{A}}(\lambda)$
for $\mathcal{E}\in\{\MNO,\SMDP^+\}$

\sqcircle\; $\pp\leftarrow\Setup(1^\lambda)$.
Initialise
$\mathcal{Q}_{\mathsf{reg}},
 \mathcal{Q}_{\mathsf{chal}},
 \mathcal{Q}_{\mathsf{auth}}\leftarrow\emptyset$
and $L_{\mathrm{spent}}\leftarrow\emptyset$.
All authorities are honest; $\mathcal{A}$ controls users and eUICCs.

\sqcircle\; $\mathcal{A}$ may adaptively query:
\[
\mathcal{O}_{\mathsf{Register}},\quad
\mathcal{O}_{\mathsf{Challenge}},\quad
\mathcal{O}_{\mathsf{Authorize}},\quad
\mathcal{O}_{\mathsf{Provision}}.
\]
These oracles run the corresponding honest ZK-eSIM procedures and record
registered EIDs, issued MNO nonces, issued authorisation tuples
$(\Hpid,\Ti,\sigma_{\cred},h_{\mathit{cert}})$, and spent tokens.

\sqcircle\; If $\mathcal{E}=\MNO$, $\mathcal{A}$ outputs
$(x^*,\PCert_U^*,\pi_{\mathrm{req}}^*)$, where
$$
x^*=(\pk_{\MNO},\pk_{\LEA},\pk_U^*,
\mathsf{nonce}^*,\pid^*,\EncEid^*).
$$

\sqcircle\; If $\mathcal{E}=\SMDP^+$, $\mathcal{A}$ outputs
$$
(\PCert_U^*,\Hpid^*,\Ti^*,\sigma_{\cred}^*,
\pi_{\mathrm{inc}}^*,\mathsf{root}_{\mathrm{auth}}^*,
\sigma^{\mathsf{root}*}_{\MNO}).
$$
Set $h_{\mathit{cert}}^*\leftarrow H''(\PCert_U^*)$.

\sqcircle\; $\mathcal{A}$ wins iff the honest verifier for role
$\mathcal{E}$ accepts an unauthorised request: for $\mathcal{E}=\MNO$,
an accepted request not backed by a fresh MNO challenge, a registered
credential, and a valid same-eUICC binding; for
$\mathcal{E}=\SMDP^+$, an accepted download not backed by valid
MNO-issued authorisation material for
$(\Hpid^*,h_{\mathit{cert}}^*)$, or using a token already in
$L_{\mathrm{spent}}$.

\sqcircle\; The game outputs $1$ iff $\mathcal{A}$ wins.

\end{minipage}
}
\caption{Unforgeability game for ZK-eSIM.}
\label{fig:uf}
\end{figure}

\noindent Figures~\ref{fig:unlink-unified} and~\ref{fig:uf} formalise
the game-based unlinkability and unforgeability properties used in
Sec.~\ref{sec:security}, respectively.

\noindent \textbf{Unlinkability} (Fig.~\ref{fig:unlink-unified}).
The challenge bit determines whether two accepted provisioning sessions
come from the same registered eUICC or from two different registered
eUICCs. The adversary observes only the selected infrastructure entity's
view. The protocol is unlinkable if no PPT adversary can distinguish
these two cases with non-negligible advantage. Timing and packet-size
metadata are excluded from the formal view, consistent with the threat
model.

\noindent \textbf{Unforgeability} (Fig.~\ref{fig:uf}).
The adversary may obtain honest registrations, MNO challenges,
authorisations, and provisioning transcripts through the four learning
oracles. For the MNO side, a successful forgery is an accepted request
that is not backed by a registered eligibility credential, a fresh MNO
challenge, and a valid proof binding the same eUICC to the credential,
pseudonym, escrow ciphertext, and pseudonym certificate. For the SM-DP+
side, a successful forgery is an accepted download that is not backed by
MNO-issued authorisation material bound to the same
$(\Hpid,h_{\mathit{cert}})$ tuple, or that reuses a token already recorded
in $L_{\mathrm{spent}}$. Therefore, a fresh $\Hpid$ generated by a
legitimate registered device with a fresh MNO challenge is not a forgery.

\section{Security Proof}
\label{app:securityproof}

\subsection{Proof of Theorem~\ref{thm:unlink}}

We prove the theorem by a standard hybrid argument. Let
$\mathsf{Game}_i$ denote the $i$-th hybrid, and let $\mathsf{Win}_i$
be the event that $\mathcal{A}$ outputs the correct challenge bit
$\mathsf{b}'=\mathsf{b}$. We present the proof for
$\mathcal{E}=\MNO$; the case $\mathcal{E}=\SMDP^+$ follows by the same
argument using $\View^{\SMDP^+}_s$ from Fig.~\ref{fig:unlink-unified},
and the case $\mathcal{E}=\PCA$ follows from the zero-knowledge of
$\pi_{\mathrm{bind}}$ and ephemeral certificate indistinguishability.

\noindent \textbf{Game$_0$: Real game.}
This is the real unlinkability game
$\UNLINK^{\MNO}_{\Pi,\mathcal{A}}$
\\$(\lambda)$ in
Fig.~\ref{fig:unlink-unified}. The challenger samples
$\mathsf{b}\xleftarrow{\$}\{0,1\}$, executes two accepted provisioning
sessions for the devices determined by $\mathrm{idx}_s(\mathsf{b})$,
and returns
$$
\View^{\MNO}_s =
(\pid_s,\EncEid_s,\pi_{\mathrm{req},s},\PCert_{U,s},
$$
$$
h_{\mathit{cert},s},\Hpid_s,\sigma_{\cred,s},{\Ti}_{s},L_{\mathrm{auth}}).
$$
By definition,
$$
\mathsf{Adv}^{\mathsf{UNLINK}\text{-}\MNO}_{\Pi,\mathcal{A}}(\lambda)
=
\Big|
\Pr[\mathsf{Win}_0]-\tfrac{1}{2}
\Big|.
$$

\noindent \textbf{Game$_1$: Simulated zero-knowledge proofs.}
Replace each real proof
$\pi_{\mathrm{req},s}\leftarrow\mathsf{ZK.Prove}(crs,x_s,w_s)$
with a simulated proof generated by the NIZK simulator. By
zero-knowledge,
$$
\Big|
\Pr[\mathsf{Win}_1]-\Pr[\mathsf{Win}_0]
\Big|
\leq
\mathsf{Adv}^{\mathsf{ZK}}_{\mathcal{B}_{\mathsf{ZK}}}(\lambda).
$$

\noindent \textbf{Game$_2$: Randomised key.}
Replace
$K_{\pid,s}\leftarrow\KDF(\sk_b,\mathsf{nonce}_s)$
with a uniform key
$K'_s\xleftarrow{\$}\{0,1\}^{\lambda}$, and compute
$\pid_s\leftarrow$\\$\PRF_{K'_s}(\EID_{\mathrm{idx}_s(\mathsf{b})})$.
By the security of $\KDF$,
$$
\Big|
\Pr[\mathsf{Win}_2]-\Pr[\mathsf{Win}_1]
\Big|
\leq
\mathsf{Adv}^{\mathsf{KDF}}_{\mathcal{B}_{\mathsf{KDF}}}(\lambda).
$$

\noindent \textbf{Game$_3$: Random pseudonyms.}
Replace
$\pid_s\leftarrow\PRF_{K'_s}(\EID_{\mathrm{idx}_s(\mathsf{b})})$
with an independently uniform
$\pid_s\xleftarrow{\$}\{0,1\}^{\lambda}$, except with negligible
collision probability. By PRF security,
$$
\Big|
\Pr[\mathsf{Win}_3]-\Pr[\mathsf{Win}_2]
\Big|
\leq
\mathsf{Adv}^{\mathsf{PRF}}_{\mathcal{B}_{\PRF}}(\lambda).
$$

\noindent \textbf{Game$_4$: Random hash outputs.}
Replace
$\Hpid_s\leftarrow H'(\pid_s)$ and
$h_{\mathit{cert},s}\leftarrow H''(\PCert_{U,s})$
with independent uniform values, unless the adversary has queried the
corresponding random-oracle inputs. Since $\pid_1,\pid_2$ are uniform
and independent in Game$_3$, the probability of hitting either
pseudonym input is at most $2q_H/2^\lambda$. Thus,
$$
\Big|
\Pr[\mathsf{Win}_4]-\Pr[\mathsf{Win}_3]
\Big|
\leq
\frac{2q_H}{2^\lambda}.
$$

\noindent \textbf{Game$_5$: Independent pseudonym certificates.}
Replace each $\PCert_{U,s}$ with a fresh pseudonym certificate whose
public metadata is distributed identically but is independent of
$\EID_{\mathrm{idx}_s(\mathsf{b})}$. By ephemeral certificate
indistinguishability,
$$
\Big|
\Pr[\mathsf{Win}_5]-\Pr[\mathsf{Win}_4]
\Big|
\leq
\mathsf{Adv}^{\mathsf{Cert}}_{\mathcal{B}_{\mathsf{Cert}}}(\lambda).
$$

\noindent \textbf{Game$_6$: Simulated escrow ciphertexts.}
Replace
$\EncEid_s\leftarrow\mathsf{Enc}_{\pk_{\LEA}}
(\EID_{\mathrm{idx}_s(\mathsf{b})})$
with
$\EncEid_s\leftarrow\mathsf{Enc}_{\pk_{\LEA}}(0^\lambda)$.
Since the request proofs have already been simulated, the ciphertext
plaintext is no longer used elsewhere in the visible transcript. By
IND-CPA security,
$$
\Big|
\Pr[\mathsf{Win}_6]-\Pr[\mathsf{Win}_5]
\Big|
\leq
\mathsf{Adv}^{\mathsf{IND}\text{-}\mathsf{CPA}}_{\mathcal{B}_{\mathsf{Enc}}}(\lambda).
$$

In Game$_6$, all fields that may depend on whether the two sessions use
the same eUICC or different eUICCs have been simulated or randomised:
the proofs, pseudonyms, hashed pseudonyms, certificate hashes,
pseudonym certificates, and escrow ciphertexts. The remaining fields,
including $\sigma_{\cred}$, $\Ti$, and $L_{\mathrm{auth}}$, are generated
from fresh session randomness and authorised session-local values, and
therefore have the same distribution for $\mathsf{b}=0$ and
$\mathsf{b}=1$. Hence
$$
\Pr[\mathsf{Win}_6]=\frac{1}{2}.
$$
Applying the triangle inequality over the hybrids gives the bound in
Theorem~\ref{thm:unlink}. The $\SMDP^+$ view is handled identically,
with $\Hpid$, $h_{\mathit{cert}}$, $\Ti$, $\sigma_{\cred}$, and
$L_{\mathrm{spent}}$ replacing the MNO-side state. For the PCA view,
the only session-visible values are
$(\pk_U,\pi_{\mathrm{bind}},\PCert_U,\mathsf{expiry})$; simulated
$\pi_{\mathrm{bind}}$, fresh $\pk_U$, and ephemeral certificate
indistinguishability remove dependence on the underlying eUICC identity.

\subsection{Proof of Theorem~\ref{thm:uf-mno}}

We prove MNO-side unforgeability for the game in Fig.~\ref{fig:uf}.
The adversary wins only if the honest MNO accepts an adversarial request
$(x^*,\PCert_U^*,\pi_{\mathrm{req}}^*)$ that is not supported by a
registered credential, a fresh MNO challenge, and a valid same-eUICC
binding.

\noindent \textbf{Case 1: No valid witness.}
Suppose the MNO accepts $\pi_{\mathrm{req}}^*$, but there is no witness
satisfying the request relation $R_{\mathrm{req}}$. Then an adversary
$\mathcal{B}_{\mathsf{Sound}}$ against NIZK knowledge soundness is
obtained directly: it runs the unforgeability game for $\mathcal{A}$ and,
upon receiving an accepting false proof, outputs
$(x^*,\pi_{\mathrm{req}}^*)$ as its soundness violation. Hence this case
is bounded by
$$
\mathsf{Adv}^{\mathsf{Sound}\text{-}\mathsf{ZK}}_{\mathcal{B}_{\mathsf{Sound}}}(\lambda).
$$

\noindent \textbf{Case 2: Credential forgery.}
By knowledge soundness, from any accepting proof with a valid witness one
can extract, among other values, an eligibility credential
$\sigma_{\EID^*}$ and an identifier $\EID^*$. If
$\EID^*\notin\mathcal{Q}_{\mathsf{reg}}$ but
$\sigma_{\EID^*}$ verifies under $\pk_{\MNO}$, then
$\mathcal{A}$ has produced a credential that was never issued by the
registration oracle. This gives a credential-forgery adversary
$\mathcal{B}_{\mathsf{Cred}}$ by forwarding registration queries to the
credential-issuance oracle and outputting
$(\EID^*,\sigma_{\EID^*})$ when $\mathcal{A}$ succeeds. The probability
of this case is bounded by
$$
\mathsf{Adv}^{\mathsf{UF}\text{-}\mathsf{Cred}}_{\mathcal{B}_{\mathsf{Cred}}}(\lambda).
$$

\noindent \textbf{Case 3: Freshness violation.}
If the extracted witness is valid and the credential was registered, then
the request can still be a forgery only if the MNO challenge is missing,
reused, or not freshly issued. Since MNO challenges are sampled from
$\{0,1\}^{\lambda}$ and consumed after successful use, the probability
that $\mathcal{A}$ guesses an unissued fresh challenge is at most
$$
\frac{q_N}{2^\lambda},
$$
where $q_N$ is the number of challenge queries. Reuse is rejected by the
stateful freshness check.

\noindent \textbf{Case 4: Binding or hash inconsistency.}
It remains to consider the case where the credential is registered and
the nonce is fresh, but the accepted request binds inconsistent values.
If the proof verifies for inconsistent values, this is already a
soundness violation. Otherwise, the only way to make two distinct
pseudonyms or request bindings map to the same accepted hashed
pseudonym is to find a collision in $H'$. Likewise, if the credential
binding through $m_{\EID}=H(\EID\Vert \sk_b)$ is changed while preserving
verification, this gives a collision in $H$. Therefore this case is
bounded by
$$
\mathsf{Adv}^{\mathsf{CR}}_{H}(\lambda)
+
\mathsf{Adv}^{\mathsf{CR}}_{H'}(\lambda).
$$

The four cases cover every instance of accepting an unauthorised MNO-side request.
Thus,
$$
\mathsf{Adv}^{\mathsf{UF}\text{-}\MNO}_{\Pi,\mathcal{A}}(\lambda)
\leq
\mathsf{Adv}^{\mathsf{Sound}\text{-}\mathsf{ZK}}_{\mathcal{B}_{\mathsf{Sound}}}(\lambda)
+
\mathsf{Adv}^{\mathsf{UF}\text{-}\mathsf{Cred}}_{\mathcal{B}_{\mathsf{Cred}}}(\lambda)
+
\mathsf{Adv}^{\mathsf{CR}}_{H}(\lambda)
$$
$$
+
\mathsf{Adv}^{\mathsf{CR}}_{H'}(\lambda)
+
\frac{q_N}{2^\lambda}.
$$

\begin{table*}[!t]
\centering
\small
\renewcommand{\arraystretch}{1.12}
\caption{Analysis of the erosion of privacy and the sustained security and privacy safeguards across different collusion scenarios.}
\label{tab:collusion-degradation}
\scalebox{0.9}{
\begin{tabular}{
  p{0.12\textwidth}
  p{0.36\textwidth}
  p{0.25\textwidth}
  p{0.29\textwidth}
}
\toprule
\textbf{Case} &
\textbf{Additional information revealed} &
\textbf{Not Supported design goals (\S\ref{sec:designgoals}}) &
\textbf{Supported design goals (\S\ref{sec:designgoals})} \\
\midrule

No collusion &
No additional information &
None &
DG1--DG6; provisioning correctness \\

\midrule

MNO--SM-DP+ &
Subscriber identity and EID linked to profile-order and provisioning
records &
DG2, DG4--DG6 &
DG1, DG3; provisioning correctness \\

\midrule

MNO--PCA &
Subscriber identity and EID linked to pseudonym-certificate issuance
and the corresponding profile request &
DG2, DG4--DG6 &
DG1, DG3; provisioning correctness \\

\midrule

SM-DP+--PCA &
Pseudonym-certificate issuance linked to the corresponding provisioning
session &
None &
DG1--DG6; provisioning correctness \\

\midrule

Full collusion &
Subscriber identity and EID linked to the complete infrastructure
record of the provisioning session &
DG2, DG4--DG6 &
DG1, DG3; provisioning correctness \\

\bottomrule
\end{tabular}
}
\label{tab:collusion-degradation}
\end{table*}

\subsection{Proof of Theorem~\ref{thm:uf-smdp}}

We prove SM-DP+-side unforgeability for the game in Fig.~\ref{fig:uf}.
The adversary wins only if the honest SM-DP+ accepts adversarial
download material that is not backed by MNO-issued authorisation
material for the same $(\Hpid,h_{\mathit{cert}})$ tuple, or if it accepts
a token that has already been spent.

Let
$$
h_{\mathit{cert}}^* \leftarrow H''(\PCert_U^*).
$$

\noindent \textbf{Case 1: Authorisation credential forgery.}
Suppose SM-DP+ accepts $\sigma_{\cred}^*$, but there is no matching
MNO-issued authorisation tuple
$$
(\Hpid^*,\Ti^*,\sigma_{\cred}^*,h_{\mathit{cert}}^*)
\in\mathcal{Q}_{\mathsf{auth}}.
$$
If $\sigma_{\cred}^*$ verifies as an MNO authorisation on the required
message containing $(\Hpid^*,h_{\mathit{cert}}^*)$, then
$\mathcal{A}$ yields an EUF-CMA forgery against the signature scheme
used for $\sigma_{\cred}$. This case is bounded by
$$
\mathsf{Adv}^{\mathsf{EUF}\text{-}\mathsf{CMA}}_{\sigma_{\cred}}(\lambda).
$$

\noindent \textbf{Case 2: Token forgery.}
Similarly, if $\Ti^*$ verifies as a valid one-time token for the required
message containing $(\Hpid^*,h_{\mathit{cert}}^*)$ but was not issued by
the MNO authorisation oracle, then $\mathcal{A}$ gives an EUF-CMA
forgery against the token-signing scheme. This case is bounded by
$$
\mathsf{Adv}^{\mathsf{EUF}\text{-}\mathsf{CMA}}_{\Ti}(\lambda).
$$

\noindent \textbf{Case 3: Replay or double spending.}
If the tuple was honestly issued, then acceptance is still unauthorised
when $\Ti^*\in L_{\mathrm{spent}}$ before the adversarial attempt is
processed. This is exactly a failure of the stateful spent-token check,
whose probability is
$$
\mathsf{Adv}^{\mathsf{DS}}(\lambda).
$$

\noindent \textbf{Case 4: Hash rebinding.}
The remaining possibility is that $\mathcal{A}$ reuses honestly issued
authorisation material but attaches it to different underlying values. If
a different pseudonym $\pid\neq\pid^*$ yields the same $\Hpid$, then this
is a collision in $H'$. If a different certificate
$\PCert_U\neq\PCert_U^*$ yields the same $h_{\mathit{cert}}$, then this
is a collision in $H''$. Hence this case is bounded by
$$
\mathsf{Adv}^{\mathsf{CR}}_{H'}(\lambda)
+
\mathsf{Adv}^{\mathsf{CR}}_{H''}(\lambda).
$$

The above cases exhaust all ways in which SM-DP+ can accept an
unauthorised or replayed download-side request. Therefore,
$$
\mathsf{Adv}^{\mathsf{UF}\text{-}\SMDP^+}_{\Pi,\mathcal{A}}(\lambda)
\leq
\mathsf{Adv}^{\mathsf{EUF}\text{-}\mathsf{CMA}}_{\sigma_{\cred}}(\lambda)
+
\mathsf{Adv}^{\mathsf{EUF}\text{-}\mathsf{CMA}}_{\Ti}(\lambda)
$$
$$
+
\mathsf{Adv}^{\mathsf{CR}}_{H'}(\lambda)
+
\mathsf{Adv}^{\mathsf{CR}}_{H''}(\lambda)
+
\mathsf{Adv}^{\mathsf{DS}}(\lambda).
$$

\section{Collusion Analysis}

\label{app:collusion}

We use an entity's \emph{view} to mean the protocol messages, records, logs, and local state available to that entity during normal operation. Two roles collude when they share these observations and analyse them together. This includes both explicit sharing between separate organisations and deployments in which one organisation operates multiple roles and can access the records of each. Colluding parties remain honest-but-curious: they follow ZK-eSIM as specified, but may combine all information available to them to infer subscriber identity or link provisioning sessions. As shown in Table \ref{tab:collusion-degradation}, we compare no collusion, pairwise collusion, and full collusion among the MNO, SM-DP+, and PCA, and examine what additional information is exposed in each case and which design goals in \S\ref{sec:designgoals} remain satisfied.


\paragraph{Case 1: No collusion.}
This is the baseline honest-but-curious model of ZK-eSIM. The MNO knows the subscriber and EID from Phase~0.a, but subsequent provisioning uses fresh $\pid$, $PCert_U$, $Hpid$, and one-time authorisation material instead of a persistent provisioning identifier. The SM-DP+ and PCA therefore receive no subscriber identity or EID through their normal protocol views, and separate provisioning sessions remain unlinkable under the analysis of Sec.~8. DG1--DG6 and provisioning correctness hold in this case.
\paragraph{Case 2: Pairwise collusion.}
The privacy degradation depends on which two parties share their records.












\smallskip
\noindent\uline{\emph{MNO--SM-DP+ collusion.}} The MNO contributes the subscriber and EID associated with the session, while the SM-DP+ contributes the corresponding profile-order and provisioning records. The shared $Hpid$ connects the MNO's $\OrderProfile$ state to the SM-DP+ state, allowing the subscriber and EID to be associated with the corresponding ICCID and provisioning session. Since the same EID can identify repeated sessions, DG2, DG4, and DG5 no longer hold. The intended MNO--LEA separation in DG6 is also bypassed. DG1, DG3, and provisioning correctness remain.

\smallskip
\noindent\uline{\emph{MNO--PCA collusion.}} The PCA observes $PCert_U$ when issuing the short-lived certificate, and the MNO observes the same certificate in $\ZKRequest$. Combining these records associates the certificate issuance and profile request with the subscriber and EID shared by the MNO. Repeated sessions can then be associated through that EID, so DG2, DG4, and DG5 fail, and DG6 is bypassed. The SM-DP+'s ICCID and provisioning records remain outside this combined view. DG1, DG3, and provisioning correctness remain.

\smallskip
\noindent\uline{\emph{SM-DP+--PCA collusion.}} The PCA records the issuance of $\PCert_U$, while the SM-DP+ observes this $\PCert_U$ when it is used for provisioning. If SMDP+ and PCA collude, by sharing these records, the SM-DP+ and PCA can \textit{link} a certificate issuance to the specific provisioning session. However, neither entity has the MNO's subscriber--EID record, so this linkage does not reveal the subscriber or EID. Since a fresh $\PCert_U$, $\Hpid$, and authorisation material are generated for each session, their combined records do not provide a persistent identifier for linking different provisioning sessions. DG1--DG6 and provisioning correctness therefore remain, while the privacy separation between certificate issuance and provisioning is lost.

\paragraph{Case 3: Full infrastructure collusion.}
When the MNO, SM-DP+, and PCA share their records, the common session values connect all provisioning stages. $\PCert_U$ connects $\CertInit$ with $\ZKRequest$ and $\MutualAuthAndProvision$, while $\Hpid$ connects the MNO's $\OrderProfile$ state with the SM-DP+ provisioning state. Together with the subscriber and EID shared by the MNO, the parties can associate the subscriber with the pseudonym certificate, profile request, profile order, ICCID, and provisioning session. DG2, DG4, DG5, and DG6 therefore fail. DG1, DG3, and provisioning correctness remain.

\paragraph{Retained security and correctness.}
The collusion considered here changes the information available to the infrastructure parties, but not their protocol behaviour. The checks in $\ZKRequest$, $\OrderProfile$, and $\MutualAuthAndProvision$ are still performed, including authorisation verification, one-time-token checking, authenticated key establishment, and profile confidentiality and integrity. Consequently, DG1 and DG3 remain satisfied in all cases. Provisioning correctness is also unchanged: an authorised profile is installed on the intended eUICC, the corresponding token is accepted only once, and the MNO and SM-DP+ maintain consistent records.

\section{Open Science}

In accordance with open science principles, we make our code available through the link: \url{https://github.com/NaivEPoi/ZK-eSim}. In this repository, we provide: (i) Our modified GSMA compliant SM-DP+ server - adapted from \cite{osmocom_osmocompysim_2026} to accommodate ZK-eSIM; (ii) A LPA implementation \cite{estkme_group_github_2025} extended to provide alternative RSP flows for our scheme; (iii) A Java Card applet that interfaces with the previous artefacts to complete the full RSP process flow. These artefacts enable the complete reproduction of our results and the extension of our work to further enhance privacy in the eSIM ecosystem.

\section{Ethical Considerations}
ZK-eSIM addresses a concrete privacy risk in eSIM provisioning: persistent device identifiers and certificate material can enable cross-session and cross-operator tracking. The main benefit of our work is therefore defensive, namely reducing unnecessary identifier exposure while preserving profile-delivery security and accountable traceability. At the same time, stronger provisioning privacy could be misused if deployed without operational safeguards. We mitigate these risks by requiring valid operator-issued eligibility credentials, one-time tokens, replay checks, and short-lived pseudonym certificates, and by designing traceability so that no single party can deanonymise a user unilaterally; identity recovery requires a scoped legal process and joint LEA--MNO cooperation. Our evaluation uses a controlled test-bed with test eUICCs and modified open-source RSP components; we do not access production carrier infrastructure, disclose carrier secrets, collect subscriber telemetry, or use human-subject data. Any real deployment should undergo operator, standards-body, and jurisdiction-specific legal review.
\end{document}